\documentclass[
    reprint,
    aps,
    pra,
    twocolumn,
    superscriptaddress,
    amssymb,
    footinbib,
    tightenlines,
    longbibliography
]{revtex4-2}
\usepackage{physics}
\usepackage{tabularx,booktabs}
\usepackage{mathtools}
\usepackage{bbold}
\AtBeginDocument{%
  \heavyrulewidth=.08em
  \lightrulewidth=.05em
  \cmidrulewidth=.03em
  \belowrulesep=.65ex
  \belowbottomsep=0pt
  \aboverulesep=.4ex
  \abovetopsep=0pt
  \cmidrulesep=\doublerulesep
  \cmidrulekern=.5em
  \defaultaddspace=.5em
}
\usepackage{csquotes}
\usepackage{comment}
\usepackage{graphicx}
\usepackage{dcolumn}
\usepackage{bm}
\usepackage{overpic}
\usepackage[colorlinks,bookmarks=false,citecolor=blue,linkcolor=red,urlcolor=blue]{hyperref}
\usepackage[capitalise]{cleveref}
\Crefname{figure}{Fig.}{Figs.}
\Crefname{section}{Sec.}{Secs.}
\usepackage{cmds}

\usepackage{stackengine,xcolor,scalerel}
\definecolor{mygreen}{rgb}{0,0.5,0}
\definecolor{myblue}{rgb}{0,0,0.75}

\graphicspath{{figures/}}
\newcolumntype{P}[1]{>{\centering\arraybackslash}p{#1}}

\newcommand{\logtwo}{\mathrm{log}_{\rm \scriptscriptstyle 2}}
\newcommand{\Tbase}{T_{\rm \scriptscriptstyle base}}
\newcommand{\nummpithreads}{n_{\rm \scriptscriptstyle MPI-threads}}

\usepackage{orcidlink}
\newcommand{\orcidgiuseppe}{\orcidlink{0000-0002-7280-445X}}
\newcommand{\orciddaniel}{\orcidlink{0000-0001-7658-3546}}
\newcommand{\orcidpietro}{\orcidlink{0000-0001-5279-7064}}
\newcommand{\orcidsimone}{\orcidlink{0000-0002-8882-2169}}
\newcommand{\orcidgiovanni}{\orcidlink{0000-0002-9073-8978}}
\newcommand{\orcidmarco}{\orcidlink{0000-0002-4544-3513}}
\newcommand{\orcidpeter}{\orcidlink{0009-0009-7020-7246}}

\newcommand{\DFA}{\affiliation{Dipartimento di Fisica e Astronomia ``G. Galilei'', Università di Padova, I-35131 Padova, Italy.}}
\newcommand{\INFNBA}{\affiliation{Istituto Nazionale di Fisica Nucleare (INFN), Sezione di Bari, I-70125 Bari, Italy.}}
\newcommand{\PQTC}{\affiliation{Padua Quantum Technologies Research Center, Università degli Studi di Padova}}
\newcommand{\INFNPD}{\affiliation{Istituto Nazionale di Fisica Nucleare (INFN), Sezione di Padova, I-35131 Padova, Italy.}}
\newcommand{\bari}{\affiliation{Dipartimento di Fisica, Università di Bari, I-70126 Bari, Italy.}}
\newcommand{\uulm}{\affiliation{Institute for Complex Quantum Systems, Ulm University, D-89069 Ulm, Germany}}

\usepackage[normalem]{ulem}

\begin{document}
\title{Tensor Networks for Lattice Gauge Theories beyond one dimension:\\ a Roadmap}

\author{Giuseppe Magnifico\orcidgiuseppe}\bari \INFNBA \DFA 
\author{Giovanni Cataldi\orcidgiovanni} \DFA \PQTC \INFNPD
\author{Marco Rigobello\orcidmarco} \DFA \PQTC \INFNPD 
\author{\\Peter~Majcen \orcidpeter}\DFA \PQTC \INFNPD
\author{Daniel Jaschke\orciddaniel} \DFA \PQTC \INFNPD \uulm
\author{Pietro Silvi\orcidpietro} \DFA \PQTC \INFNPD
\author{Simone Montangero\orcidsimone} \DFA \PQTC \INFNPD

\date{\today}
\begin{abstract}
  Tensor network methods are a class of numerical tools and algorithms to study many-body quantum systems in and out of equilibrium, based on tailored variational wave functions. 
  They have found significant applications in simulating lattice gauge theories approaching relevant problems in high-energy physics.
  Compared to Monte Carlo methods, they do not suffer from the sign problem, allowing them to explore challenging regimes such as finite chemical potentials and real-time dynamics.
  Further development is required to tackle fundamental challenges, such as accessing continuum limits or computations of large-scale quantum chromodynamics. 
  In this work, we review the state-of-the-art of Tensor Network methods and discuss a possible roadmap for algorithmic development and strategies to enhance their capabilities and extend their applicability to open high-energy problems.
  We provide tailored estimates of the theoretical and computational resource scaling for attacking large-scale lattice gauge theories.
\end{abstract}

\maketitle
Gauge theories play a role of paramount importance in our understanding and description of the fundamental constituents of matter, their spectrum, and their interactions.
At low energies, gauge theories characterize a large variety of collective phases of matter and phenomena, such as ferromagnetic superconductivity, spin liquids, topological order, and the fractional quantum Hall effect \cite{Kleinert1989GaugeFieldsCondensed, Fradkin2013FieldTheoriesCondensed, Ichinose2014LatticeGaugeTheory}.
At high energies, as elegantly summarised in the Standard Model of particle physics, they are at the heart of the microscopical description of the building blocks of our universe, i.e. quarks, leptons and their interactions mediated by gauge bosons \cite{Peskin1995IntroductionQuantumField, Schwartz2013QuantumFieldTheory}.

A powerful approach to studying and simulating gauge theories in nonperturbative regimes lies in Lattice Gauge Theories (LGTs), in which the matter and the gauge degrees of freedom are discretized and regularized on a finite lattice.  LGTs were originally introduced by Wilson to encode Quantum Chromodynamics (QCD) on a lattice, as a model for quark confinement beyond the perturbative regime \cite{Wilson1974ConfinementQuarks, Kogut1979IntroductionLatticeGauge}. In LGTs, matter and antimatter fermionic fields are defined on lattice sites, whereas the gauge fields live on the links connecting nearest-neighbor sites. This approach has opened the doors to the application of powerful numerical methods, such as Monte Carlo (MC), to the simulations of LGTs on classical computers \cite{Creutz1983MonteCarloComputations}.
In the last decades, MC methods have provided a wide variety of significant results on LGTs in the context of high-energy physics, such as phases diagrams at equilibrium, characterization of the quark-gluon plasma, precise determination of the masses of quarks, mesons, and baryons, hadronic and nucleon form factors, hadronic spectra, and predictions for dark-matter models \cite{Davoudi2022ReportSnowmass2021}. Furthermore, MC simulations of LGTs currently represent a powerful numerical tool to predict and interpret data from multiple large-scale high-energy experiments, such as the ones performed at the Large Hadron Collider (LHC).

Despite their impressive success, MC sampling methods are limited in some regimes of parameters of LGTs, such as in the presence of finite baryon chemical potentials, topological $\theta$-terms, or for simulating out-of-equilibrium dynamics in real-time. 
In these cases, the notorious sign problems make the MC numerical approach ineffective and inaccurate \cite{Nagata2022FiniteDensityLattice}.

Ranging from these problems, sign-problem-free methods based on Tensor Networks (TNs) have found significant applications in the simulations of LGTs in the last years \cite{Banuls2019TensorNetworksTheir}.
TNs were originally introduced as a class of variational wave functions in the field of quantum many-body physics.
They provide a compressed representation of physical states based on their entanglement content, capable of efficiently reproducing equilibrium properties, such as phase diagrams, and real-time dynamics of interacting quantum systems \cite{Montangero2018IntroductionTensorNetwork, Orus2019TensorNetworksComplex, Banuls2023TensorNetworkAlgorithms}.
Nowadays, they represent one of the state-of-the-art numerical tools for simulating quantum many-body systems, even with strong correlations.

In the context of high-energy physics, TN methods have proven noteworthy achievements in simulating LTGs in (1+1) dimensions, for both Abelian and non-Abelian gauge groups \cite{Byrnes2002DensityMatrixRenormalization, Silvi2014LatticeGaugeTensor, Pichler2016RealTimeDynamics,Silvi2019TensorNetworksAnthology, Funcke2020TopologicalVacuumStructure, Buyens2014MatrixProductStates, Rico2014TensorNetworksLattice, Haegeman2015GaugingQuantumStates, Silvi2017FiniteDensityPhase, Buyens2017FiniteRepresentationApproximation, Banuls2017DensityInducedPhase, Ercolessi2018PhaseTransitionsZn, Magnifico2019SymmetryProtectedTopological, Magnifico2019ZnGaugeTheories, Sala2018GaussianStatesVariational, Magnifico2020RealTimeDynamics, Banuls2017EfficientBasisFormulation, Rigobello2021EntanglementGeneration11d, Funcke2023ExploringCpViolating, Angelides2023ComputingMassShift, Chanda2023SpectralPropertiesCritical,schmoll2023hamiltonian,hayata2023dense,florio2023mass,osborne2023probing,kebric2024confinement,Belyansky2024PhysRevLett,Papaefstathiou2024realtime,
Calajo2024QuantumManyBodyScarring}. 
They have recently found applications to Abelian LGTs up to (3+1) dimensions \cite{Felser2020TwoDimensionalQuantum, Emonts2023FindingGroundState, Magnifico2021LatticeQuantumElectrodynamics, Knaute2024TransferPEPS, Pradhan2024QEDLadder, Su2024coldatomcollider}, and also in simulating (2+1)-dimensional non-Abelian SU(2) models \cite{Cataldi2023a}.
Despite their effectiveness in these first important applications, further and intensive developments are still required to tackle, with TN methods, high-energy physics problems at the center of current research efforts, such as large-scale non-Abelian LGTs and their continuum limits.

In this work, we present a general overview of TN methods for LGTs, and we discuss a possible roadmap in terms of algorithmic development and strategies to improve TN capabilities, toward the ambitious long-term goal of applying TNs to (3+1)-dimensional QCD.

Importantly, TN methods share a common language with quantum computers and simulators. 
Thus, these developments could also be relevant for encoding, validating, and benchmarking the current and future quantum computations and simulations of LGTs on experimental quantum hardware \cite{Villalonga2019FlexibleHighPerformance, Mathis2020ScalableSimulationsLattice, Zhou2020WhatLimitsSimulation, Huang2021EfficientParallelizationTensor, Haghshenas2022VariationalPowerQuantum, Zohar2021QuantumSimulationLattice, Catterall2022ReportSnowmass2021, Pomarico2023DynamicalQuantumPhase, Funcke2023ReviewQuantumComputing, Mariani2023HamiltoniansGaugeInvariant, Zache2023FermionQuditQuantuma}.

The paper is organized into sections as follows. In \cref{sec_TNs}, we give a general overview of TN methods, particularly the computational complexity of the most used algorithms for ground state computations and real-time dynamics. In \cref{sec_TN_LGTs}, we introduce the main concepts related to LGTs and the main strategies for simulating them with TN methods. In \cref{sec_TN_roadmap}, we present a possible roadmap of algorithmic development and optimization strategies crucial for making the TN approach competitive as a complementary method to MC techniques for simulating challenging LGT models. In \cref{sec_Conclusions}, we draw our conclusions.

\section{\label{sec_TNs}Tensor Networks Overview}
\begin{figure*}
  \includegraphics[width=1\textwidth]{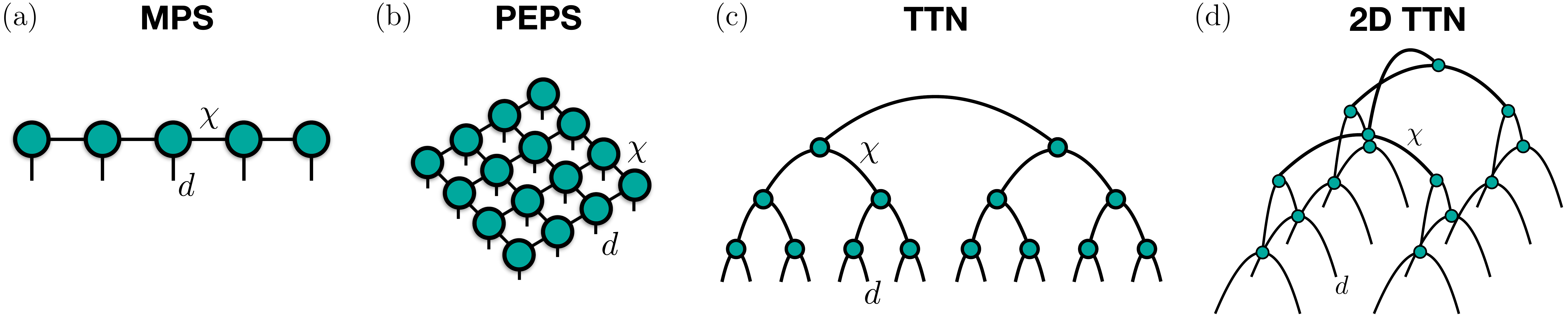}
  \caption{Examples of tensor network structures: (a)~Matrix Product States (MPS); (b)~Projected Entangled Pair States (PEPS); (c)~Tree Tensor Networks (TTN) for an underlying one-dimensional system; (d)~TTN for an underlying two-dimensional square lattice. 
  The physical links with local dimension $d$ and the virtual links with bond dimension $\chi$ are highlighted in all the figures.}
  \label{fig_TN_structures}
\end{figure*}

In this section, we present a general overview of TN methods with a particular focus on Tree Tensor Networks (TTNs), the latter being particularly useful for LTGs in higher dimensions \cite{Felser2020TwoDimensionalQuantum, Magnifico2021LatticeQuantumElectrodynamics, Cataldi2023a}. 
However, this work's main concepts and ideas can be easily generalized and applied to other TN structures.

Consider a quantum many-body (QMB) system defined on a lattice of $N$ sites. The generic site $j$ is described by a $d$-dimensional local Hilbert space $\mathcal{H}_j$, spanned by a local basis of vectors ${\ket{i}}_{1 \leq i \leq d}$.
The quantum states of the whole system live on the total Hilbert space $\mathcal{H}=\mathcal{H}_1 \otimes \mathcal{H}_2 \otimes \dots \otimes \mathcal{H}_N$, that is the tensor product of the local Hilbert space of the lattice sites, with dimension $d^N$.
Then, any pure state of the system $\ket{\psi}$ can be exactly expanded in terms of a complete basis set of $\mathcal{H}$:
\begin{equation}
  \label{eq_exact_expansion}
  \ket{\psi} = \sum_{i_1, i_2, \dots, i_N = 1}^{d} c_{i_1, i_2, ..., i_N}\ket{ i_1, i_2, ..., i_N } \, ,
\end{equation}
where $\ket{ i_1, i_2, ..., i_N } $ represents the tensor product of the local basis vectors, i.e. $\ket{i_1} \otimes\ket{ i_2 } \otimes\ket{ ... } \otimes\ket{ i_N }$. The coefficients of the linear combination $c_{i_1, i_2, ..., i_N}$ are in general complex scalars; their number scales as $d^N$, i.e. they scale exponentially with the system size $N$.

This is a fundamental limitation when solving a QMB problem on a classical computer since an exponential scaling with the number of degrees of freedom implies that the exact representation described in \cref{eq_exact_expansion} is completely unfeasible from a computational and numerical point of view.

Nonetheless, a great variety of natural QMB systems happens to be described by ground and thermal equilibrium states that own little-to-moderate entanglement content.
The physical states of these systems, instead of exploring the exponentially large dimension of the Hilbert space, live in a small corner of it, which can be efficiently targeted and parameterized.
This property is formally described by the entanglement area law, fulfilled by low-energy states of local Hamiltonians \cite{Eisert2010ColloquiumAreaLaws}: the entanglement between a partition of the system and the rest is proportional to the area of the boundary between them, instead of its volume, as happens for the majority of states in the Hilbert space.
Thus, obeying area law implies that the state contains much fewer quantum correlations than expected for a generic (or random) QMB state.
Small corrections to the area law exist for instance close to a transition point for one-dimensional quantum systems (logarithmic corrections).
However, the entanglement remains overall moderate \cite{Calabrese2004EntanglementEntropyQuantum}.
From a theoretical point of view, the entanglement area law has been rigorously proven for (i) one-dimensional gapped local Hamiltonians, where the locality means that a lattice site interacts only with neighboring sites, without two-body all-to-all interactions \cite{Hastings2007AreaLawOne, Kuwahara2020AreaLawNoncritical, Cho2018RealisticAreaLaw}; (ii) for quantum states at thermal equilibrium, independently from the dimensionality of the system \cite{Wolf2008AreaLawsQuantum}.
Even though rigorous proof for QMB systems in higher dimensions is lacking, several numerical and phenomenological shreds of evidence suggest that area law still holds in the presence of local interactions \cite{Masanes2009AreaLaw, Hamza2009ApproximatingtheGroundState, Eisert2010ColloquiumAreaLaws, Kastoryano2019LocalityattheBoundary, Cirac2019MathematicalOpenProblems}.

The area law has important implications on the TN simulation of quantum lattice models: indeed, it is possible to obtain an approximate but efficient representation capable of describing the main properties of these states if the entanglement content is low-to-moderate \cite{Eisert2010ColloquiumAreaLaws}; the condition applies for example to ground-states and first excited states of local Hamiltonians.
TNs give a natural language for this representation, by replacing the complete tensor of rank-$N$ $c_{i_1, i_2, ..., i_N}$ of \cref{eq_exact_expansion} with a chain of smaller tensors interconnected using auxiliary bond indices.
The network keeps a number $N$ of physical indices of dimension $d$ (one for each lattice site), whereas the dimension of the bond indices (called bond dimension) is a control parameter $\chi$ that can be tuned in the numerical simulations and is related to the Schmidt decomposition \cite{Cirac2021MatrixProductStates, Silvi2019TensorNetworksAnthology}.

The key advantage of passing from the exact representation of \cref{eq_exact_expansion} to a TN representation is that the number of parameters in the TN is of the order $O(\mathrm{poly}(d)\mathrm{poly}(N)\mathrm{poly}(\chi))$, e.g., $O(N \chi \max(\chi, d)^2)$ for the TTN. The scaling with the system size is now polynomial and not exponential.
In this way, we obtain an efficient representation of the quantum state in terms of computational complexity.
It is worth noting that the bond dimension $\chi$ determines the degree of entanglement and quantum correlations encoded in the TN, e.g. for $\chi{=}1$ the TN describes a product state (no entanglement), whereas one recovers the exact but inefficient representation in the limit $\chi {\lesssim} d^{N}$.
Tuning $\chi$ properly allows interpolating between these two extreme regimes, efficiently reproducing the entanglement of the quantum state.

The most widely used TN architectures are represented in \cref{fig_TN_structures}.
Matrix product states (MPS) are an established ansatz for one-dimensional systems, in which the structure of the lattice is reproduced with a network of tensors, one for each lattice site \cite{PerezGarcia2007MatrixProductState}.
As shown in \cref{fig_TN_structures}(a), each tensor in the bulk of the network has three indices: one physical leg of dimension $d$ representing the local degrees of freedom, and two virtual legs of dimension $\chi$ connected to the neighboring sites.
In open boundary conditions, the tensors at the boundaries have one trivial leg together with one physical and one virtual leg.
MPS intrinsically satisfies area law and allows for efficient computation of scalar products between two states and physical observables.
Currently, the MPS-based Density Matrix Renormalization Group (DMRG) stands as one of the most consolidated and accurate techniques for numerical simulations of one-dimensional QMB systems as well as quasi-one-dimensional systems such as ladder structures \cite{Schollwoeck2011DensityMatrixRenormalization}.
The computational complexity of this ground state searching algorithm for MPS is of the order $O(N d \chi^3 + N d^{2} \chi^{2})$ or $O(Nd^3\chi^3)$, depending if the algorithm optimizes a single MPS tensor at a time, or two-tensors simultaneously \cite{Schollwoeck2011DensityMatrixRenormalization, Chan2016MatrixProductOperators, CommentScalingA}.
Two-site optimization is usually important to avoid getting stuck in local minima or meta-stable configurations during the energy variational minimization \cite{Gleis2023ControlledBondExpansion}. The subspace expansion is an intermediate approach with the benefits of the two-tensor update and a tunable computational cost in between the two approaches \cite{Silvi2019TensorNetworksAnthology}.

The generalization of the MPS ansatz to two- or higher-dimensional lattices is represented by Projected Entangled Pair State (PEPS) \cite{Cirac2021MatrixProductStates}. In this case, each tensor in the bulk has a physical leg of dimension $d$, and a number of $\chi$-dimensional virtual legs depending on the coordination number of the considered lattice.
For example, this coordination number is four in the case of a two-dimensional square lattice, as shown in \cref{fig_TN_structures}(b). PEPS directly encode in their structure the area law of entanglement, however, their exact contraction is an exponentially hard problem, meaning that PEPS can not be efficiently contracted for numerical computing, e.g. scalar products of states or physical observables \cite{Schuch2007ComputationalComplexityProjected}.
To circumvent this problem, approximate contraction methods have been developed during the last years, and are still at the center of current research efforts \cite{Cirac2021MatrixProductStates}.
But even with exploiting these approximate techniques, the computational complexity for ground state optimization remains quite high, e.g. of the order of $O(Nd^2\chi^8)$ \cite{Eisert2013EntanglementTensorNetwork, Vanderstraeten2022VariationalMethodsContracting}, limiting the maximum reachable bond dimensions (typical values are of the order $\chi \approx 10$).

Another important family of TN ansätze is represented by Tree Tensor Networks (TTN), in which the wave function is decomposed into a hierarchical network of tensors that do not contain internal loops \cite{Shi2006ClassicalSimulationQuantum, Silvi2019TensorNetworksAnthology}.
This way, the network can be efficiently contracted and manipulated in polynomial time.
A particular class of TTN is represented by binary tree tensor networks, reported in \cref{fig_TN_structures}(c)-(d) for one- and two-dimensional lattices.
In these structures, tensors in the lowest layer have two physical legs of dimension $d$ (representing two lattice sites) and a virtual leg of dimension $\chi$, whereas, in the upper layers, they have three virtual legs of dimensions up to $\chi$.
The network intrinsically encodes a renormalization procedure, in which, at each layer, two sites are mapped into a single effective one.
In finite-range models, ground searching algorithms for binary TTN architectures display a numerical complexity of the order $O(N d^2 \chi^2 + N \chi^4)$, see \cite{Silvi2019TensorNetworksAnthology, CommentScalingA}.
This is a much more favorable scaling concerning equivalent algorithms for other TN structures, such as PEPS, that allows reaching quite large values of bond dimensions ($\chi \approx 500$) \cite{Qian2022TreeTensorNetwork}.
The drawback of loopless structures, such as binary TTN, is that the area law may not be explicitly reproduced in dimensions higher than one \cite{Ferris2013AreaLawReal}, which becomes a limiting factor when large systems are addressed.
The convergence and the precision of the numerical results obtained via a variational optimization of TTN can be analyzed and kept under control, e.g. by exploiting the large range of available bond dimensions.
Furthermore, it is possible to explicitly encode the area law of high dimensional systems in the TTN ansatz by introducing an additional layer of independent disentanglers, acting on different couples of lattice sites and connected to the corresponding physical legs.
This process augments the expressive power of TTN, and the resulting ansatz is known as augmented Tree Tensor Network (aTTN) \cite{Felser2021EfficientTensorNetwork}.
The computational complexity of variationally optimizing an aTTN structure, which means optimizing both the tensors and the disentanglers, is of the order
$O(N \chi^{4} d^{4} + N \chi d^{7})$.
We point out that the scaling of the computational costs with the local dimension $d$ is particularly severe in the case of aTTN due to the presence of the disentanglers layer.

Besides variational optimization for ground state searching, the previous TN families can also be exploited to simulate the real-time dynamics of local Hamiltonians via at least four methods \cite{Jaschke2018OpenSourceMatrix}.
One of the most widely used approaches, the Time Evolved Block Decimation (TEBD) algorithm, is based on a Suzuki-Trotter decomposition of the time evolution exponential \cite{Vidal2004EfficientSimulationOne}.
The total evolution time is discretized in small time steps.
The corresponding evolution operator is computed as products of local terms, such as two-body operators, and repeatedly applied to the TN wave function to generate the time-evolved state.
Each application can determine an increase in the bond dimension of the network, so an optimized truncation is needed to maintain an efficient and manageable description of the quantum state.
This truncation reduces the bond dimension back to $\chi$ and is performed through a singular value decomposition that minimizes the distance between the evolved and the truncated state.
In general, the TEBD method allows the simulation of the real-time dynamics for nearest-neighbor or finite-range Hamiltonians; one time-step with an MPS for a one-dimensional system with local interactions comes with a computational cost that is below a two-tensor sweep for the ground-state search algorithm.

Another method for simulating the evolution of quantum states via TN is the Time-Dependent Variational Principle (TDVP), which does not rely on the Suzuki-Trotter decomposition \cite{Haegeman2011TimeDependentVariational, Bauernfeind2020TimeDependentVariational}.
In general, TDVP constrains the time evolution to the specific TN manifold considered, such as MPS or TTN, of a given initial bond dimension \cite{PhysRevA.101.023617}.
This is obtained by projecting the action of the Hamiltonian into the tangent space of the TN manifold and then solving the time-dependent Schr{\"o}dinger equation within this manifold.
This approach automatically preserves the energy and the norm of the quantum states during the time evolution. The TDVP algorithm and the variational ground state search rely both on a set of Krylov vectors and therefore have the same computational scaling for one time step compared to one sweep.

These algorithms represent important and efficient tools for simulating with TN the real-time dynamics of QMB systems.
While equilibrium states satisfy the aforementioned area law, out-of-equilibrium time evolution can generate a linear growth of entanglement.
In this case, the time-evolved state requires an exponential growth of the bond dimension as a function of the total time \cite{Schuch2008EntropyScalingSimulability}.
For this reason, TN methods are currently limited to studying the dynamics for low-to-moderate times, or close-to-equilibrium phenomena \cite{Paeckel2019TimeEvolutionMethods}.
In this framework, further developments are extremely important to avoid or at least mitigate this barrier, by devising new algorithms or optimizing existing strategies \cite{White2018QuantumDynamicsThermalizing, Surace2019SimulatingOutEquilibrium}.

\section{Tensor Networks for Hamiltonian Lattice Gauge Theories}
\label{sec_TN_LGTs}
In the traditional MC approach to LGT, the action of a continuum gauge theory is regularized by working on a finite and discrete Euclidean spacetime | \idest{}, both space and (imaginary) time are discretized \cite{Creutz1983MonteCarloComputations}.
Instead, TN (and quantum) simulations typically rely on the Hamiltonian formalism, where time remains a real, continuous variable while $D$-dimensional space is replaced by a cubic lattice $\Lambda$.
Simultaneously, Hamiltonian LGTs present features that distinguish them from other lattice models commonly simulated via TN \cite{Kogut1979IntroductionLatticeGauge}.
We now discuss these properties in more detail and outline the main steps that have to be taken to exploit TN algorithms in LGT.
We focus on matter-coupled LGTs of the Yang-Mills type, such as those routinely employed in high-energy physics to describe nature's fundamental interactions;
the paradigmatic example is lattice QCD, the SU(3) LGT describing quarks, gluons and their strong interactions.

\subsection{LGT building blocks}
\begin{figure}
  \includegraphics[width=1\columnwidth]{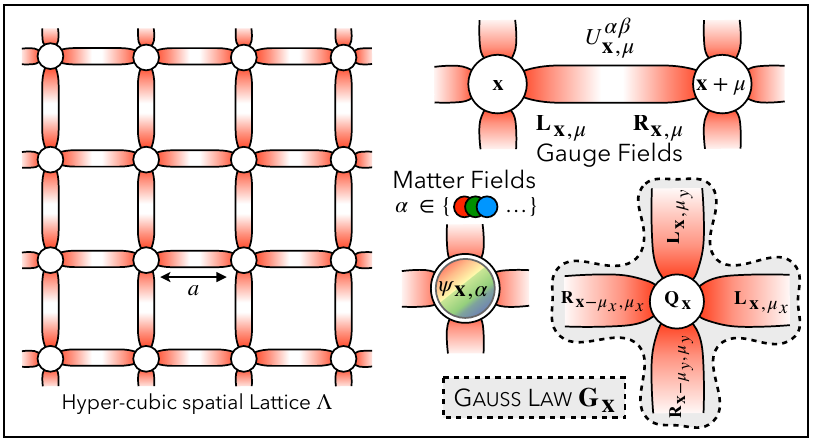}
  \caption{Graphical representation of the degrees of freedom of a 2D LGT:
    fermionic matter fields $\matt$, defined on lattice sites,
    and gauge fields (the parallel transporter, $\para$, and the chromo-electric fields, $\elel$ and $\eler$), living on lattice links. A local gauge transformation at $\vb{x}$ acts on a matter site and all its attached links.
  }
  \label{fig_lattice_LGT}
\end{figure}
As depicted in \cref{fig_lattice_LGT}, LGTs involve two types of degrees of degrees of freedom, matter and gauge, hosted respectively on lattice sites $\vb{x}\in\Lambda$ and links ($\vb{x}, \vb*{\mu}$), where $\vb*{\mu}$ denotes one of the lattice basis vectors.

\emph{Matter fields} |
such as the electron or quark fields.
As for these examples, we assume matter consists of Dirac fermions and employ staggered fermions \cite{Susskind1977LatticeFermions} to tame the fermion doubling problem, \idest{}, the proliferation of propagating fermionic degrees of freedom on the lattice.
Then, matter is represented by a multiplet $\matt$ of anti-commuting fields,
\begin{math}
  \{\matt, \matt*{y}[\beta]\} = \delta_{\alpha\beta}\delta_{\vb{xy}}
\end{math}.
Assuming a single matter flavor in the fundamental representation of the gauge group $\gaugegroup$,
gauge transformations are generated by
\begin{math}
  Q^{\nu}_{\vb{x}} = \sum_{\alpha,\beta} \matt* \lambda^{\nu}_{\alpha\beta} \matt[\beta]
\end{math},
where $\{\lambda^{\nu}\}_{\nu=1}^{\dim \gaugegroup}$ are the hermitian generators of $\gaugegroup$ |
\eg{}, the electric charge for U(1) or the three spin matrices for SU(2).
The generalization to multiple matter flavors is straightforward.

\emph{Gauge fields} | such as the photon or gluon fields.
They are represented by bosonic operators, namely the parallel transporter $\para$ and the generators of its left and right gauge transformations, $\elel$ and $\eler$,
\begin{subequations}
\label{eq:gauge_operator_algebra}
  \begin{align}
    \comm*{\elel}{\para{y}<\mup>} & = -\delta_{\vb{xy}} \delta_{\vb*{\mu\mup}}\sum\nolimits_{\gamma} \lambda^{\nu}_{\alpha \gamma} \para[\gamma \beta] \,, \\
    \comm*{\eler}{\para{y}<\mup>} & = +\delta_{\vb{xy}} \delta_{\vb*{\mu\mup}}\sum\nolimits_{\gamma} \para[\alpha\gamma]\lambda^{\nu}_{\gamma \beta}   \,;
  \end{align}
\end{subequations}
In the Abelian $\mathrm{U}(1)$ case there is only one generator: the electric field,
\begin{math}
  \elee\mathop=\elel[]\mathop=\eler[]
\end{math},
for which parallel transporters act as raising operators,
\begin{math}
  \comm*{\elee}{\para[]} \mathop= \para[]
\end{math}.

In terms of the above ingredients, a possible discretization of a matter coupled Yang-Mills theory is defined by the \emph{Kogut-Susskind Hamiltonian} \cite{Kogut1979IntroductionLatticeGauge}
\begin{align}\label{eq_Ham_LGT}
  \hamlgt =
   & -\frac{c\hbar}{2a} \sum_{\vb{x},\vb*{\mu}} \sum_{\alpha,\beta} \qty(
  s_{\vb{x},\vb*{\mu}}
  \matt* \para \matt{x+\vb*{\mu}}[\beta] + \text{h.c.})
  \nonumber                                                                   \\
   & + mc^2 \sum_{\vb{x}} \sum_{\alpha} s_{\vb{x}} \matt*\matt
  \nonumber                                                                   \\
   & + \frac{c\hbar g^2}{2a^{D-2}} \sum_{\vb{x},\vb*{\mu}} \casimir
  \nonumber                                                                   \\
   & - \frac{c\hbar}{2g^2a^{4-D}} \sum_{\square} \mathrm{Tr} (\plaq + \plaq*)
  \,,
\end{align}
where $c$ is the speed of light, $\hbar$ is the Plank constant, $a$ is the lattice spacing, $g$ is the gauge coupling, and $m$ is the mass parameter.
The first $\hamlgt$ term describes matter hopping between neighboring lattice sites, mediated by the gauge field; the second term is the mass-energy; $s_{\vb{x},\vb*{\mu}}$ and $s_{\vb{x}}$ are phases that arise due to the use of staggered fermions.
The last two terms represent the (chromo)electric and (chromo)magnetic energy of the gauge field, respectively.
The electric energy density is given by the Casimir operator:
\begin{equation}\label{Casimir}
  \casimir = \sum_{\nu}(\elel)^2 = \sum_{\nu}(\eler)^2
  \,.
\end{equation}
The magnetic energy density is a plaquette term that, on a cubic lattice, corresponds to a four-body interaction:
\begin{equation}
  \plaq = \;\sum_{\mathclap{\alpha,\beta,\gamma,\delta}}\;
  \para
  \para{x+\vb*{\mu}}<\mu'>[\beta\gamma]
  \para*{x+\vb*{\mu'}\!}<\mu>[\gamma\delta]
  \para*<\mu'>[\delta\alpha]
\end{equation}
where $\vb*{\mu}$ and $\vb*{\mu'}$ span the plaquette's plane.
Plaquette terms only exist in $D>1$, contributing to the increased complexity of quantum and TN simulations of LGTs in higher dimensions \cite{Zohar2021QuantumSimulationLattice}.

\subsection{Gauge field truncation}\label{sec:gauge_truncation}
The link Hilbert space is the space of square-integrable functions on the gauge group, $L^2(\gaugegroup)$, which is infinite-dimensional for a continuous $\gaugegroup$ \cite{Zohar2015FormulationLatticeGauge}.

In one space dimension, gauge degrees of freedom are unphysical (absence of transverse polarizations) and can thus be integrated out, albeit at the price of introducing non-local interactions \cite{Sala2018VariationalStudyU1}.

Beyond one dimension, the removal is much more delicate, because it requires first decoupling the gauge field's longitudinal component \cite{Bender2020GaugeRedundancyFree}.
When impossible or inconvenient to remove, gauge degrees of freedom might have to be truncated to perform TN or quantum simulation.
Among known truncation recipes are Quantum Link Models (QLM) \cite{Horn1981FiniteMatrixModels, Orland1990LatticeGaugeMagnets, Chandrasekharan1997QuantumLinkModels, Brower1999QcdAsQuantum}, which have been already adopted for quantum simulation of LGTs \cite{Byrnes2006SimulatingLatticeGauge, Mathis2020ScalableSimulationsLattice, Davoudi2020TowardsAnalogQuantum, Mazzola2021GaugeInvariantQuantum, Kan2021Investigating31mathrmdTopological, Zohar2021QuantumSimulationLattice, Mariani2023HamiltoniansGaugeInvariant, Pomarico2023DynamicalQuantumPhase, Bauer2023QuantumSimulationFundamental}, finite subgroups \cite{Ercolessi2018PhaseTransitionsZn, Magnifico2020RealTimeDynamics, Haase2021ResourceEfficientApproach}, digitization of gauge fields \cite{Hackett2019DigitizingGaugeFields}, and fusion-algebra deformation \cite{Zache2023}.

Another adopted solution is truncating the spectrum of the electric energy density operator
\begin{math}
  \norm{\casimir}\leq\casimircutoff
\end{math} on each link \cite{Cataldi2023a, Rigobello2023a}.
The cutoff is conveniently imposed in the irreducible representation (irrep) basis $\{\ket{j m n}\}$ \cite{Zohar2015FormulationLatticeGauge} of $L^2(\gaugegroup)$, where $\casimir$ is diagonal:
\begin{equation}
  \casimir \ket{j m n} = C_2(j) \ket{j m n}
  \,.
\end{equation}
Here $m$ and $n$ are indices in the $j$-irrep of $\gaugegroup$ and $C_2(j)$ is the quadratic Casimir of $j$ \cite{Zohar2015FormulationLatticeGauge}.
In the strong coupling limit, where the electric energy term dominates $\hamlgt$, this truncation is equivalent to an energy cutoff \cite{Rigobello2023a}.

\subsection{Gauss law and the dressed site}\label{sec:dressed_site}
The most distinctive feature of gauge theories is arguably the presence of local constraints, analogous to the Gauss law of classical electrodynamics, relating the configuration of the gauge field to the spatial distribution of charges \cite{Strocchi2013IntroductionNonPerturbative}.
At the quantum level, Gauss law is the statement that only gauge invariant states are physical, namely,
\begin{math}
  \gauss \ket*{\Psi_{\text{phys}}} = 0
  \;\forall
  \vb{x}, \nu
\end{math},
where $\gauss$ are the generators of local gauge transformations at $\vb{x}$:
\begin{equation}
  \gauss = \dynq + \bkgq + \sum_{\vb*{\mu}} [\elel + \eler{x-\vb*{\mu}}]
  \,,
  \label{eq_gauss_law_op}
\end{equation}
with $\bkgq$ representing eventual static background charges (typically vanishing).
On the lattice, Gauss law provides a set of \emph{vertex constraints}, each involving a lattice site and its $2D$ neighboring links.

Due to Gauss law, the physical Hilbert space of LGTs is much smaller than the tensor product of all local sites and link Hilbert spaces.
Properly exploiting gauge symmetries can thus significantly speed up numerical simulations \cite{Silvi2014LatticeGaugeTensor}.
Strategies that solve Gauss law by eliminating (partially or entirely) either the gauge fields or the matter fields have been developed.
Nonetheless, such approaches come with specific limitations: the range of interaction has to be extended, moreover, integrating-out gauge fields become problematic in $D>1$ \cite{Bender2020GaugeRedundancyFree}, while the recipe for removing matter is a model (matter content) dependent \cite{Zohar2019RemovingStaggeredFermionic}.

\begin{figure}
  \includegraphics[width=1\columnwidth]{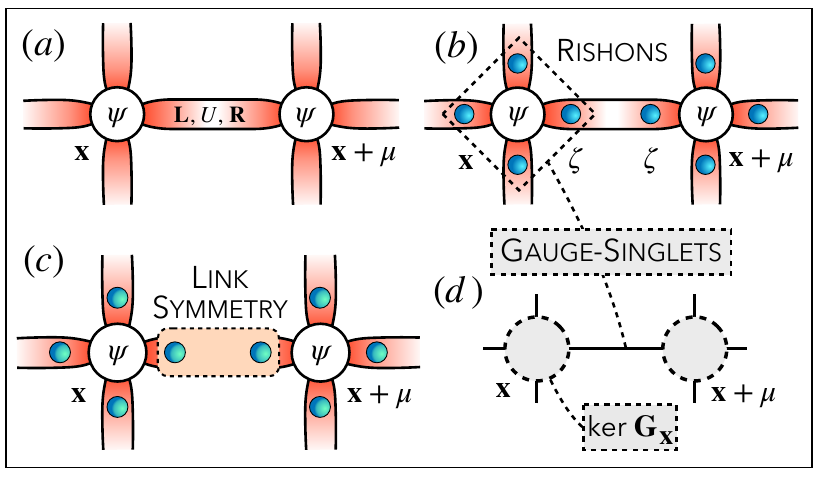}
  \caption{Pictorial representation of the dressed site formalism adopted for TN simulations of LGTs:
    (a) starting from the original formulation of matter and gauge fields,
    (b) each truncated gauge link is split into two representations, one per half-link; each is equipped with a proper fermionic rishon mode $\zeta$.
    (c) All the half-links are absorbed into the attached matter site, forming a gauge-invariant dressed site (d) whose Hilbert space spans all the possible gauge singlets.}
  \label{fig_rishons}
\end{figure}

Another possibility is to enforce Gauss law using a dressed site construction, sketched in \cref{fig_rishons} and outlined below.
Dressed sites have local dimensions typically larger than those resulting from the aforementioned approaches, but they are obtained from a model-independent prescription which has the advantage of preserving the locality of the interactions \cite{Silvi2014LatticeGaugeTensor, Rigobello2023a}.

\paragraph{Rishons.}
As a first step, every link is decomposed into a pair of left ($L$) and right ($R$) \emph{rishon} degrees of freedom, each associated with a Hilbert space spanned by the basis states $\ket{j m}$,
identifying
\begin{math}
  \ket{j m n} \hookrightarrow \ket{j m}_L \mathop\otimes \ket{j n}_R
\end{math},
and writing parallel transporters as rishon bilinears \cite{Cataldi2023a}:
\begin{equation}
  \para \to \sum_{k}
  \zeta^{L(k)\alpha}_{\vb{x},+\vb*{\mu}}\zeta^{R(k)\beta\,\dagger}_{\vb{x}+\vb*{\mu},-\vb*{\mu}} \,
  \label{eq_from_U_to_rishons}
  \,.
\end{equation}

\paragraph{Abelian link symmetries.}
Physical gauge link configurations, \idest{} those with the left and right rishons in the same irrep, are selected introducing a local link symmetry at the TN simulation level \cite{Silvi2014LatticeGaugeTensor, Rigobello2023a}.
Notice this is always Abelian, regardless of the Abelian or non-Abelian nature of the gauge group.

\paragraph{Gauge invariant computational basis.}
Crucially, the gauge generators $\gauss$ now involve only the matter site at $\vb{x}$ and its $2D$ neighboring rishons.
Fusing these degrees of freedom in a composite site, Gauss law becomes an internal constraint that singles out the dressed site Hilbert space as their gauge invariant subspace:
\begin{equation}\label{eq:dressed_site}
  \hildress = \ker G \subset \hilsite \otimes (\hilsemi)^{\otimes 2D}
  \,.
\end{equation}
The expansion of the gauge singlet basis states of $\hildress$ in terms of the matter and rishon bases is computed via Clebsch-Gordan decomposition \cite{Rigobello2023a}.

\paragraph{Defermionization.}
It is possible to effectively eliminate fermionic matter, and treat dressed sites as large spins, for any gauge theory where the gauge field has a well-defined parity. 
Specifically, a local parity operator $P_{\vb{x},\vb*{\mu}} = P_{\vb{x},\vb*{\mu}}^{\dagger}$ such that $P_{\vb{x},\vb*{\mu}}^2 = 1$ must satisfy $\{\para,P_{\vb{x},\vb*{\mu}}\} = 0$, as it happens for $\mathbb{Z}_{2N}$, U(N), and SU(2N) \cite{Zohar2018EliminatingFermionicMatter, Zohar2019RemovingStaggeredFermionic, Cataldi2023a}. 
In this case, it is possible to take fermionic rishons, and as a result, all physical (gauge invariant) dressed site operators are genuinely local, i.e.~they mutually commute at a nonzero distance (as spins or bosons) \cite{Ballarin2023ScalableDigitalQuantum}.
In particular, this applies to the Hamiltonian $\hamlgt$, making gauge-defermionization particularly convenient for higher-dimensional systems, where Jordan-Wigner strings result in long-range interactions \cite{Verstraete2008MatrixProductStates}.

\subsection{Scaling of the local basis dimension}
\label{sec_basis_dimension}
\Cref{tab:local_dims} lists the dressed site dimension $d=\dim\hildress$ associated with the first few nontrivial gauge truncations of three representative LGT models in $D=2,3$ space dimensions.
All the considered LGTs include dynamical matter, represented by one fermionic field multiplet in the fundamental representation of the gauge group.
The $\gtl$-th row of \cref{tab:local_dims} is obtained keeping only the first $\gtl$ nonzero electric energy levels (\idest{}, using the $\gtl$-th smallest quadratic Casimir eigenvalue as cutoff $\casimircutoff$, see \cref{sec:gauge_truncation}).
As \Cref{tab:local_dims} shows, the local dimension increases rapidly with $\gtl$.
For 3-dimensional non-Abelian LGTs, $d \sim O(10^4)$ is reached already within the first two truncations, making the study of the untruncated limit prohibitive.

Differently from the models typically encountered in condensed matter physics, the local dimension of LGTs can thus be a limiting factor for TN simulation | especially when $d$ becomes comparable to commonly used TN bond dimensions ($100 \lesssim \chi \lesssim 500$ for TTN).
In these cases, strategies aimed at further compressing the local computational basis are needed (see \cref{sec:opt:basistruncation}).
As just discussed, such truncation strategies are particularly crucial for high-dimensional LGTs.

Several numerical analyses suggest that, in some cases, a small-to-moderate truncation of the gauge group is enough for accurately approximating the continuum limits, at least for low-energy states \cite{Buyens2017FiniteRepresentationApproximation, Ercolessi2018PhaseTransitionsZn, Rigobello2021EntanglementGeneration11d, Ciavarella2021TrailheadQuantumSimulation, Davoudi2021SearchEfficientFormulations, Tong2022ProvablyAccurateSimulation}.
However, the optimal gauge truncation depends on the Hamiltonian parameters $m$ and $g$.

\newcolumntype{R}{>{\raggedleft\arraybackslash}X}
\begin{table}
  \centering
  \setlength{\tabcolsep}{4pt}
  \renewcommand{\arraystretch}{1.05}
  \begin{tabularx}{\columnwidth}{>{\centering\arraybackslash}X|rrr|rrr}
    \toprule
    $\gtl$ & \multicolumn{6}{c}{$d$}                                                                                                     \\ \midrule
           & \multicolumn{3}{c|}{$(2+1)$-dimensions} & \multicolumn{3}{c}{$(3+1)$-dimensions}                                            \\[1pt]
           & U(1)                                    & SU(2)                                  & SU(3)  & U(1)    & SU(2)    & SU(3)      \\
    \midrule
    1      & 35                                      & 30                                     & 164    & 267     & 178      & 3096       \\
    2      & 165                                     & 168                                    & 752    & 3437    & 3670     & 52476      \\
    3      & 455                                     & 600                                    & 3738   & 18487   & 35280    & 813438     \\
    4      & 969                                     & 1650                                   & 19878  & 64953   & 214958   & 17490134   \\
    5      & 1771                                    & 3822                                   & 43698  & 177155  & 967466   & 69232482   \\
    6      & 2925                                    & 7840                                   & 82128  & 408421  & 3509062  & 228461186  \\
    7      & 4495                                    & 14688                                  & 212496 & 835311  & 10828494 & 1245755754 \\
    8      & 6545                                    & 25650                                  & 333538 & 1561841 & 29473038 & 2782999996 \\ \bottomrule
  \end{tabularx}
  \caption{Dressed site Hilbert space dimension $d$ for increasing number $\gtl$ of allowed electric energy density levels in some 2- and 3-dimensional paradigmatic LGTs with dynamical matter and gauge groups U(1), SU(2), and SU(3).}
  \label{tab:local_dims}
\end{table}
As an example, we consider the (2+1)D U(1) LGT including dynamical matter (QED), whose Hamiltonian can be obtained from \cref{eq_Ham_LGT}:
\begin{equation}
  \begin{aligned}
    H_{\text{QED}}&=
    \frac{c\hbar}{2a}\sum_{\vb{x}} \Big[\text{-i}  \mattU*[] \para<\mu_x>[] \mattU{x+\vb*{\mu_x}}[] \\
    & - (-1)^{j_{\vb{x}} + j_{\vb{y}}}\mattU*[] \para<\mu_y>[] \mattU{x+\vb*{\mu_y}}[] + \text{H.c.} \Big]    \\
    & + m c^{2} \sum_{\vb{x}} (-1)^{j_{\vb{x}} + j_{\vb{y}}}\mattU*[]\mattU[] \\
    & +\frac{g^2 c \hbar}{2a} \sum_{\vb{x},\vb*{\mu}} \casimir
    - \frac{c \hbar}{2 g^{2}a}\sum_{\square} \mathrm{Tr} (\plaq + \plaq*),
  \end{aligned}
  \label{eq_H_QED}
\end{equation}
Since the Abelian U(1) group has only one generator, the gauge fields $E_{\vb{x}}$ and $\para[]$ can be represented as 
\begin{align}
    E_{\vb{x},\vb*{\mu}}&=\sigma^{z}_{\vb{x},\vb*{\mu}}(\ell)&
    \para[]&=\zeta_{\vb{x},+\vb*{\mu}}(\ell)\cdot \zeta_{\vb{x}+\vb*{\mu}, -\vb*{\mu}}(\ell)
\end{align}
where $\sigma^{z}(\ell)$ is the spin $z$-operator in the $\ell$ SU(2) irrep, while $\zeta(\ell)$ is a $\ell\times \ell$ ladder operator with Fermi statistics.

\begin{figure}
  \includegraphics[width=1\columnwidth]{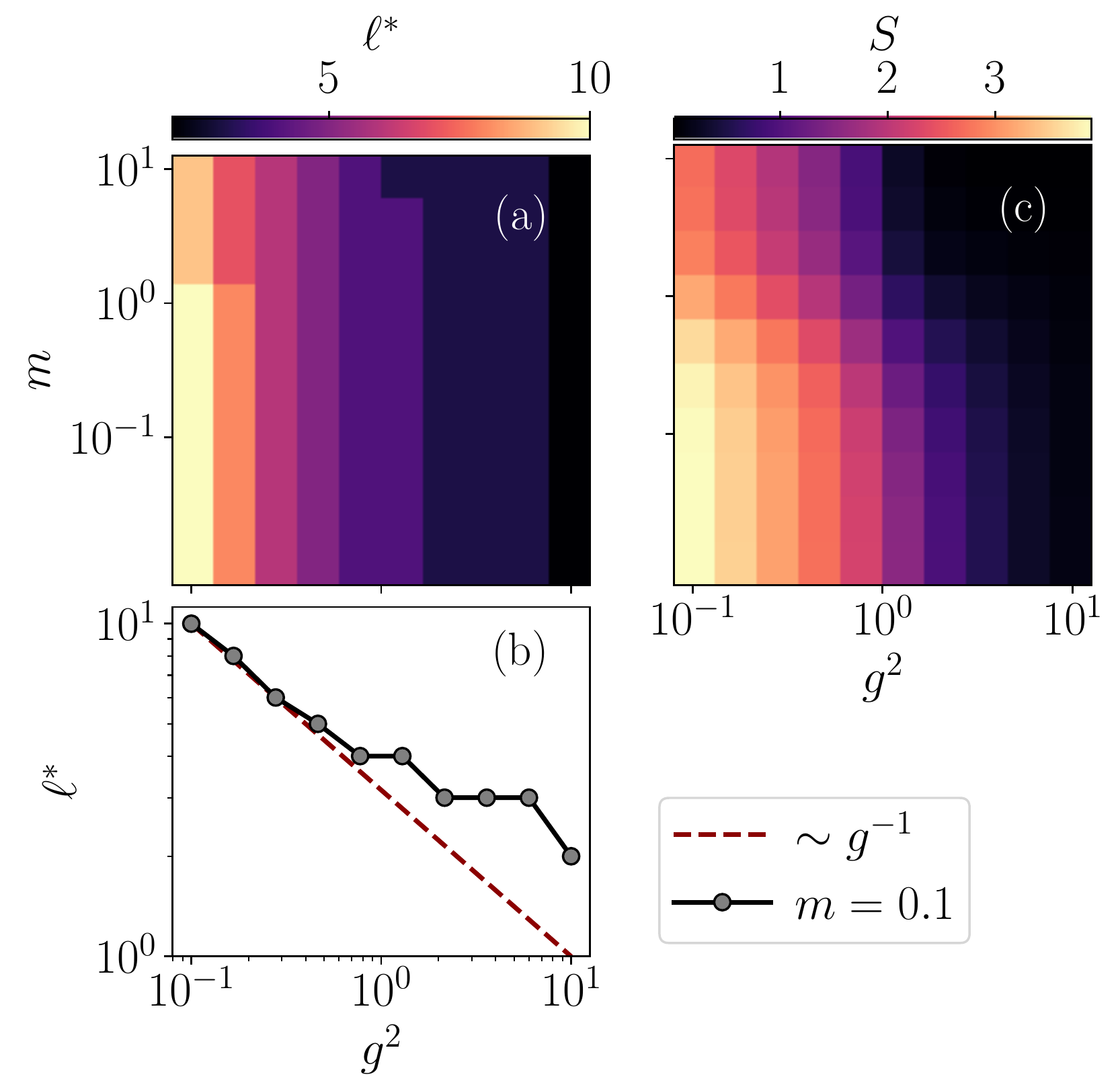}
  \label{fig_QED_convergence}
  \caption{Exact diagonalization of a QED plaquette for a grid of masses and couplings,  $m\in [10^{-2},10^{1}]$ and $g^{2}\in [10^{-1},10^{1}]$.
  (a,b) Minimal gauge truncation $\gtlstar$ required to reach a precision $\gterr=10^{-5}$ in the magnetic energy $\ev*{\Re\plaq}$.
  (c) Corresponding entanglement entropy $S$ associated with a symmetric bipartition of the plaquette.}
\end{figure}

We focus on a single plaquette in open boundary conditions, as it provides the minimal setting allowing for both electric and magnetic effects.
Then, to characterize the convergence in the gauge truncation $\gtl$, we consider a candidate observable $O$ and compute its ground state expectation value $\langle\gtobs\rangle_{\gtl}$ for increasing $\gtl$.
We iterate until the relative deviation between consecutive truncations drops below some threshold $\gterr$:
\begin{math}
  |\langle \gtobs\rangle_{\gtlstar} - \langle\gtobs\rangle_{\gtlstar-1}|
  <
  \gterr |\langle \gtobs \rangle_{\gtlstar}|
\end{math} for some $\gtlstar$.
\Cref{fig_QED_convergence}(a) shows the minimal truncation $\gtlstar$ at which the magnetic energy operator $\gtobs = \Re \plaq$ is converged to $\gterr=10^{-5}$.
We explore a grid of model parameters, whose extent has been chosen according to standard MC literature \cite{Fiore2005Qed3SpaceTime, Raviv2014NonperturbativeBetaFunction, Svetitsky2015BetaFunctionThree, Xu2019MonteCarloStudy, Creutz1983MonteCarloComputations, Creutz1988LatticeGaugeTheory, Creutz1989LatticeGaugeTheories, Loan2003PathIntegralMonte, Rothe2012LatticeGaugeTheories,  Funcke2023HamiltonianLimitLattice, Strouthos2008, Bender2023QuantumClassicalMethods, Clemente2022StrategiesDeterminationRunning}.
$\Re\plaq$ is used as a benchmark due to its relevance in the weak coupling regime, where the continuum limit of $D<3$ lattice QED is located \cite{Creutz1983MonteCarloComputations}.
As \cref{fig_QED_convergence}(b) shows, $\gtlstar$ depends heavily on the coupling, growing asymptotically like $\gtlstar\sim g^{-1}$ as $g$ is decreased, while $m$ plays almost no role.
An analogous inverse dependence of the minimal gauge truncation on the coupling is expected for non-Abelian LGT in arbitrary dimensions.
Moreover, the continuum limit of non-Abelian LGTs in $D\leq3$ is also expected at $g\to0$ \cite{Creutz1983MonteCarloComputations}, further substantiating the need to compress the local dimension in TN simulations of LGT whenever extrapolation to the continuum is in order.

Apart from the growth of the local dimension, extrapolation to the continuum is further complicated by the fact that the continuum limit of a lattice quantum field theory corresponds to a critical point of the underlying lattice model \cite{Hernandez2011LatticeFieldTheory}.
Close to criticality, quantum correlations are boosted and violations of the entanglement area law are expected \cite{Eisert2013EntanglementTensorNetwork}.
The higher entanglement entropy in the proximity of the continuum ($g,m\ll1$ regime) is already captured by the single-plaquette analysis of \cref{fig_QED_convergence}(c).
Continuum limits of LGTs are thus an area of potential advantage for quantum computation over classical methods, as the former is not limited by entanglement.
Nonetheless, quantum computation is also affected by the need to relax gauge truncations when $g \to 0$, either by increasing the number of qubits used to encode a dressed site, which is at least $\lceil\log_2(d)\rceil$, or by using hybrid devices, that have both qubits and bosonic modes \cite{kang2023leveraging}.

\section{Roadmap for advanced LGT simulations via tensor networks}
\label{sec_TN_roadmap}
As detailed in the previous sections, LGT models present some peculiar features that make TN simulations particularly challenging, especially for large system sizes and for studying the continuum limits in terms of gauge field truncation, lattice spacing, and volume.

State-of-the-art techniques, such as TTN algorithms, have been applied for simulating ground state properties of QED in (2+1)- and (3+1)-dimensions for small-to-intermediate sizes \cite{Felser2020TwoDimensionalQuantum, Magnifico2021LatticeQuantumElectrodynamics}. Very recently, they have been also applied to the SU(2) Yang-Mills model in (2+1)-dimensions \cite{Cataldi2023a}.
In all these simulations, small non-trivial representations of the gauge groups have been exploited, e.g. three electric field levels for QED and the first two irreducible representations of SU(2) for the Yang-Mills model.

Nowadays, lattice computations with MC-based techniques are performed on large lattices, e.g. of the order $64^{3} \times 128 $ sites (space
and time discretization), and with no truncation of the gauge fields.
These large sizes are required to control finite-volume effects and to perform extrapolations toward the continuum limits \cite{Workman2022}.
In the last decades, the impressive progress in algorithmic development, high-performance optimizations, and the availability of increasingly powerful supercomputer facilities have played a major role in the advancement of MC-based LGT computations.
Indeed, this progress has opened the doors to large-scale simulations, that currently represent the standard approach for studying non-perturbative phenomena in quantum field theory.

However, MC techniques are generally based on computations of path integrals in which the integrand functions are overall positive.
Many physically relevant scenarios, such as finite baryon density regimes or real-time dynamics of quarks, give rise to a change in the sign of the integrands and highly oscillating behaviors.
Thus, numerical evaluations suffer from the near cancellation of the opposite-sign contributions to the integrals.
This is the essence of the infamous sign problem, a long-standing issue of LGT simulations with MC methods \cite{Nagata2022FiniteDensityLattice}.

Hence the quest of conceiving, developing, and optimizing alternative approaches that enable simulating these regimes, being the latter at the heart of many open problems related to our understanding of high-energy physics.

As described in the previous sections, TNs represent a promising complementary method, which found the first applications in simulating non-trivial instances of high-dimensional LGTs on small-to-intermediate lattice sizes. TNs are intrinsically sign-problem-free, enabling the simulation of both static properties at equilibrium, such as low-energy states, and real-time dynamics, even in the presence of finite chemical potentials or non-trivial topological terms. It is worth noting that, in addition to local observables and correlation functions, TNs allow the numerical computations of entanglement properties, such as entanglement entropies and central charges, that could potentially shield new light on LGT phenomena \cite{Rigobello2021EntanglementGeneration11d}.

Nevertheless, TN simulations of high-dimensional LGTs still represent a challenging problem, especially for large lattice sizes or higher representations of gauge groups needed for analyzing continuum limits. 
In this framework, further and intensive developments are required to tackle TNs' current problems related to LGTs, such as QCD's non-perturbative effects on lattices of sizes comparable with MC simulations. In this regard, we note that sign-problem-free TN ansätze can also be used in combination with variational MC methods to tackle high-dimensional lattice gauge theories with arbitrary gauge groups \cite{kelman2024gauged}.

In the following, we present a possible roadmap in terms of algorithmic development and optimization strategies that we foresee to be crucial for making the TN approach competitive as a complementary method to MC techniques. Therein, the first two sections, i.e. \cref{sec:opt:basistruncation,sec:opt:initialstates}, can be approached with existing TN algorithms and a good intuition on LGT problems; then, Secs.~\ref{sec:opt:hpclocal} and \ref{sec:opt:hpcmpi} outline optimization for existing algorithms to leverage HPC systems; finally, we discuss new classes of ans{\"a}tze to tackle finite temperature problems in Sec.~\ref{sec:opt:finitet}. In some parts, we focus on TTNs, but the vast majority of the presented concepts and techniques can be straightforwardly applied to other TN ansätze.

\subsection{Local basis truncation}                
\label{sec:opt:basistruncation}
\begin{figure}
  \includegraphics[width=0.73\columnwidth]{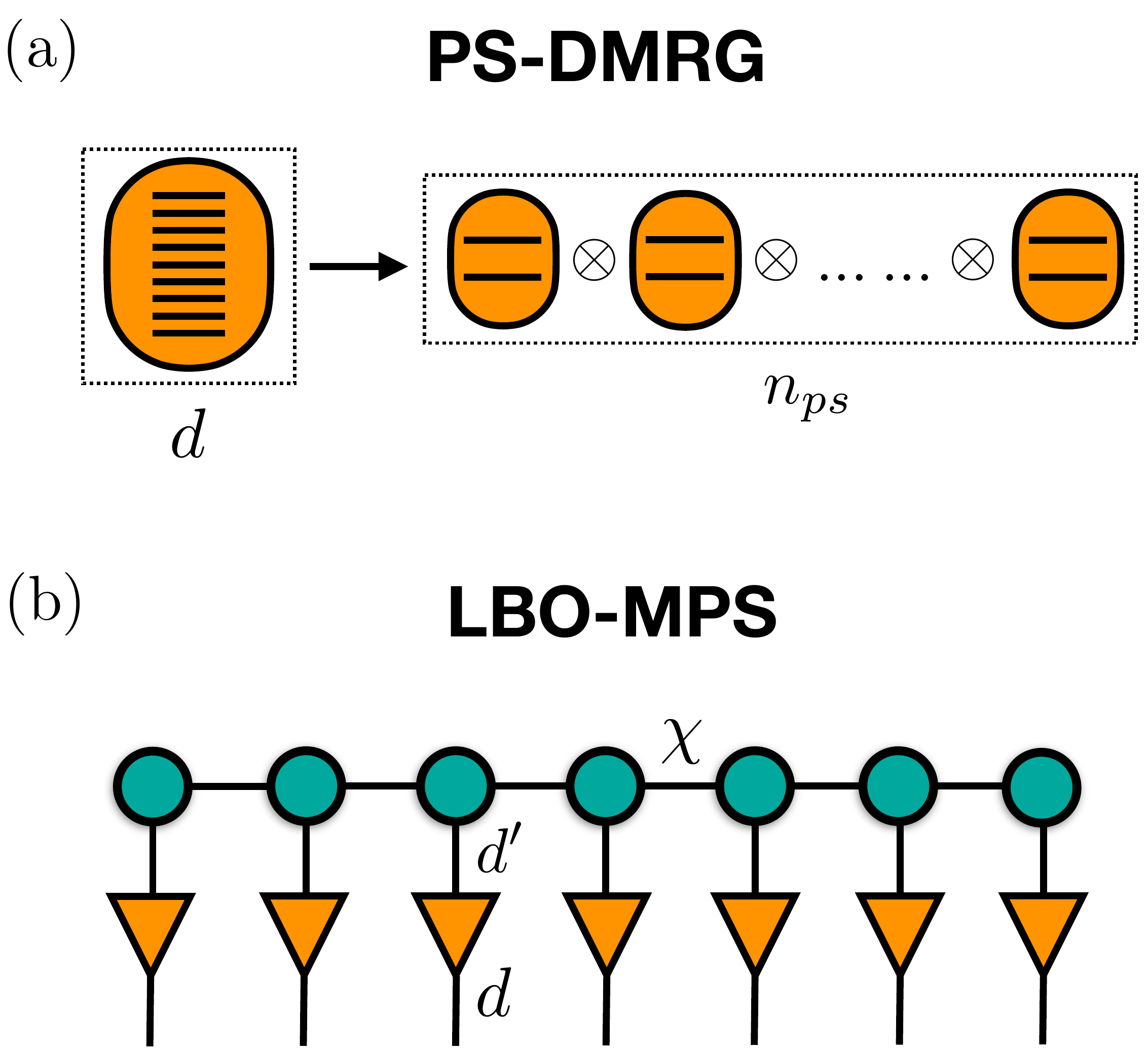}
  \caption{Graphical representation of (a) the pseudo site DMRG (PS-DMRG) approach, in which a single site having large local dimension $d$ is replaced with $n_{ps}$ pseudo sites with a smaller local dimension, e.g two; (b) the local basis optimization embedded in the MPS ansatz (LBO-MPS), so that an additional layer of tensors connected to the physical legs of the MPS performs the reduction from a large local basis with dimension $d$ into an effective basis with smaller dimension $d'$. }
  \label{fig_panel_PS_LBO}
\end{figure}

One of the main issues in simulating LGTs with TN methods lies in the very large local basis dimension $d$ that one needs to handle to represent matter and gauge-field degrees of freedom properly.
To some extent, this situation is very similar to some condensed matter models, e.g. the Holstein model, involving lattice fermions coupled with phonons \cite{Holstein1959StudiesPolaronMotion}, or bosonic systems coupled to optical cavities \cite{Bloch2005UltracoldQuantumGases, Schlawin2022CavityQuantumMaterials}. Also in these cases, the local Hilbert space is in principle infinite-dimensional, and a truncation to a fixed cutoff is needed for performing TN simulations. For one-dimensional systems of this type, some efficient algorithms based on MPS have been developed in the last years \cite{Stolpp2021ComparativeStudyState}, e.g. the pseudosite DMRG (PS-DMRG) method \cite{Jeckelmann1998DensityMatrixRenormalization}, and the DMRG with local basis optimization (DMRG-LBO) \cite{Guo2012CriticalStrongCoupling, Brockt2015MatrixProductState, Stolpp2020ChargeDensityWave}.

The key idea of the PS-DMRG is to replace a single site having large local dimension $d$ with $n_{ps} \approx \logtwo(d)$ pseudo sites of local dimension 2, as shown Fig. \ref{fig_panel_PS_LBO}(a). Since a large class of optimization algorithms for TNs scale at least quadratically with the local dimension but linearly with the total number of sites, as detailed in \cref{sec_TNs}, one obtains a more efficient and manageable representation according to this procedure.

The price to pay lies in the range of the interactions: short-range interactions in the Hamiltonian are transformed into long-range operators due to the pseudo-site encoding of the local degrees of freedom.
As a consequence, PS-DMRG should require a larger bond dimension and a larger number of variational steps to converge, and the benefits offered by the pseudo-sites could progressively fade for increasing values of $n_{ps}$ at fixed bond dimension. PS-DMRG has been applied to one-dimensional QMB systems with $d$ up to $O(100)$ \cite{Stolpp2021ComparativeStudyState}.

The core of the DMRG-LBO algorithm, instead, lies in a local basis optimization protocol which enables a controlled and efficient truncation of the local Hilbert space \cite{Zhang1998DensityMatrixApproach}. For each site $\vb{x'}$ of the lattice, the optimized local basis is computed by starting from its reduced density matrix, i.e.
\begin{equation}\label{eq_RDM}
  \rho_{\vb{x'}} = \mathrm{Tr}_{\vb{x}\ne \vb{x'}} \ketbra{\psi},
\end{equation}
where $\ket{\psi}$ is a general state of the whole system and the trace is over all the degrees of freedom which do not involve the site $\vb{x'}$.
By performing the diagonalization procedure, the eigenvalues $\lambda_{\alpha}$ and the eigenvectors $v_{\alpha}$ of $\rho_{\vb{x'}}$ can be easily determined.
The set of values $\lambda_{\alpha}$ represent the probabilities associated with the states $v_{\alpha}$.
If $\lambda_{\alpha}$ is small, e.g. below a certain numerical threshold, the related eigenvector $v_{\alpha}$ can be discarded from the local basis of the site $\vb{x'}$, with a controllable loss of accuracy for the state $\ket{\psi}$.
Thus, for reducing the local basis dimension of $\vb{x'}$ from $d$ to a smaller value $d'$, an optimal choice is keeping the $d'$ eigenvectors of $\rho_{\vb{x'}}$ with the largest probabilities.
Then, the original state $\ket{\psi}$ can be projected on the new basis, without losing the relevant physical information. Let us notice that the site $\vb{x'}$ can also be a generic unit cell of the system, composed of a certain number of lattice sites.

A crucial point of the LBO procedure is the knowledge of the original state $\ket{\psi}$, generally the ground state of the system, that is not known prior.
This issue can be overcome in several ways: by performing exact diagonalization of the system's Hamiltonian with local dimension $d$ for small lattice sizes, to determine $\ket{\psi}$, and then truncating the basis from $d$ to $d'$ for increasing values of the cutoff $d'$.
From this procedure, we can obtain an optimized local basis ensuring a controlled approximation of the original ground state.
This optimized basis can then be exploited in optimization algorithms for simulating larger lattice sizes, such as DMRG.
Another strategy directly incorporates the LBO procedure in the tensor network ansatz, as shown \cref{fig_panel_PS_LBO}(b): by inserting an additional tensor on each physical leg of the MPS, the large local basis with dimension $d$ is transformed into an effective basis with smaller dimension $d'$.
In this way, the MPS tensors only see the effective basis in the optimization procedures, with a significant reduction of the computational costs described in \cref{sec_TNs}.
This method has been used for both static DMRG and time evolution algorithms, such as TEBD, for the study of one-dimensional quantum impurity models and correlated electron-phonon systems \cite{Guo2012CriticalStrongCoupling, Brockt2015MatrixProductState, Schroeder2016SimulatingOpenQuantum}.

In the context of LGTs, the dimension of the local basis $d$ can easily go beyond values of the order of $10^6$, especially for higher-dimensional non-Abelian models and large representations of the gauge groups, as shown in \cref{sec_TN_LGTs}.
In this scenario, numerical simulations with TNs are practically infeasible without an optimized scheme for truncating the local degrees of freedom.
Techniques like PS-DMRG might offer some benefits for small system sizes, such as unveiling the most relevant degrees of freedom in the low-energy states, but it is difficult to scale them up for large sizes due to the long-range interactions induced between the pseudo-sites.
Also, the local constraints imposed by Gauss law would become highly non-local when splitting a single site into multiple pseudo-sites representing the matter and link fields.

In principle, LBO-based procedures could instead represent a well-grounded route for addressing the problem.
Their use in condensed matter, e.g. bosonic systems, is a rather consolidated approach, whereas their application to TN simulations of LGTs currently is an uncharted but promising territory.
The main steps that we foresee as needed and important in this direction are the following:

\begin{itemize}
  \item[(i)]
        Employing exact diagonalization, testing the convergence of LBO procedures for one-dimensional systems, such as the $\varphi^4$-theory or the Schwinger model, for which analytical solutions, at least in some regimes of the phase diagram, and numerical results are widely available, also in the limit of no gauge field truncations.
        In this way, we can obtain valuable information about the scaling of the basis cutoffs concerning the final accuracy of the state representation for small system sizes.
  \item[(ii)]
        Performing the same analysis on (2+1)-dimensional LGTs, such as QED or $SU(N)$ models, for one unit cell, like a single lattice-plaquette, to systematically study the effects of the magnetic interactions on the local degrees of freedom.
        Indeed, the QLM representation of LGTs, detailed in \cref{sec_TN_LGTs}, generally exploits the ``electric field" basis, in which the electric field terms of the Hamiltonian and Gauss law are diagonal.
        In this scheme, the magnetic interactions correspond to non-diagonal operators which can increase the number of local electric states to include for an accurate description of the system, especially for small values of the coupling constant $g$ of \cref{eq_Ham_LGT}, as highlighted in the numerical analysis in Sec. \ref{sec_basis_dimension}.
  \item[(iii)]
        Exploiting the optimized local bases obtained from exact diagonalization as input of TN simulations for larger sizes, testing the effects on the global ground state accuracy and in computing physically relevant quantities, such as the mass gap \cite{Clemente2022StrategiesDeterminationRunning}.
  \item[(iv)]
        In the same spirit of LBO-MPS, implementing LBO protocols directly in TN ansätze that are more suitable for simulating high-dimensional LGT models, such as TTN.
        This step could be of great benefit in particular for aTTNs, which encode the area law for the entanglement but are severely limited by large local bases (see \cref{sec_TNs}).
\end{itemize}
By following and combining all these steps, we expect to reduce the effective local basis of LGT models, potentially enabling TN simulations for large representations of Abelian and non-Abelian gauge groups. 

It is worth noting that constructing optimal bases for numerical and quantum computation of LGTs is an active area of research. 
Several approaches that have been recently proposed involve performing canonical transformations of the gauge degrees of freedom before truncation \cite{Haase2021ResourceEfficientApproach, mathur2023exact, bauer2023new}. 
By exploiting a resource-efficient protocol of this type, Ref. \cite{Haase2021ResourceEfficientApproach} has shown that the number of local states required to reach a $1\%$ accuracy level when computing the expectation value of the plaquette operators in two-dimensional (2+1)-dimensional QED can be reduced by more than $94\%$ compared to the unoptimized truncation.
Integrating these approaches into TN algorithms could greatly benefit LGT simulations.

\subsection{Tailored initial states}
\label{sec:opt:initialstates}
In TN algorithms for ground state searching, the optimization procedure generally starts from a random TN initial state $\ket*{\psi_{\mathrm{init}} }$, \idest{} the tensors in the network are filled with random coefficients at the beginning.
This strategy usually guarantees that the probability of overlapping with the true ground state $ |\braket*{\psi_{\mathrm{init}}}{\psi_{\mathrm{gs}}}| ^2 $ is not vanishing.
To reach a small error in the final energy, this procedure typically requires from 10 to 50 optimization sweeps for LGTs simulations, depending on the specific models, the Hamiltonian parameters, and the lattice size.
Since the time for completing a sweep can be very long, especially for large bond dimensions, strategies for reducing the number of needed sweeps could be beneficial for scaling up system sizes.
From this perspective, constructing appropriate states to be used as initial guesses can speed up the convergence, similar to the choice of the trial wave function for variational Monte Carlo simulations. We consider the following options:

\begin{itemize}
  \item[(i)] \emph{Physical insight:}
        Initial states can be constructed by following physical intuition, at least in those regimes in which analytical or partial numerical results are available.
        For instance, initial guesses can be constructed by starting from the TN ground states numerically obtained for a lower representation to simulate large spin representations of the gauge fields.
  \item[(ii)] \emph{Machine learning:}
        Machine learning-assisted protocols can improve the construction of tailored initial states in the different regimes of the model parameters.
        For instance, feed-forward neural networks have been proposed as trial wave functions for quantum Monte Carlo simulations \cite{Kessler2021ArtificialNeuralNetworks}, and machine learning techniques have been used to feed TN simulations \cite{Schroeder2019TensorNetworkSimulation}. Similarly, neural networks might reveal great potential in constructing initial states for LGTs to be used in large-scale TN simulations, in which reducing the number of sweeps is a key point for feasibility.
  \item[(iii)] \emph{Tensor network results:}
        Following the idea of the physical insight, it is also possible to feed neighboring ground states as initial guesses into a ground state search.
        This option exists especially when scanning a phase diagram and varying parameters in a small increment such that the overlap between neighboring wave functions is sufficient; this overlap decreases for two points on the opposite sides of a quantum critical point.
        The same idea can be implemented by preparing an initial guess quenching from an easily accessible ground state to the target parameters; the quench itself does not have to be adiabatic or free of numerical errors, but must only have sufficient overlap with the ground state.
        The advantage here is that one quench can generate multiple initial guesses along the quench for different parameters.
\end{itemize}

\subsection{Leverage HPC techniques for local optimization}                     
\label{sec:opt:hpclocal}
We dedicate the two following sections to the numerical optimization of the TN algorithms. 
In this section, we give an overview of the topic and we discuss possible strategies to improve the optimization. 
The more technical steps are discussed in the following Sec.~\ref{sec:opt:hpcmpi}. 

To scale up TN simulations of LGTs, in particular regarding lattice sizes, another important factor is the number of optimization steps to be carried out. 
The number of optimization steps scales linearly with the number of sweeps as well as with the system size for MPS, PEPS, and TTN. 
The choice of the number of sweeps is set such that the algorithm reaches convergence when computing ground or low energy states. 
Let us briefly describe the general procedure for ground state searches: we focus on the TTN optimization, however, the main points described here can be applied to other TN structures, such as MPS or PEPS. 
For a complete and technical description of the algorithms and implementation details, see Ref. \cite{Silvi2019TensorNetworksAnthology}.

Consider a generic QMB Hamiltonian $H$ and a generic normalized state $\ket{\psi}$, defined on the same Hilbert space. 
To numerically determine the ground state of $H$, the following global minimization problem has to be solved:
\begin{equation}
\label{eq_min_problem}
  \min_{\ket{\psi}} \left \{ E ( {\ket{\psi}} ) \right \} = \min_{\ket{\psi}} \bra{\psi} H\ket{\psi} \, .
\end{equation}
If $\ket{\psi}$ is written in terms of the TTN ansatz introduced in \cref{sec_TNs}, the variational
parameters are the coefficients in the TTN. In this case, the global optimization problem of \cref{eq_min_problem} is broken down into a sequence of smaller optimizations, each of which involves only a minimal subset of tensors in the TTN. 
The algorithm solves the optimization via an eigenproblem searching for the eigenvector with the lowest eigenvalue.
Without any loss of generality, one single tensor at a time is optimized in the simplest case, as shown in \cref{fig_TTN_optimization}(a). 
In detail, the energy is computed by contracting the Hamiltonian between the TTN and its complex conjugate. 
Then, we start the optimization procedure from a target tensor $T$, by computing its environment, i.e. the network without the tensors $T$ and $T^\dagger$, which represents the effective Hamiltonian $H_{\text{eff}}$ for the local problem. 
At this stage, an eigenproblem of $H_{\text{eff}}$ is solved, and the tensor $T$ is updated with the newly found ground state. 
The whole procedure is sequentially iterated for all the tensors in the network, performing an optimization sweep.

For each operation, efficient algorithms from linear algebra are typically used, e.g. the Arnoldi algorithm implemented in the ARPACK library \cite{Arnoldi1951PrincipleMinimizedIterations, Lehoucq1997ArpackSolutionLarge}. 
We recall that the numerical complexity of this procedure for a single TTN-tensor is $O(d^2 \chi^2 + \chi^4)$ \cite{CommentScalingA}.
Therefore, a single optimization can be time-consuming when $\chi$ is very large, e.g.
$\chi \approx 1000$, as for simulating high-dimensional LGTs. 
We point out the established and promising future parallelization schemes for the single tensor optimization:

\begin{itemize}
    \item[(i)] \emph{opemMP:} 
        An efficient openMP implementation of the contraction between the effective operators with the tensor can speed up simulations. 
        Moreover, the Arnoldi algorithm of ARPACK is optimized for large-scale linear algebra operations and supports intra-node multi-core parallelization based on openMP \cite{Dagum1998OpenmpIndustryStandard}; thus, ARPACK does not become a bottleneck in the openMP implementation. 
        Nonetheless, many simulations remain expensive in computation time even with 64 or more cores available in HPC facilities; therefore, we consider more approaches beyond the well-established openMP path.
    \item[(ii)] \emph{Accelerators:} Graphics Processing Units (GPU) and Tensor
        Processing Units (TPU) offer a path to accelerate linear algebra routines,
        where both have demonstrated their usefulness: GPUs have reported speedups
        of up to a factor of 10 due to the efficient tensor
        manipulations \cite{Shi2016TensorContractionsExtended, Abdelfattah2016HighPerformanceTensor, Vincent2022JetFastQuantum, Pan2022SimulationQuantumCircuits};
        tensor processing units have shown great potential in large-scale simulations
        of several quantum systems, e.g. drastically reducing the computational time
        of DMRG calculations with very large bond dimensions from months to
        hours \cite{jouppi2017indatacenter,hauru2021simulation, Morningstar2022SimulationQuantumMany, Ganahl2023DensityMatrixRenormalization}.
        Tensor processing units are application-specific integrated circuits originally introduced
        for machine learning; we consider the integration and tuning of TPUs therefore as a
        step after the successful integration of GPUs. While single GPUs can solve
        TTN-problems up to a bond dimension of $\chi < 1000$, multi-GPU support
        is available for libraries; HPC systems typically provide hardware with four GPUs
        per node.
    \item[(iii)] \emph{Multi-node approaches to local tensor optimizations:} Both CPU and
        GPU algorithms can be further scaled by using multiple nodes. The underlying linear
        algebra routines of the local eigenvalue problem are parallelizable via libraries
        such as ScaLAPACK or MAGMA. Both libraries provide routines for distributed memory
        machines \cite{Blackford1997ScalapackUsersGuide}; MAGMA supports moreover CPU and GPUs. In this way, the workload
        of the eigenproblem procedure can be split into several computation nodes. Then, it
        is important to analyze the performances as a function of the bond dimension $\chi$,
        to test the effectiveness of this approach against the latency of the inter-node
        communications.
    \item[(iv)] \emph{Tuning of parameters and algorithms:} accelerators developed for machine learning
        applications have excellent support for lower and real precision. Tuning parameters
        over the different sweeps is beneficial, e.g. increasing the precision towards the end of the sweep. This approach profits from faster single-precision implementations during the first sweeps. Selecting algorithms like random singular value decompositions can also bring benefits \cite{Lu2017HighPerformance}.
\end{itemize}

\subsection{Sweeps and HPC parallelization}        
\label{sec:opt:hpcmpi}
So far, we have parallelized single tensor optimizations within a sweep, but the sweep itself was sequential, i.e. serial. 
Recent works formulated parallel versions of MPS algorithms for ground state search and time evolution, e.g. via the Message Passing Interface (MPI) \cite{Gabriel2004OpenMpiGoals, Stoudenmire2013RealSpaceParallel, Secular2020ParallelTimeDependent}.
The main difference between the serial and parallel algorithms is the effective operators used in the optimization. 
In the serial version, the effective operators contain the information of the most recent version of all other tensors. 
This update is delayed in the parallel version, i.e. the tensor that entered the effective operator is not necessarily the one of the current sweep but can be the version of an earlier sweep.

\begin{figure*}
  \includegraphics[width=1\textwidth]{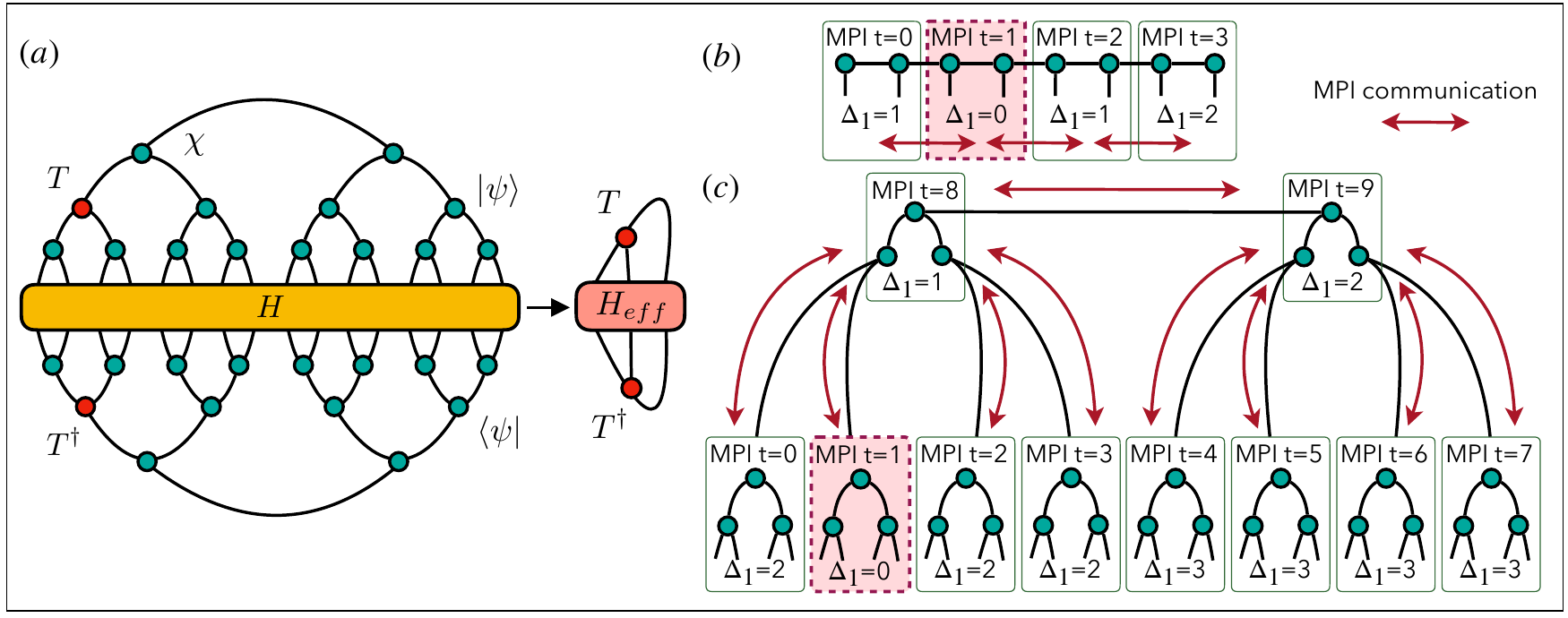}
  \caption{\emph{Effective operators and parallel tensor networks.}
    (a)~Procedure for optimizing a TTN to find the ground state of a QMB system: the energy is computed by contracting the Hamiltonian $H$ (yellow tensor) with the TTN, representing the state $\ket{\psi}$, and its hermitian conjugate, representing $\bra{\psi}$.
    The variational optimization starts from a target tensor $T$ (red tensor), by computing its effective Hamiltonian $H_{\text{eff}}$ and then solving the local eigenvalue problem for the latter. 
    The tensor $T$ is then updated with the newly found ground state, and the procedure iterates over all the tensors in the network (sweep).
    (b) The workload itself consists of optimizing each tensor held by the MPI thread $t$, which requires effective operators calculated by other MPI threads.
    We dub \emph{delays} $\Delta_{i}$ the number of optimization cycles needed to obtain the information of tensors in the i-th MPI thread in another MPI thread via MPI communication.
    Matrix product states naturally split into sub-chains which communicate with one or two neighboring MPI threads to obtain updated effective operators. Delays for updates scale with the distance between two MPI threads along the chain. 
    Each MPI thread can use threading or openMP, e.g. in a hybrid openMP-MPI approach.
    (c)~Similarly, TTNs can be split into sub-trees for each MPI thread allowing for optimizing the sub-tree without communication with other MPI threads.
    Delays due to updating scale logarithmically as any distance in a tree network.}
  \label{fig_TTN_optimization}
\end{figure*}

If the delay becomes an obstacle to convergence, there is the option to modify parameters during the sweeps. 
Typically, ten to fifty sweeps are necessary to converge to a solution. 
As the initial state is usually random, MPI can be used especially at the beginning. 
To ensure convergence, one can consider serial steps at the end; even a gradual reduction of the MPI processes as the sweeps proceed is possible and gradually reduces the delays.

Considering the MPS scenario of a chain in \cref{fig_TTN_optimization}(b), we split the chain into equal parts of $(N / \nummpithreads)$ sites. 
Each part of the chain communicates with its two neighbors apart from the two boundaries.
The effective operators take into account the tensors of the same MPI process with zero delay as in the serial case.
The tensors of the i-th neighboring MPI process have a delay of $i$. 
The worst-case scenario of the delay scales linearly with the number of MPI processes.
The delay can be avoided by communicating the effective operators after each update through the chain, which is a quasi-serial step with no more than two MPI processes active at the same time.

The problem becomes more complicated for the TTNs suggested for LGT, but we expect a benefit for the parallelization of a TTN versus an MPS for higher dimensional systems.
\cref{fig_TTN_optimization}(c) shows an example of how each MPI process gets assigned a sub-tree within the complete TTN.
Unlike the MPS, the number of neighboring MPI processes for communication is at least three and increases with $(N / \nummpithreads)$ tensors per thread.
For simplicity, we assume equally shaped subtrees for all MPI processes. 
The delay of the tensor update is now $2 \cdot \logtwo(N)$ in the worst-case scenario.

One-dimensional systems with nearest-neighbor interactions thus exhibit a delay of $1$ in the worst-case scenario in the MPS, while the delay is up to $2 \cdot \logtwo(N)$ for the nearest neighbors in the center of the TTN. 
Rather, higher dimensional systems change this aspect, e.g., for an $N \times N$ two-dimensional system mapped to 1D via a zig-zag mapping. 
The MPS has a worst-case delay of the tensor update of $N$ for the slow index. 
In contrast, the TTN has the same log behavior and a maximum delay at the center of the TTN as $2 \cdot \logtwo(N^2)$. 
Thus, the worst-case delay is equal for $16 \times 16$ systems; increasing $N$ further, TTNs exhibit smaller maximum delays during parallel sweeps. 
Moreover, the TTN is unaffected by the type of mapping used; in contrast, the worst-case delay for the MPS grows to $2N$ for the snake mapping and to at least $N^2 / 2$ for the Hilbert curve \cite{Cataldi2021hilbertcurvevs}. 
Equal arguments hold for 3D systems and delays of $N^2$ (MPS, zig-zag) versus $2 \logtwo(N^3)$ (TTN, any mapping).

To get an intuition of what parallelized simulations can solve, we sketch out the specifications for a parallel simulation on the pre-exascale cluster \emph{Leonardo} hosted by \emph{Cineca}.
We choose an MPI approach together with the GPUs. Leonardo has 3456 nodes with four GPUs totaling 13824 GPUs available for the complete cluster. 
Bond dimensions on the order of $\chi = 450$ consume 54GB of memory without effective operators (assuming double complex precision, 40 Lanczos vectors) and allow to solve the eigenproblem on the GPU without temporarily storing data on the CPU.
We use the single-tensor per MPI-thread with $\chi=450$ as a baseline where we extract a rough empirical estimate with Leonardo; in detail, we use a 2D quantum Ising model with $\mathbb{Z}_{2}$ symmetry in the vicinity of the quantum critical point and the initial tensor optimizations \cite{qtealeaves,dibona}.
Due to the delay of the tensor in the effective operators, the minimum number of sweeps must be beyond 24. 
Then, we consider the scaling of the TTN previously introduced and generate \cref{hpctime} with an overview of different system sizes and bond dimensions. 
These results provide a coarse-grained estimate, since plaquette terms, different symmetries, entanglement generation, and optimization time within later sweeps could further impact the computational time.

Our estimate predicts that a system of $256 \times 256$ sites takes about two months for bond dimension $\chi=450$ on 1024 GPU nodes of Cineca's \emph{leonardo}.
Future improvements are likely to bring this simulation time down, e.g., the next-generation GPUs in comparison to the A100 or further optimization in data movement.
In contrast, a three-dimensional system with large entanglement and many sites requires three to four orders of magnitude improvement, where cluster size and other improvements have to come together to reach this challenge.
Furthermore, \cref{hpctime} provides an estimate of the boundary for a potential quantum advantage in simulating lattice gauge theories with quantum computers or simulators.
\begin{table}[t]
  \begin{center}
    \begin{tabular}{@{} lcrr @{}}
      \toprule
      System size              & $\chi$       & Factor               & Estimated walltime    \\
      \cmidrule(r){1-1} \cmidrule(r){2-2} \cmidrule(rl){3-3} \cmidrule(l){4-4}
      $64 \times 64$           & $ 450$ & $\Tbase$             & $4.16~\mathrm{days}$  \\
      $64 \times 64$           & $ 900$ & $16 \cdot \Tbase$    & $66.6~\mathrm{days}$  \\
      $256 \times 256$         & $ 450$ & $28 \cdot \Tbase$    & $116.5~\mathrm{days}$ \\
      $256 \times 256$         & $ 900$ & $448 \cdot \Tbase$   & $5.1~\mathrm{years}$  \\
      $16 \times 16 \times 16$ & $ 450$ & $4 \cdot \Tbase$     & $16.6~\mathrm{days}$  \\
      $16 \times 16 \times 16$ & $ 900$ & $64 \cdot \Tbase$    & $266~\mathrm{days}$   \\
      $64 \times 64 \times 64$ & $ 450$ & $1984 \cdot \Tbase$  & $23~\mathrm{years}$   \\
      $64 \times 64 \times 64$ & $ 900$ & $31744 \cdot \Tbase$ & $362~\mathrm{years}$  \\
      \bottomrule
    \end{tabular}
    \caption{\label{hpctime}%
      \emph{Estimated simulation time.} We derive the baseline from
      a single-tensor optimization of a $64 \times 64$ quantum Ising simulation with
      $\mathbb{Z}_{2}$ symmetry taking $7192\mathrm{s}$ on a A100 GPU. Further, we assume
      that single-tensor update, one tensor and one GPU per MPI thread, and 50 sweeps for
      the baseline. To extrapolate to larger systems, we assume a scaling with $\mathcal{O}(\chi^4 N^{D-1})$
      as well as seven (thirty-one) tensors per MPI thread for $256 \times 256$ ($64 \times 64 \times 64$) systems. The empirical scalings
      are approximately a factor of $2.3$ for doubling the system size and $13$ for doubling
      the bond dimension, which we obtain from smaller
      simulations with $\chi = 225$ and for $32 \times 32$ qubits. The times are valid for any $d < \chi$.
    }
  \end{center}
\end{table}

\subsection{Finite temperature regime}
\label{sec:opt:finitet}
To date, TN simulations of high-dimensional LGTs including dynamical matter are exploring zero temperature regimes, which are important to understand the low-energy properties of the models. 
To explore finite temperature phenomena, particularly relevant for open research problems such as the QCD phase diagram, technical and numerical challenges have to be tackled, e.g. devising and testing efficient TN algorithms for targeting quantum states at equilibrium. 
As suggested in the next paragraph, we foresee two possible paths toward finite temperature TN states.

Matrix product density operators (MPDOs) and locally purified tensor networks (LPTNs) provide already today the option to tackle finite temperature regimes via an imaginary time evolution \cite{Verstraete2004MatrixProductDensity, Zwolak2004MixedStateDynamics, Werner2016PositiveTensorNetwork}.
Herein, the algorithm starts at the infinite temperature state and starts \enquote{cooling} the system via a specified number of time steps and specified step size to reach a given temperature. 
In its original formulation, both approaches are one-dimensional chains. 
Matrix product density operator can be formulated as TTN but faces some challenges in terms of the question of positivity \cite{Kliesch2014MatrixProductOperators} or integrating symmetries. 
In contrast, tree tensor operators (TTO) are the tree-equivalent of an LPTN; they are also a positive loopless representation of density matrices, recently introduced in \cite{Arceci2022EntanglementFormationMixed}.
However, TTOs cannot represent the infinite temperature state necessary for the imaginary time evolution approach.
Instead, a possible use of the TTO employed in LGT simulations consists of a variational algorithm to target finite-temperature states or reconstruct open-system quantum dynamics, by efficiently compressing the relevant information.
The TTO enables useful measures, e.g. computing the entanglement of formation as already shown for representative one-dimensional models at finite temperature \cite{Arceci2022EntanglementFormationMixed}.

\section{Conclusions}
\label{sec_Conclusions}
The field of TN methods for LGTs has shown great potential in the last decade, during which significant efforts have been devoted to developing numerical algorithms and strategies that can be seen as a complementary approach to MC simulations for high-energy physics.
The validity of sign-problem-free TN algorithms has been proven for one-dimensional LGT models for both Abelian and non-Abelian scenarios.
Recently, TN methods have also been applied to higher-dimensional LGTs with simple truncated gauge groups and small-to-intermediate lattice sizes.
On the one hand, these results prove the effectiveness of TN methods for simulating LGTs even in regimes that are problematic for other numerical methods; on the other hand, they highlight the challenges that one needs to tackle to address state-of-the-art research problems, such as accessing the continuum limits or simulating high-dimensional QCD on large lattices.

In this work, after a general overview of TN methods and their use for LGT simulations, we have described these challenges, starting from the problem of the very large local basis required for complex gauge groups and then discussing several computational bottlenecks in terms of bond dimensions and system sizes.

To mitigate and possibly overcome these problems soon, we have presented a feasible roadmap, in terms of algorithmic developments and numerical strategies that might have a concrete impact in extending the range of applicability of TN algorithms to current research problems in high-energy physics. We foresee that all the presented steps could potentially have a key role in making the TN approach competitive as a complementary method to MC techniques for simulating high-dimensional LGTs, such as large-scale QCD.

\begin{acknowledgments}
The authors acknowledge financial support from the following institutions:
the European Union (EU)
via the Horizon 2020 research and innovation program (Quantum Flagship) projects PASQuanS2 and Euryqa,
via the NextGenerationEU project CN00000013 - Italian Research Center on HPC, Big Data and Quantum Computing (ICSC),
and via the QuantERA projects T-NiSQ and QuantHEP;
the Italian Ministry of University and Research (MUR)
via the Departments of Excellence grants 2023-2027 projects Quantum Frontiers and Quantum Sensing and Modelling for One-Health (QuaSiModO),
and via PRIN2022 project TANQU;
the Italian Istituto Nazionale di Fisica Nucleare (INFN)
via specific initiatives IS-QUANTUM and IS-NPQCD;
the German Federal Ministry of Education and Research (BMBF)
via project QRydDemo;
the University of Bari 
via the 2023-UNBACLE-0244025 grant;
the World-Class Research Infrastructure $-$ Quantum Computing and Simulation Center (QCSC) of Padova University.
We acknowledge computational resources by the Cloud Veneto, CINECA, the BwUniCluster, the University of Padova Strategic Research Infrastructure Grant 2017: “CAPRI: Calcolo ad Alte Prestazioni per la Ricerca e l’Innovazione”, and ReCaS Bari.
\end{acknowledgments}

\bibliography{references}

\providecommand{\noopsort}[1]{}\providecommand{\singleletter}[1]{#1}%
\begin{thebibliography}{172}%
\makeatletter
\providecommand \@ifxundefined [1]{%
 \@ifx{#1\undefined}
}%
\providecommand \@ifnum [1]{%
 \ifnum #1\expandafter \@firstoftwo
 \else \expandafter \@secondoftwo
 \fi
}%
\providecommand \@ifx [1]{%
 \ifx #1\expandafter \@firstoftwo
 \else \expandafter \@secondoftwo
 \fi
}%
\providecommand \natexlab [1]{#1}%
\providecommand \enquote  [1]{``#1''}%
\providecommand \bibnamefont  [1]{#1}%
\providecommand \bibfnamefont [1]{#1}%
\providecommand \citenamefont [1]{#1}%
\providecommand \href@noop [0]{\@secondoftwo}%
\providecommand \href [0]{\begingroup \@sanitize@url \@href}%
\providecommand \@href[1]{\@@startlink{#1}\@@href}%
\providecommand \@@href[1]{\endgroup#1\@@endlink}%
\providecommand \@sanitize@url [0]{\catcode `\\12\catcode `\$12\catcode
  `\&12\catcode `\#12\catcode `\^12\catcode `\_12\catcode `\%12\relax}%
\providecommand \@@startlink[1]{}%
\providecommand \@@endlink[0]{}%
\providecommand \url  [0]{\begingroup\@sanitize@url \@url }%
\providecommand \@url [1]{\endgroup\@href {#1}{\urlprefix }}%
\providecommand \urlprefix  [0]{URL }%
\providecommand \Eprint [0]{\href }%
\providecommand \doibase [0]{https://doi.org/}%
\providecommand \selectlanguage [0]{\@gobble}%
\providecommand \bibinfo  [0]{\@secondoftwo}%
\providecommand \bibfield  [0]{\@secondoftwo}%
\providecommand \translation [1]{[#1]}%
\providecommand \BibitemOpen [0]{}%
\providecommand \bibitemStop [0]{}%
\providecommand \bibitemNoStop [0]{.\EOS\space}%
\providecommand \EOS [0]{\spacefactor3000\relax}%
\providecommand \BibitemShut  [1]{\csname bibitem#1\endcsname}%
\let\auto@bib@innerbib\@empty
\bibitem [{\citenamefont {Kleinert}(1989)}]{Kleinert1989GaugeFieldsCondensed}%
  \BibitemOpen
  \bibfield  {author} {\bibinfo {author} {\bibfnamefont {H.}~\bibnamefont
  {Kleinert}},\ }\href {https://doi.org/10.1142/0356} {\emph {\bibinfo {title}
  {Gauge {{Fields}} in {{Condensed Matter}}}}}\ (\bibinfo  {publisher} {{WORLD
  SCIENTIFIC}},\ \bibinfo {year} {1989})\BibitemShut {NoStop}%
\bibitem [{\citenamefont {Fradkin}(2013)}]{Fradkin2013FieldTheoriesCondensed}%
  \BibitemOpen
  \bibfield  {author} {\bibinfo {author} {\bibfnamefont {E.}~\bibnamefont
  {Fradkin}},\ }\href {https://doi.org/10.1017/CBO9781139015509} {\emph
  {\bibinfo {title} {Field {{Theories}} of {{Condensed Matter Physics}}}}},\
  \bibinfo {edition} {2nd}\ ed.\ (\bibinfo  {publisher} {Cambridge University
  Press},\ \bibinfo {address} {{Cambridge}},\ \bibinfo {year}
  {2013})\BibitemShut {NoStop}%
\bibitem [{\citenamefont {Ichinose}\ and\ \citenamefont
  {Matsui}(2014)}]{Ichinose2014LatticeGaugeTheory}%
  \BibitemOpen
  \bibfield  {author} {\bibinfo {author} {\bibfnamefont {I.}~\bibnamefont
  {Ichinose}}\ and\ \bibinfo {author} {\bibfnamefont {T.}~\bibnamefont
  {Matsui}},\ }\bibfield  {title} {\bibinfo {title} {Lattice gauge theory for
  condensed matter physics: ferromagnetic superconductivity as its example},\
  }\href {https://doi.org/10.1142/S0217984914300129} {\bibfield  {journal}
  {\bibinfo  {journal} {Modern Physics Letters B}\ }\textbf {\bibinfo {volume}
  {28}},\ \bibinfo {pages} {1430012} (\bibinfo {year} {2014})}\BibitemShut
  {NoStop}%
\bibitem [{\citenamefont {Peskin}\ and\ \citenamefont
  {Schroeder}(1995)}]{Peskin1995IntroductionQuantumField}%
  \BibitemOpen
  \bibfield  {author} {\bibinfo {author} {\bibfnamefont {M.~E.}\ \bibnamefont
  {Peskin}}\ and\ \bibinfo {author} {\bibfnamefont {D.~V.}\ \bibnamefont
  {Schroeder}},\ }\href {https://cds.cern.ch/record/257493} {\emph {\bibinfo
  {title} {{An introduction to quantum field theory}}}}\ (\bibinfo  {publisher}
  {Westview},\ \bibinfo {address} {Boulder, CO},\ \bibinfo {year}
  {1995})\BibitemShut {NoStop}%
\bibitem [{\citenamefont {Schwartz}(2013)}]{Schwartz2013QuantumFieldTheory}%
  \BibitemOpen
  \bibfield  {author} {\bibinfo {author} {\bibfnamefont {M.~D.}\ \bibnamefont
  {Schwartz}},\ }\href {https://doi.org/10.1017/9781139540940} {\emph {\bibinfo
  {title} {Quantum {{Field Theory}} and the {{Standard Model}}}}}\ (\bibinfo
  {publisher} {Cambridge University Press},\ \bibinfo {year}
  {2013})\BibitemShut {NoStop}%
\bibitem [{\citenamefont {Wilson}(1974)}]{Wilson1974ConfinementQuarks}%
  \BibitemOpen
  \bibfield  {author} {\bibinfo {author} {\bibfnamefont {K.~G.}\ \bibnamefont
  {Wilson}},\ }\bibfield  {title} {\bibinfo {title} {Confinement of quarks},\
  }\href {https://doi.org/10.1103/PhysRevD.10.2445} {\bibfield  {journal}
  {\bibinfo  {journal} {Phys. Rev. D}\ }\textbf {\bibinfo {volume} {10}},\
  \bibinfo {pages} {2445} (\bibinfo {year} {1974})}\BibitemShut {NoStop}%
\bibitem [{\citenamefont {Kogut}(1979)}]{Kogut1979IntroductionLatticeGauge}%
  \BibitemOpen
  \bibfield  {author} {\bibinfo {author} {\bibfnamefont {J.~B.}\ \bibnamefont
  {Kogut}},\ }\bibfield  {title} {\bibinfo {title} {An introduction to lattice
  gauge theory and spin systems},\ }\href
  {https://doi.org/10.1103/RevModPhys.51.659} {\bibfield  {journal} {\bibinfo
  {journal} {Reviews of Modern Physics}\ }\textbf {\bibinfo {volume} {51}},\
  \bibinfo {pages} {659} (\bibinfo {year} {1979})}\BibitemShut {NoStop}%
\bibitem [{\citenamefont {Creutz}\ \emph {et~al.}(1983)\citenamefont {Creutz},
  \citenamefont {Jacobs},\ and\ \citenamefont
  {Rebbi}}]{Creutz1983MonteCarloComputations}%
  \BibitemOpen
  \bibfield  {author} {\bibinfo {author} {\bibfnamefont {M.}~\bibnamefont
  {Creutz}}, \bibinfo {author} {\bibfnamefont {L.}~\bibnamefont {Jacobs}},\
  and\ \bibinfo {author} {\bibfnamefont {C.}~\bibnamefont {Rebbi}},\ }\bibfield
   {title} {\bibinfo {title} {Monte {{Carlo}} computations in lattice gauge
  theories},\ }\href {https://doi.org/10.1016/0370-1573(83)90016-9} {\bibfield
  {journal} {\bibinfo  {journal} {Physics Reports}\ }\textbf {\bibinfo {volume}
  {95}},\ \bibinfo {pages} {201} (\bibinfo {year} {1983})}\BibitemShut
  {NoStop}%
\bibitem [{\citenamefont {Davoudi}\ \emph {et~al.}(2022)\citenamefont
  {Davoudi}, \citenamefont {Neil}, \citenamefont {Bauer}, \citenamefont
  {Bhattacharya}, \citenamefont {Blum}, \citenamefont {Boyle}, \citenamefont
  {Brower}, \citenamefont {Catterall}, \citenamefont {Christ}, \citenamefont
  {Cirigliano}, \citenamefont {Colangelo}, \citenamefont {DeTar}, \citenamefont
  {Detmold}, \citenamefont {Edwards}, \citenamefont {El-Khadra}, \citenamefont
  {Gottlieb}, \citenamefont {Gupta}, \citenamefont {Hackett}, \citenamefont
  {Hasenfratz}, \citenamefont {Izubuchi}, \citenamefont {Jay}, \citenamefont
  {Jin}, \citenamefont {Kelly}, \citenamefont {Kronfeld}, \citenamefont
  {Lehner}, \citenamefont {Lin}, \citenamefont {Lin}, \citenamefont {Lytle},
  \citenamefont {Meinel}, \citenamefont {Meurice}, \citenamefont {Mukherjee},
  \citenamefont {Nicholson}, \citenamefont {Prelovsek}, \citenamefont {Savage},
  \citenamefont {Shanahan}, \citenamefont {Van De~Water}, \citenamefont
  {Wagman},\ and\ \citenamefont {Witzel}}]{Davoudi2022ReportSnowmass2021}%
  \BibitemOpen
  \bibfield  {author} {\bibinfo {author} {\bibfnamefont {Z.}~\bibnamefont
  {Davoudi}}, \bibinfo {author} {\bibfnamefont {E.~T.}\ \bibnamefont {Neil}},
  \bibinfo {author} {\bibfnamefont {C.~W.}\ \bibnamefont {Bauer}}, \bibinfo
  {author} {\bibfnamefont {T.}~\bibnamefont {Bhattacharya}}, \bibinfo {author}
  {\bibfnamefont {T.}~\bibnamefont {Blum}}, \bibinfo {author} {\bibfnamefont
  {P.}~\bibnamefont {Boyle}}, \bibinfo {author} {\bibfnamefont {R.~C.}\
  \bibnamefont {Brower}}, \bibinfo {author} {\bibfnamefont {S.}~\bibnamefont
  {Catterall}}, \bibinfo {author} {\bibfnamefont {N.~H.}\ \bibnamefont
  {Christ}}, \bibinfo {author} {\bibfnamefont {V.}~\bibnamefont {Cirigliano}},
  \bibinfo {author} {\bibfnamefont {G.}~\bibnamefont {Colangelo}}, \bibinfo
  {author} {\bibfnamefont {C.}~\bibnamefont {DeTar}}, \bibinfo {author}
  {\bibfnamefont {W.}~\bibnamefont {Detmold}}, \bibinfo {author} {\bibfnamefont
  {R.~G.}\ \bibnamefont {Edwards}}, \bibinfo {author} {\bibfnamefont {A.~X.}\
  \bibnamefont {El-Khadra}}, \bibinfo {author} {\bibfnamefont {S.}~\bibnamefont
  {Gottlieb}}, \bibinfo {author} {\bibfnamefont {R.}~\bibnamefont {Gupta}},
  \bibinfo {author} {\bibfnamefont {D.~C.}\ \bibnamefont {Hackett}}, \bibinfo
  {author} {\bibfnamefont {A.}~\bibnamefont {Hasenfratz}}, \bibinfo {author}
  {\bibfnamefont {T.}~\bibnamefont {Izubuchi}}, \bibinfo {author}
  {\bibfnamefont {W.~I.}\ \bibnamefont {Jay}}, \bibinfo {author} {\bibfnamefont
  {L.}~\bibnamefont {Jin}}, \bibinfo {author} {\bibfnamefont {C.}~\bibnamefont
  {Kelly}}, \bibinfo {author} {\bibfnamefont {A.~S.}\ \bibnamefont {Kronfeld}},
  \bibinfo {author} {\bibfnamefont {C.}~\bibnamefont {Lehner}}, \bibinfo
  {author} {\bibfnamefont {H.-W.}\ \bibnamefont {Lin}}, \bibinfo {author}
  {\bibfnamefont {M.}~\bibnamefont {Lin}}, \bibinfo {author} {\bibfnamefont
  {A.~T.}\ \bibnamefont {Lytle}}, \bibinfo {author} {\bibfnamefont
  {S.}~\bibnamefont {Meinel}}, \bibinfo {author} {\bibfnamefont
  {Y.}~\bibnamefont {Meurice}}, \bibinfo {author} {\bibfnamefont
  {S.}~\bibnamefont {Mukherjee}}, \bibinfo {author} {\bibfnamefont
  {A.}~\bibnamefont {Nicholson}}, \bibinfo {author} {\bibfnamefont
  {S.}~\bibnamefont {Prelovsek}}, \bibinfo {author} {\bibfnamefont {M.~J.}\
  \bibnamefont {Savage}}, \bibinfo {author} {\bibfnamefont {P.~E.}\
  \bibnamefont {Shanahan}}, \bibinfo {author} {\bibfnamefont {R.~S.}\
  \bibnamefont {Van De~Water}}, \bibinfo {author} {\bibfnamefont {M.~L.}\
  \bibnamefont {Wagman}},\ and\ \bibinfo {author} {\bibfnamefont
  {O.}~\bibnamefont {Witzel}},\ }\bibfield  {title} {\bibinfo {title} {{Report
  of the Snowmass 2021 Topical Group on Lattice Gauge Theory}},\ }\href
  {https://doi.org/10.48550/arXiv.2209.10758} {\bibfield  {journal} {\bibinfo
  {journal} {arXiv e-prints}\ ,\ \bibinfo {eid} {arXiv:2209.10758}} (\bibinfo
  {year} {2022})},\ \Eprint {https://arxiv.org/abs/2209.10758}
  {arXiv:2209.10758 [hep-lat]} \BibitemShut {NoStop}%
\bibitem [{\citenamefont {Nagata}(2022)}]{Nagata2022FiniteDensityLattice}%
  \BibitemOpen
  \bibfield  {author} {\bibinfo {author} {\bibfnamefont {K.}~\bibnamefont
  {Nagata}},\ }\bibfield  {title} {\bibinfo {title} {Finite-density lattice qcd
  and sign problem: Current status and open problems},\ }\href
  {https://doi.org/10.1016/j.ppnp.2022.103991} {\bibfield  {journal} {\bibinfo
  {journal} {Progress in Particle and Nuclear Physics}\ }\textbf {\bibinfo
  {volume} {127}},\ \bibinfo {pages} {103991} (\bibinfo {year}
  {2022})}\BibitemShut {NoStop}%
\bibitem [{\citenamefont {Banuls}\ \emph {et~al.}(2019)\citenamefont {Banuls},
  \citenamefont {Cichy}, \citenamefont {Cirac}, \citenamefont {Jansen},\ and\
  \citenamefont {K{\"{u}}hn}}]{Banuls2019TensorNetworksTheir}%
  \BibitemOpen
  \bibfield  {author} {\bibinfo {author} {\bibfnamefont {M.~C.}\ \bibnamefont
  {Banuls}}, \bibinfo {author} {\bibfnamefont {K.}~\bibnamefont {Cichy}},
  \bibinfo {author} {\bibfnamefont {J.~I.}\ \bibnamefont {Cirac}}, \bibinfo
  {author} {\bibfnamefont {K.}~\bibnamefont {Jansen}},\ and\ \bibinfo {author}
  {\bibfnamefont {S.}~\bibnamefont {K{\"{u}}hn}},\ }\bibfield  {title}
  {\bibinfo {title} {Tensor networks and their use for lattice gauge
  theories},\ }in\ \href {https://doi.org/10.22323/1.334.0022} {\emph {\bibinfo
  {booktitle} {Proceedings of The 36th Annual International Symposium on
  Lattice Field Theory {\textemdash} {PoS}({LATTICE}2018)}}},\ Vol.\ \bibinfo
  {volume} {334}\ (\bibinfo  {publisher} {Sissa Medialab},\ \bibinfo {year}
  {2019})\ p.\ \bibinfo {pages} {022}\BibitemShut {NoStop}%
\bibitem [{\citenamefont
  {Montangero}(2018)}]{Montangero2018IntroductionTensorNetwork}%
  \BibitemOpen
  \bibfield  {author} {\bibinfo {author} {\bibfnamefont {S.}~\bibnamefont
  {Montangero}},\ }\href {https://doi.org/10.1007/978-3-030-01409-4} {\emph
  {\bibinfo {title} {Introduction to {{Tensor Network Methods}}: {{Numerical}}
  Simulations of Low-Dimensional Many-Body Quantum Systems}}}\ (\bibinfo
  {publisher} {Springer International Publishing},\ \bibinfo {address}
  {{Cham}},\ \bibinfo {year} {2018})\BibitemShut {NoStop}%
\bibitem [{\citenamefont {Or{\'u}s}(2019)}]{Orus2019TensorNetworksComplex}%
  \BibitemOpen
  \bibfield  {author} {\bibinfo {author} {\bibfnamefont {R.}~\bibnamefont
  {Or{\'u}s}},\ }\bibfield  {title} {\bibinfo {title} {Tensor networks for
  complex quantum systems},\ }\href {https://doi.org/10.1038/s42254-019-0086-7}
  {\bibfield  {journal} {\bibinfo  {journal} {Nature Reviews Physics}\ }\textbf
  {\bibinfo {volume} {1}},\ \bibinfo {pages} {538} (\bibinfo {year}
  {2019})}\BibitemShut {NoStop}%
\bibitem [{\citenamefont
  {Ba\~{n}uls}(2023)}]{Banuls2023TensorNetworkAlgorithms}%
  \BibitemOpen
  \bibfield  {author} {\bibinfo {author} {\bibfnamefont {M.~C.}\ \bibnamefont
  {Ba\~{n}uls}},\ }\bibfield  {title} {\bibinfo {title} {Tensor network
  algorithms: A route map},\ }\href
  {https://doi.org/10.1146/annurev-conmatphys-040721-022705} {\bibfield
  {journal} {\bibinfo  {journal} {Annual Review of Condensed Matter Physics}\
  }\textbf {\bibinfo {volume} {14}},\ \bibinfo {pages} {null} (\bibinfo {year}
  {2023})}\BibitemShut {NoStop}%
\bibitem [{\citenamefont {Byrnes}\ \emph {et~al.}(2002)\citenamefont {Byrnes},
  \citenamefont {Sriganesh}, \citenamefont {Bursill},\ and\ \citenamefont
  {Hamer}}]{Byrnes2002DensityMatrixRenormalization}%
  \BibitemOpen
  \bibfield  {author} {\bibinfo {author} {\bibfnamefont {T.~M.~R.}\
  \bibnamefont {Byrnes}}, \bibinfo {author} {\bibfnamefont {P.}~\bibnamefont
  {Sriganesh}}, \bibinfo {author} {\bibfnamefont {R.~J.}\ \bibnamefont
  {Bursill}},\ and\ \bibinfo {author} {\bibfnamefont {C.~J.}\ \bibnamefont
  {Hamer}},\ }\bibfield  {title} {\bibinfo {title} {Density matrix
  renormalization group approach to the massive schwinger model},\ }\href
  {https://doi.org/10.1103/PhysRevD.66.013002} {\bibfield  {journal} {\bibinfo
  {journal} {Phys. Rev. D}\ }\textbf {\bibinfo {volume} {66}},\ \bibinfo
  {pages} {013002} (\bibinfo {year} {2002})}\BibitemShut {NoStop}%
\bibitem [{\citenamefont {Silvi}\ \emph {et~al.}(2014)\citenamefont {Silvi},
  \citenamefont {Rico}, \citenamefont {Calarco},\ and\ \citenamefont
  {Montangero}}]{Silvi2014LatticeGaugeTensor}%
  \BibitemOpen
  \bibfield  {author} {\bibinfo {author} {\bibfnamefont {P.}~\bibnamefont
  {Silvi}}, \bibinfo {author} {\bibfnamefont {E.}~\bibnamefont {Rico}},
  \bibinfo {author} {\bibfnamefont {T.}~\bibnamefont {Calarco}},\ and\ \bibinfo
  {author} {\bibfnamefont {S.}~\bibnamefont {Montangero}},\ }\bibfield  {title}
  {\bibinfo {title} {Lattice gauge tensor networks},\ }\href
  {https://doi.org/10.1088/1367-2630/16/10/103015} {\bibfield  {journal}
  {\bibinfo  {journal} {New Journal of Physics}\ }\textbf {\bibinfo {volume}
  {16}},\ \bibinfo {pages} {103015} (\bibinfo {year} {2014})}\BibitemShut
  {NoStop}%
\bibitem [{\citenamefont {Pichler}\ \emph {et~al.}(2016)\citenamefont
  {Pichler}, \citenamefont {Dalmonte}, \citenamefont {Rico}, \citenamefont
  {Zoller},\ and\ \citenamefont {Montangero}}]{Pichler2016RealTimeDynamics}%
  \BibitemOpen
  \bibfield  {author} {\bibinfo {author} {\bibfnamefont {T.}~\bibnamefont
  {Pichler}}, \bibinfo {author} {\bibfnamefont {M.}~\bibnamefont {Dalmonte}},
  \bibinfo {author} {\bibfnamefont {E.}~\bibnamefont {Rico}}, \bibinfo {author}
  {\bibfnamefont {P.}~\bibnamefont {Zoller}},\ and\ \bibinfo {author}
  {\bibfnamefont {S.}~\bibnamefont {Montangero}},\ }\bibfield  {title}
  {\bibinfo {title} {Real-{Time Dynamics} in {U}(1) {Lattice Gauge Theories}
  with {Tensor Networks}},\ }\href {https://doi.org/10.1103/PhysRevX.6.011023}
  {\bibfield  {journal} {\bibinfo  {journal} {Physical Review X}\ }\textbf
  {\bibinfo {volume} {6}},\ \bibinfo {pages} {011023} (\bibinfo {year}
  {2016})}\BibitemShut {NoStop}%
\bibitem [{\citenamefont {Silvi}\ \emph {et~al.}(2019)\citenamefont {Silvi},
  \citenamefont {Tschirsich}, \citenamefont {Gerster}, \citenamefont
  {J{\"u}nemann}, \citenamefont {Jaschke}, \citenamefont {Rizzi},\ and\
  \citenamefont {Montangero}}]{Silvi2019TensorNetworksAnthology}%
  \BibitemOpen
  \bibfield  {author} {\bibinfo {author} {\bibfnamefont {P.}~\bibnamefont
  {Silvi}}, \bibinfo {author} {\bibfnamefont {F.}~\bibnamefont {Tschirsich}},
  \bibinfo {author} {\bibfnamefont {M.}~\bibnamefont {Gerster}}, \bibinfo
  {author} {\bibfnamefont {J.}~\bibnamefont {J{\"u}nemann}}, \bibinfo {author}
  {\bibfnamefont {D.}~\bibnamefont {Jaschke}}, \bibinfo {author} {\bibfnamefont
  {M.}~\bibnamefont {Rizzi}},\ and\ \bibinfo {author} {\bibfnamefont
  {S.}~\bibnamefont {Montangero}},\ }\bibfield  {title} {\bibinfo {title} {The
  {{Tensor Networks Anthology}}: {{Simulation}} techniques for many-body
  quantum lattice systems},\ }\href
  {https://doi.org/10.21468/SciPostPhysLectNotes.8} {\bibfield  {journal}
  {\bibinfo  {journal} {SciPost Physics Lecture Notes}\ ,\ \bibinfo {pages}
  {008}} (\bibinfo {year} {2019})}\BibitemShut {NoStop}%
\bibitem [{\citenamefont {Funcke}\ \emph {et~al.}(2020)\citenamefont {Funcke},
  \citenamefont {Jansen},\ and\ \citenamefont
  {K\"uhn}}]{Funcke2020TopologicalVacuumStructure}%
  \BibitemOpen
  \bibfield  {author} {\bibinfo {author} {\bibfnamefont {L.}~\bibnamefont
  {Funcke}}, \bibinfo {author} {\bibfnamefont {K.}~\bibnamefont {Jansen}},\
  and\ \bibinfo {author} {\bibfnamefont {S.}~\bibnamefont {K\"uhn}},\
  }\bibfield  {title} {\bibinfo {title} {Topological vacuum structure of the
  schwinger model with matrix product states},\ }\href
  {https://doi.org/10.1103/PhysRevD.101.054507} {\bibfield  {journal} {\bibinfo
   {journal} {Phys. Rev. D}\ }\textbf {\bibinfo {volume} {101}},\ \bibinfo
  {pages} {054507} (\bibinfo {year} {2020})}\BibitemShut {NoStop}%
\bibitem [{\citenamefont {Buyens}\ \emph {et~al.}(2014)\citenamefont {Buyens},
  \citenamefont {Haegeman}, \citenamefont {Van~Acoleyen}, \citenamefont
  {Verschelde},\ and\ \citenamefont
  {Verstraete}}]{Buyens2014MatrixProductStates}%
  \BibitemOpen
  \bibfield  {author} {\bibinfo {author} {\bibfnamefont {B.}~\bibnamefont
  {Buyens}}, \bibinfo {author} {\bibfnamefont {J.}~\bibnamefont {Haegeman}},
  \bibinfo {author} {\bibfnamefont {K.}~\bibnamefont {Van~Acoleyen}}, \bibinfo
  {author} {\bibfnamefont {H.}~\bibnamefont {Verschelde}},\ and\ \bibinfo
  {author} {\bibfnamefont {F.}~\bibnamefont {Verstraete}},\ }\bibfield  {title}
  {\bibinfo {title} {Matrix {Product States} for {Gauge Field Theories}},\
  }\href {https://doi.org/10.1103/PhysRevLett.113.091601} {\bibfield  {journal}
  {\bibinfo  {journal} {Physical Review Letters}\ }\textbf {\bibinfo {volume}
  {113}},\ \bibinfo {pages} {091601} (\bibinfo {year} {2014})}\BibitemShut
  {NoStop}%
\bibitem [{\citenamefont {Rico}\ \emph {et~al.}(2014)\citenamefont {Rico},
  \citenamefont {Pichler}, \citenamefont {Dalmonte}, \citenamefont {Zoller},\
  and\ \citenamefont {Montangero}}]{Rico2014TensorNetworksLattice}%
  \BibitemOpen
  \bibfield  {author} {\bibinfo {author} {\bibfnamefont {E.}~\bibnamefont
  {Rico}}, \bibinfo {author} {\bibfnamefont {T.}~\bibnamefont {Pichler}},
  \bibinfo {author} {\bibfnamefont {M.}~\bibnamefont {Dalmonte}}, \bibinfo
  {author} {\bibfnamefont {P.}~\bibnamefont {Zoller}},\ and\ \bibinfo {author}
  {\bibfnamefont {S.}~\bibnamefont {Montangero}},\ }\bibfield  {title}
  {\bibinfo {title} {Tensor {Networks} for {Lattice Gauge Theories} and {Atomic
  Quantum Simulation}},\ }\href
  {https://doi.org/10.1103/PhysRevLett.112.201601} {\bibfield  {journal}
  {\bibinfo  {journal} {Physical Review Letters}\ }\textbf {\bibinfo {volume}
  {112}},\ \bibinfo {pages} {201601} (\bibinfo {year} {2014})}\BibitemShut
  {NoStop}%
\bibitem [{\citenamefont {Haegeman}\ \emph {et~al.}(2015)\citenamefont
  {Haegeman}, \citenamefont {Van~Acoleyen}, \citenamefont {Schuch},
  \citenamefont {Cirac},\ and\ \citenamefont
  {Verstraete}}]{Haegeman2015GaugingQuantumStates}%
  \BibitemOpen
  \bibfield  {author} {\bibinfo {author} {\bibfnamefont {J.}~\bibnamefont
  {Haegeman}}, \bibinfo {author} {\bibfnamefont {K.}~\bibnamefont
  {Van~Acoleyen}}, \bibinfo {author} {\bibfnamefont {N.}~\bibnamefont
  {Schuch}}, \bibinfo {author} {\bibfnamefont {J.~I.}\ \bibnamefont {Cirac}},\
  and\ \bibinfo {author} {\bibfnamefont {F.}~\bibnamefont {Verstraete}},\
  }\bibfield  {title} {\bibinfo {title} {Gauging quantum states: From global to
  local symmetries in many-body systems},\ }\href
  {https://doi.org/10.1103/PhysRevX.5.011024} {\bibfield  {journal} {\bibinfo
  {journal} {Phys. Rev. X}\ }\textbf {\bibinfo {volume} {5}},\ \bibinfo {pages}
  {011024} (\bibinfo {year} {2015})}\BibitemShut {NoStop}%
\bibitem [{\citenamefont {Silvi}\ \emph {et~al.}(2017)\citenamefont {Silvi},
  \citenamefont {Rico}, \citenamefont {Dalmonte}, \citenamefont {Tschirsich},\
  and\ \citenamefont {Montangero}}]{Silvi2017FiniteDensityPhase}%
  \BibitemOpen
  \bibfield  {author} {\bibinfo {author} {\bibfnamefont {P.}~\bibnamefont
  {Silvi}}, \bibinfo {author} {\bibfnamefont {E.}~\bibnamefont {Rico}},
  \bibinfo {author} {\bibfnamefont {M.}~\bibnamefont {Dalmonte}}, \bibinfo
  {author} {\bibfnamefont {F.}~\bibnamefont {Tschirsich}},\ and\ \bibinfo
  {author} {\bibfnamefont {S.}~\bibnamefont {Montangero}},\ }\bibfield  {title}
  {\bibinfo {title} {Finite-density phase diagram of a {$(1+1)-d$} non-abelian
  lattice gauge theory with tensor networks},\ }\href
  {https://doi.org/10.22331/q-2017-04-25-9} {\bibfield  {journal} {\bibinfo
  {journal} {{Quantum}}\ }\textbf {\bibinfo {volume} {1}},\ \bibinfo {pages}
  {9} (\bibinfo {year} {2017})}\BibitemShut {NoStop}%
\bibitem [{\citenamefont {Buyens}\ \emph {et~al.}(2017)\citenamefont {Buyens},
  \citenamefont {Montangero}, \citenamefont {Haegeman}, \citenamefont
  {Verstraete},\ and\ \citenamefont
  {Van~Acoleyen}}]{Buyens2017FiniteRepresentationApproximation}%
  \BibitemOpen
  \bibfield  {author} {\bibinfo {author} {\bibfnamefont {B.}~\bibnamefont
  {Buyens}}, \bibinfo {author} {\bibfnamefont {S.}~\bibnamefont {Montangero}},
  \bibinfo {author} {\bibfnamefont {J.}~\bibnamefont {Haegeman}}, \bibinfo
  {author} {\bibfnamefont {F.}~\bibnamefont {Verstraete}},\ and\ \bibinfo
  {author} {\bibfnamefont {K.}~\bibnamefont {Van~Acoleyen}},\ }\bibfield
  {title} {\bibinfo {title} {Finite-representation approximation of lattice
  gauge theories at the continuum limit with tensor networks},\ }\href
  {https://doi.org/10.1103/PhysRevD.95.094509} {\bibfield  {journal} {\bibinfo
  {journal} {Phys. Rev. D}\ }\textbf {\bibinfo {volume} {95}},\ \bibinfo
  {pages} {094509} (\bibinfo {year} {2017})}\BibitemShut {NoStop}%
\bibitem [{\citenamefont {Ba\~nuls}\ \emph
  {et~al.}(2017{\natexlab{a}})\citenamefont {Ba\~nuls}, \citenamefont {Cichy},
  \citenamefont {Cirac}, \citenamefont {Jansen},\ and\ \citenamefont
  {K\"uhn}}]{Banuls2017DensityInducedPhase}%
  \BibitemOpen
  \bibfield  {author} {\bibinfo {author} {\bibfnamefont {M.~C.}\ \bibnamefont
  {Ba\~nuls}}, \bibinfo {author} {\bibfnamefont {K.}~\bibnamefont {Cichy}},
  \bibinfo {author} {\bibfnamefont {J.~I.}\ \bibnamefont {Cirac}}, \bibinfo
  {author} {\bibfnamefont {K.}~\bibnamefont {Jansen}},\ and\ \bibinfo {author}
  {\bibfnamefont {S.}~\bibnamefont {K\"uhn}},\ }\bibfield  {title} {\bibinfo
  {title} {Density induced phase transitions in the schwinger model: A study
  with matrix product states},\ }\href
  {https://doi.org/10.1103/PhysRevLett.118.071601} {\bibfield  {journal}
  {\bibinfo  {journal} {Phys. Rev. Lett.}\ }\textbf {\bibinfo {volume} {118}},\
  \bibinfo {pages} {071601} (\bibinfo {year} {2017}{\natexlab{a}})}\BibitemShut
  {NoStop}%
\bibitem [{\citenamefont {Ercolessi}\ \emph {et~al.}(2018)\citenamefont
  {Ercolessi}, \citenamefont {Facchi}, \citenamefont {Magnifico}, \citenamefont
  {Pascazio},\ and\ \citenamefont {Pepe}}]{Ercolessi2018PhaseTransitionsZn}%
  \BibitemOpen
  \bibfield  {author} {\bibinfo {author} {\bibfnamefont {E.}~\bibnamefont
  {Ercolessi}}, \bibinfo {author} {\bibfnamefont {P.}~\bibnamefont {Facchi}},
  \bibinfo {author} {\bibfnamefont {G.}~\bibnamefont {Magnifico}}, \bibinfo
  {author} {\bibfnamefont {S.}~\bibnamefont {Pascazio}},\ and\ \bibinfo
  {author} {\bibfnamefont {F.~V.}\ \bibnamefont {Pepe}},\ }\bibfield  {title}
  {\bibinfo {title} {Phase transitions in ${Z}_{n}$ gauge models: Towards
  quantum simulations of the schwinger-weyl qed},\ }\href
  {https://doi.org/10.1103/PhysRevD.98.074503} {\bibfield  {journal} {\bibinfo
  {journal} {Physical Review D}\ }\textbf {\bibinfo {volume} {98}},\ \bibinfo
  {pages} {074503} (\bibinfo {year} {2018})}\BibitemShut {NoStop}%
\bibitem [{\citenamefont {Magnifico}\ \emph
  {et~al.}(2019{\natexlab{a}})\citenamefont {Magnifico}, \citenamefont
  {Vodola}, \citenamefont {Ercolessi}, \citenamefont {Kumar}, \citenamefont
  {M\"uller},\ and\ \citenamefont
  {Bermudez}}]{Magnifico2019SymmetryProtectedTopological}%
  \BibitemOpen
  \bibfield  {author} {\bibinfo {author} {\bibfnamefont {G.}~\bibnamefont
  {Magnifico}}, \bibinfo {author} {\bibfnamefont {D.}~\bibnamefont {Vodola}},
  \bibinfo {author} {\bibfnamefont {E.}~\bibnamefont {Ercolessi}}, \bibinfo
  {author} {\bibfnamefont {S.~P.}\ \bibnamefont {Kumar}}, \bibinfo {author}
  {\bibfnamefont {M.}~\bibnamefont {M\"uller}},\ and\ \bibinfo {author}
  {\bibfnamefont {A.}~\bibnamefont {Bermudez}},\ }\bibfield  {title} {\bibinfo
  {title} {Symmetry-protected topological phases in lattice gauge theories:
  Topological ${\mathrm{qed}}_{2}$},\ }\href
  {https://doi.org/10.1103/PhysRevD.99.014503} {\bibfield  {journal} {\bibinfo
  {journal} {Phys. Rev. D}\ }\textbf {\bibinfo {volume} {99}},\ \bibinfo
  {pages} {014503} (\bibinfo {year} {2019}{\natexlab{a}})}\BibitemShut
  {NoStop}%
\bibitem [{\citenamefont {Magnifico}\ \emph
  {et~al.}(2019{\natexlab{b}})\citenamefont {Magnifico}, \citenamefont
  {Vodola}, \citenamefont {Ercolessi}, \citenamefont {Kumar}, \citenamefont
  {M\"uller},\ and\ \citenamefont {Bermudez}}]{Magnifico2019ZnGaugeTheories}%
  \BibitemOpen
  \bibfield  {author} {\bibinfo {author} {\bibfnamefont {G.}~\bibnamefont
  {Magnifico}}, \bibinfo {author} {\bibfnamefont {D.}~\bibnamefont {Vodola}},
  \bibinfo {author} {\bibfnamefont {E.}~\bibnamefont {Ercolessi}}, \bibinfo
  {author} {\bibfnamefont {S.~P.}\ \bibnamefont {Kumar}}, \bibinfo {author}
  {\bibfnamefont {M.}~\bibnamefont {M\"uller}},\ and\ \bibinfo {author}
  {\bibfnamefont {A.}~\bibnamefont {Bermudez}},\ }\bibfield  {title} {\bibinfo
  {title} {${\mathbb{z}}_{N}$ gauge theories coupled to topological fermions:
  ${\mathrm{qed}}_{2}$ with a quantum mechanical $\ensuremath{\theta}$ angle},\
  }\href {https://doi.org/10.1103/PhysRevB.100.115152} {\bibfield  {journal}
  {\bibinfo  {journal} {Phys. Rev. B}\ }\textbf {\bibinfo {volume} {100}},\
  \bibinfo {pages} {115152} (\bibinfo {year} {2019}{\natexlab{b}})}\BibitemShut
  {NoStop}%
\bibitem [{\citenamefont {Sala}\ \emph
  {et~al.}(2018{\natexlab{a}})\citenamefont {Sala}, \citenamefont {Shi},
  \citenamefont {K{\"u}hn}, \citenamefont {Banuls}, \citenamefont {Demler},\
  and\ \citenamefont {Cirac}}]{Sala2018GaussianStatesVariational}%
  \BibitemOpen
  \bibfield  {author} {\bibinfo {author} {\bibfnamefont {P.}~\bibnamefont
  {Sala}}, \bibinfo {author} {\bibfnamefont {T.}~\bibnamefont {Shi}}, \bibinfo
  {author} {\bibfnamefont {S.}~\bibnamefont {K{\"u}hn}}, \bibinfo {author}
  {\bibfnamefont {M.~C.}\ \bibnamefont {Banuls}}, \bibinfo {author}
  {\bibfnamefont {E.}~\bibnamefont {Demler}},\ and\ \bibinfo {author}
  {\bibfnamefont {J.~I.}\ \bibnamefont {Cirac}},\ }\bibfield  {title} {\bibinfo
  {title} {{Gaussian states for the variational study of (1+1)-dimensional
  lattice gauge models}},\ }in\ \href {https://doi.org/10.22323/1.334.0230}
  {\emph {\bibinfo {booktitle} {The 36th Annual International Symposium on
  Lattice Field Theory. 22-28 July}}}\ (\bibinfo {year} {2018})\ p.\ \bibinfo
  {pages} {230}\BibitemShut {NoStop}%
\bibitem [{\citenamefont {Magnifico}\ \emph {et~al.}(2020)\citenamefont
  {Magnifico}, \citenamefont {Dalmonte}, \citenamefont {Facchi}, \citenamefont
  {Pascazio}, \citenamefont {Pepe},\ and\ \citenamefont
  {Ercolessi}}]{Magnifico2020RealTimeDynamics}%
  \BibitemOpen
  \bibfield  {author} {\bibinfo {author} {\bibfnamefont {G.}~\bibnamefont
  {Magnifico}}, \bibinfo {author} {\bibfnamefont {M.}~\bibnamefont {Dalmonte}},
  \bibinfo {author} {\bibfnamefont {P.}~\bibnamefont {Facchi}}, \bibinfo
  {author} {\bibfnamefont {S.}~\bibnamefont {Pascazio}}, \bibinfo {author}
  {\bibfnamefont {F.~V.}\ \bibnamefont {Pepe}},\ and\ \bibinfo {author}
  {\bibfnamefont {E.}~\bibnamefont {Ercolessi}},\ }\bibfield  {title} {\bibinfo
  {title} {Real {T}ime {D}ynamics and {C}onfinement in the {$\mathbb{Z}_{n}$
  S}chwinger-{W}eyl lattice model for 1+1 {QED}},\ }\href
  {https://doi.org/10.22331/q-2020-06-15-281} {\bibfield  {journal} {\bibinfo
  {journal} {{Quantum}}\ }\textbf {\bibinfo {volume} {4}},\ \bibinfo {pages}
  {281} (\bibinfo {year} {2020})}\BibitemShut {NoStop}%
\bibitem [{\citenamefont {Ba\~nuls}\ \emph
  {et~al.}(2017{\natexlab{b}})\citenamefont {Ba\~nuls}, \citenamefont {Cichy},
  \citenamefont {Cirac}, \citenamefont {Jansen},\ and\ \citenamefont
  {K\"uhn}}]{Banuls2017EfficientBasisFormulation}%
  \BibitemOpen
  \bibfield  {author} {\bibinfo {author} {\bibfnamefont {M.~C.}\ \bibnamefont
  {Ba\~nuls}}, \bibinfo {author} {\bibfnamefont {K.}~\bibnamefont {Cichy}},
  \bibinfo {author} {\bibfnamefont {J.~I.}\ \bibnamefont {Cirac}}, \bibinfo
  {author} {\bibfnamefont {K.}~\bibnamefont {Jansen}},\ and\ \bibinfo {author}
  {\bibfnamefont {S.}~\bibnamefont {K\"uhn}},\ }\bibfield  {title} {\bibinfo
  {title} {Efficient basis formulation for $(1+1)$-dimensional su(2) lattice
  gauge theory: Spectral calculations with matrix product states},\ }\href
  {https://doi.org/10.1103/PhysRevX.7.041046} {\bibfield  {journal} {\bibinfo
  {journal} {Phys. Rev. X}\ }\textbf {\bibinfo {volume} {7}},\ \bibinfo {pages}
  {041046} (\bibinfo {year} {2017}{\natexlab{b}})}\BibitemShut {NoStop}%
\bibitem [{\citenamefont {Rigobello}\ \emph {et~al.}(2021)\citenamefont
  {Rigobello}, \citenamefont {Notarnicola}, \citenamefont {Magnifico},\ and\
  \citenamefont {Montangero}}]{Rigobello2021EntanglementGeneration11d}%
  \BibitemOpen
  \bibfield  {author} {\bibinfo {author} {\bibfnamefont {M.}~\bibnamefont
  {Rigobello}}, \bibinfo {author} {\bibfnamefont {S.}~\bibnamefont
  {Notarnicola}}, \bibinfo {author} {\bibfnamefont {G.}~\bibnamefont
  {Magnifico}},\ and\ \bibinfo {author} {\bibfnamefont {S.}~\bibnamefont
  {Montangero}},\ }\bibfield  {title} {\bibinfo {title} {Entanglement
  generation in $(1+1)\mathrm{D}$ qed scattering processes},\ }\href
  {https://doi.org/10.1103/PhysRevD.104.114501} {\bibfield  {journal} {\bibinfo
   {journal} {Phys. Rev. D}\ }\textbf {\bibinfo {volume} {104}},\ \bibinfo
  {pages} {114501} (\bibinfo {year} {2021})}\BibitemShut {NoStop}%
\bibitem [{\citenamefont {Funcke}\ \emph
  {et~al.}(2023{\natexlab{a}})\citenamefont {Funcke}, \citenamefont {Jansen},\
  and\ \citenamefont {K{\"u}hn}}]{Funcke2023ExploringCpViolating}%
  \BibitemOpen
  \bibfield  {author} {\bibinfo {author} {\bibfnamefont {L.}~\bibnamefont
  {Funcke}}, \bibinfo {author} {\bibfnamefont {K.}~\bibnamefont {Jansen}},\
  and\ \bibinfo {author} {\bibfnamefont {S.}~\bibnamefont {K{\"u}hn}},\
  }\bibfield  {title} {\bibinfo {title} {Exploring the {{CP-violating Dashen}}
  phase in the {{Schwinger}} model with tensor networks},\ }\href
  {https://doi.org/10.1103/PhysRevD.108.014504} {\bibfield  {journal} {\bibinfo
   {journal} {Physical Review D}\ }\textbf {\bibinfo {volume} {108}},\ \bibinfo
  {pages} {014504} (\bibinfo {year} {2023}{\natexlab{a}})},\ \Eprint
  {https://arxiv.org/abs/2303.03799} {arXiv:2303.03799 [hep-lat]} \BibitemShut
  {NoStop}%
\bibitem [{\citenamefont {Angelides}\ \emph {et~al.}(2023)\citenamefont
  {Angelides}, \citenamefont {Funcke}, \citenamefont {Jansen},\ and\
  \citenamefont {K{\"u}hn}}]{Angelides2023ComputingMassShift}%
  \BibitemOpen
  \bibfield  {author} {\bibinfo {author} {\bibfnamefont {T.}~\bibnamefont
  {Angelides}}, \bibinfo {author} {\bibfnamefont {L.}~\bibnamefont {Funcke}},
  \bibinfo {author} {\bibfnamefont {K.}~\bibnamefont {Jansen}},\ and\ \bibinfo
  {author} {\bibfnamefont {S.}~\bibnamefont {K{\"u}hn}},\ }\bibfield  {title}
  {\bibinfo {title} {Computing the {{Mass Shift}} of {{Wilson}} and {{Staggered
  Fermions}} in the {{Lattice Schwinger Model}} with {{Matrix Product
  States}}},\ }\href {https://doi.org/10.1103/PhysRevD.108.014516} {\bibfield
  {journal} {\bibinfo  {journal} {Physical Review D}\ }\textbf {\bibinfo
  {volume} {108}},\ \bibinfo {pages} {014516} (\bibinfo {year} {2023})},\
  \Eprint {https://arxiv.org/abs/2303.11016} {arXiv:2303.11016 [hep-lat]}
  \BibitemShut {NoStop}%
\bibitem [{\citenamefont {Chanda}\ \emph {et~al.}(2023)\citenamefont {Chanda},
  \citenamefont {Dalmonte}, \citenamefont {Lewenstein}, \citenamefont
  {Zakrzewski},\ and\ \citenamefont
  {Tagliacozzo}}]{Chanda2023SpectralPropertiesCritical}%
  \BibitemOpen
  \bibfield  {author} {\bibinfo {author} {\bibfnamefont {T.}~\bibnamefont
  {Chanda}}, \bibinfo {author} {\bibfnamefont {M.}~\bibnamefont {Dalmonte}},
  \bibinfo {author} {\bibfnamefont {M.}~\bibnamefont {Lewenstein}}, \bibinfo
  {author} {\bibfnamefont {J.}~\bibnamefont {Zakrzewski}},\ and\ \bibinfo
  {author} {\bibfnamefont {L.}~\bibnamefont {Tagliacozzo}},\ }\href
  {https://doi.org/10.48550/arXiv.2304.01030} {\emph {\bibinfo {title}
  {Spectral Properties of Critical 1+{{1D Abelian-Higgs}} Model}}},\ \bibinfo
  {type} {Tech. Rep.}\ \bibinfo {number} {arXiv:2304.01030}\ (\bibinfo {year}
  {2023})\ \Eprint {https://arxiv.org/abs/2304.01030} {arXiv:2304.01030
  [hep-th]} \BibitemShut {NoStop}%
\bibitem [{\citenamefont {Schmoll}\ \emph {et~al.}(2023)\citenamefont
  {Schmoll}, \citenamefont {Naumann}, \citenamefont {Nietner}, \citenamefont
  {Eisert},\ and\ \citenamefont {Sotiriadis}}]{schmoll2023hamiltonian}%
  \BibitemOpen
  \bibfield  {author} {\bibinfo {author} {\bibfnamefont {P.}~\bibnamefont
  {Schmoll}}, \bibinfo {author} {\bibfnamefont {J.}~\bibnamefont {Naumann}},
  \bibinfo {author} {\bibfnamefont {A.}~\bibnamefont {Nietner}}, \bibinfo
  {author} {\bibfnamefont {J.}~\bibnamefont {Eisert}},\ and\ \bibinfo {author}
  {\bibfnamefont {S.}~\bibnamefont {Sotiriadis}},\ }\href@noop {} {\bibinfo
  {title} {Hamiltonian truncation tensor networks for quantum field theories}}
  (\bibinfo {year} {2023}),\ \Eprint {https://arxiv.org/abs/2312.12506}
  {arXiv:2312.12506 [quant-ph]} \BibitemShut {NoStop}%
\bibitem [{\citenamefont {Hayata}\ \emph {et~al.}(2023)\citenamefont {Hayata},
  \citenamefont {Hidaka},\ and\ \citenamefont {Nishimura}}]{hayata2023dense}%
  \BibitemOpen
  \bibfield  {author} {\bibinfo {author} {\bibfnamefont {T.}~\bibnamefont
  {Hayata}}, \bibinfo {author} {\bibfnamefont {Y.}~\bibnamefont {Hidaka}},\
  and\ \bibinfo {author} {\bibfnamefont {K.}~\bibnamefont {Nishimura}},\
  }\href@noop {} {\bibinfo {title} {Dense $\textrm{QCD}_2$ with matrix product
  states}} (\bibinfo {year} {2023}),\ \Eprint
  {https://arxiv.org/abs/2311.11643} {arXiv:2311.11643 [hep-lat]} \BibitemShut
  {NoStop}%
\bibitem [{\citenamefont {Florio}\ \emph {et~al.}(2023)\citenamefont {Florio},
  \citenamefont {Weichselbaum}, \citenamefont {Valgushev},\ and\ \citenamefont
  {Pisarski}}]{florio2023mass}%
  \BibitemOpen
  \bibfield  {author} {\bibinfo {author} {\bibfnamefont {A.}~\bibnamefont
  {Florio}}, \bibinfo {author} {\bibfnamefont {A.}~\bibnamefont
  {Weichselbaum}}, \bibinfo {author} {\bibfnamefont {S.}~\bibnamefont
  {Valgushev}},\ and\ \bibinfo {author} {\bibfnamefont {R.~D.}\ \bibnamefont
  {Pisarski}},\ }\href@noop {} {\bibinfo {title} {Mass gaps of a $\mathbb{Z}_3$
  gauge theory with three fermion flavors in 1 + 1 dimensions}} (\bibinfo
  {year} {2023}),\ \Eprint {https://arxiv.org/abs/2310.18312} {arXiv:2310.18312
  [hep-th]} \BibitemShut {NoStop}%
\bibitem [{\citenamefont {Osborne}\ \emph {et~al.}(2023)\citenamefont
  {Osborne}, \citenamefont {McCulloch},\ and\ \citenamefont
  {Halimeh}}]{osborne2023probing}%
  \BibitemOpen
  \bibfield  {author} {\bibinfo {author} {\bibfnamefont {J.}~\bibnamefont
  {Osborne}}, \bibinfo {author} {\bibfnamefont {I.~P.}\ \bibnamefont
  {McCulloch}},\ and\ \bibinfo {author} {\bibfnamefont {J.~C.}\ \bibnamefont
  {Halimeh}},\ }\href@noop {} {\bibinfo {title} {Probing confinement through
  dynamical quantum phase transitions: From quantum spin models to lattice
  gauge theories}} (\bibinfo {year} {2023}),\ \Eprint
  {https://arxiv.org/abs/2310.12210} {arXiv:2310.12210 [cond-mat.quant-gas]}
  \BibitemShut {NoStop}%
\bibitem [{\citenamefont {Kebrič}\ \emph {et~al.}(2024)\citenamefont
  {Kebrič}, \citenamefont {Halimeh}, \citenamefont {Schollwöck},\ and\
  \citenamefont {Grusdt}}]{kebric2024confinement}%
  \BibitemOpen
  \bibfield  {author} {\bibinfo {author} {\bibfnamefont {M.}~\bibnamefont
  {Kebrič}}, \bibinfo {author} {\bibfnamefont {J.~C.}\ \bibnamefont
  {Halimeh}}, \bibinfo {author} {\bibfnamefont {U.}~\bibnamefont
  {Schollwöck}},\ and\ \bibinfo {author} {\bibfnamefont {F.}~\bibnamefont
  {Grusdt}},\ }\href@noop {} {\bibinfo {title} {Confinement in 1+1d
  $\mathbb{Z}_2$ lattice gauge theories at finite temperature}} (\bibinfo
  {year} {2024}),\ \Eprint {https://arxiv.org/abs/2308.08592} {arXiv:2308.08592
  [cond-mat.quant-gas]} \BibitemShut {NoStop}%
\bibitem [{\citenamefont {Belyansky}\ \emph {et~al.}(2024)\citenamefont
  {Belyansky}, \citenamefont {Whitsitt}, \citenamefont {Mueller}, \citenamefont
  {Fahimniya}, \citenamefont {Bennewitz}, \citenamefont {Davoudi},\ and\
  \citenamefont {Gorshkov}}]{Belyansky2024PhysRevLett}%
  \BibitemOpen
  \bibfield  {author} {\bibinfo {author} {\bibfnamefont {R.}~\bibnamefont
  {Belyansky}}, \bibinfo {author} {\bibfnamefont {S.}~\bibnamefont {Whitsitt}},
  \bibinfo {author} {\bibfnamefont {N.}~\bibnamefont {Mueller}}, \bibinfo
  {author} {\bibfnamefont {A.}~\bibnamefont {Fahimniya}}, \bibinfo {author}
  {\bibfnamefont {E.~R.}\ \bibnamefont {Bennewitz}}, \bibinfo {author}
  {\bibfnamefont {Z.}~\bibnamefont {Davoudi}},\ and\ \bibinfo {author}
  {\bibfnamefont {A.~V.}\ \bibnamefont {Gorshkov}},\ }\bibfield  {title}
  {\bibinfo {title} {High-energy collision of quarks and mesons in the
  schwinger model: From tensor networks to circuit qed},\ }\href
  {https://doi.org/10.1103/PhysRevLett.132.091903} {\bibfield  {journal}
  {\bibinfo  {journal} {Phys. Rev. Lett.}\ }\textbf {\bibinfo {volume} {132}},\
  \bibinfo {pages} {091903} (\bibinfo {year} {2024})}\BibitemShut {NoStop}%
\bibitem [{\citenamefont {Papaefstathiou}\ \emph {et~al.}(2024)\citenamefont
  {Papaefstathiou}, \citenamefont {Knolle},\ and\ \citenamefont
  {Bañuls}}]{Papaefstathiou2024realtime}%
  \BibitemOpen
  \bibfield  {author} {\bibinfo {author} {\bibfnamefont {I.}~\bibnamefont
  {Papaefstathiou}}, \bibinfo {author} {\bibfnamefont {J.}~\bibnamefont
  {Knolle}},\ and\ \bibinfo {author} {\bibfnamefont {M.~C.}\ \bibnamefont
  {Bañuls}},\ }\href@noop {} {\bibinfo {title} {Real-time scattering in the
  lattice schwinger model}} (\bibinfo {year} {2024}),\ \Eprint
  {https://arxiv.org/abs/2402.18429} {arXiv:2402.18429 [hep-lat]} \BibitemShut
  {NoStop}%
\bibitem [{\citenamefont {Calaj{\`o}}\ \emph {et~al.}(2024)\citenamefont
  {Calaj{\`o}}, \citenamefont {Cataldi}, \citenamefont {Rigobello},
  \citenamefont {Wanisch}, \citenamefont {Magnifico}, \citenamefont {Silvi},
  \citenamefont {Montangero},\ and\ \citenamefont
  {Halimeh}}]{Calajo2024QuantumManyBodyScarring}%
  \BibitemOpen
  \bibfield  {author} {\bibinfo {author} {\bibfnamefont {G.}~\bibnamefont
  {Calaj{\`o}}}, \bibinfo {author} {\bibfnamefont {G.}~\bibnamefont {Cataldi}},
  \bibinfo {author} {\bibfnamefont {M.}~\bibnamefont {Rigobello}}, \bibinfo
  {author} {\bibfnamefont {D.}~\bibnamefont {Wanisch}}, \bibinfo {author}
  {\bibfnamefont {G.}~\bibnamefont {Magnifico}}, \bibinfo {author}
  {\bibfnamefont {P.}~\bibnamefont {Silvi}}, \bibinfo {author} {\bibfnamefont
  {S.}~\bibnamefont {Montangero}},\ and\ \bibinfo {author} {\bibfnamefont
  {J.~C.}\ \bibnamefont {Halimeh}},\ }\href
  {https://doi.org/10.48550/arXiv.2405.13112} {\bibinfo {title} {Quantum
  {{Many-Body Scarring}} in a {{Non-Abelian Lattice Gauge Theory}}}} (\bibinfo
  {year} {2024}),\ \Eprint {https://arxiv.org/abs/2405.13112} {arxiv:2405.13112
  [cond-mat, physics:hep-lat, physics:quant-ph]} \BibitemShut {NoStop}%
\bibitem [{\citenamefont {Felser}\ \emph {et~al.}(2020)\citenamefont {Felser},
  \citenamefont {Silvi}, \citenamefont {Collura},\ and\ \citenamefont
  {Montangero}}]{Felser2020TwoDimensionalQuantum}%
  \BibitemOpen
  \bibfield  {author} {\bibinfo {author} {\bibfnamefont {T.}~\bibnamefont
  {Felser}}, \bibinfo {author} {\bibfnamefont {P.}~\bibnamefont {Silvi}},
  \bibinfo {author} {\bibfnamefont {M.}~\bibnamefont {Collura}},\ and\ \bibinfo
  {author} {\bibfnamefont {S.}~\bibnamefont {Montangero}},\ }\bibfield  {title}
  {\bibinfo {title} {Two-{{Dimensional Quantum-Link Lattice Quantum
  Electrodynamics}} at {{Finite Density}}},\ }\href
  {https://doi.org/10.1103/PhysRevX.10.041040} {\bibfield  {journal} {\bibinfo
  {journal} {Physical Review X}\ }\textbf {\bibinfo {volume} {10}},\ \bibinfo
  {pages} {041040} (\bibinfo {year} {2020})}\BibitemShut {NoStop}%
\bibitem [{\citenamefont {Emonts}\ \emph {et~al.}(2023)\citenamefont {Emonts},
  \citenamefont {Kelman}, \citenamefont {Borla}, \citenamefont {Moroz},
  \citenamefont {Gazit},\ and\ \citenamefont
  {Zohar}}]{Emonts2023FindingGroundState}%
  \BibitemOpen
  \bibfield  {author} {\bibinfo {author} {\bibfnamefont {P.}~\bibnamefont
  {Emonts}}, \bibinfo {author} {\bibfnamefont {A.}~\bibnamefont {Kelman}},
  \bibinfo {author} {\bibfnamefont {U.}~\bibnamefont {Borla}}, \bibinfo
  {author} {\bibfnamefont {S.}~\bibnamefont {Moroz}}, \bibinfo {author}
  {\bibfnamefont {S.}~\bibnamefont {Gazit}},\ and\ \bibinfo {author}
  {\bibfnamefont {E.}~\bibnamefont {Zohar}},\ }\bibfield  {title} {\bibinfo
  {title} {Finding the ground state of a lattice gauge theory with fermionic
  tensor networks: a $2+1d$ $\mathbb{Z}_2$ demonstrationfinding the ground
  state of a lattice gauge theory with fermionic tensor networks: a $2+1d$
  $\mathbb{Z}_2$ demonstration},\ }\href
  {https://doi.org/10.1103/PhysRevD.107.014505} {\bibfield  {journal} {\bibinfo
   {journal} {Physical Review D}\ }\textbf {\bibinfo {volume} {107}},\ \bibinfo
  {pages} {014505} (\bibinfo {year} {2023})}\BibitemShut {NoStop}%
\bibitem [{\citenamefont {Magnifico}\ \emph {et~al.}(2021)\citenamefont
  {Magnifico}, \citenamefont {Felser}, \citenamefont {Silvi},\ and\
  \citenamefont {Montangero}}]{Magnifico2021LatticeQuantumElectrodynamics}%
  \BibitemOpen
  \bibfield  {author} {\bibinfo {author} {\bibfnamefont {G.}~\bibnamefont
  {Magnifico}}, \bibinfo {author} {\bibfnamefont {T.}~\bibnamefont {Felser}},
  \bibinfo {author} {\bibfnamefont {P.}~\bibnamefont {Silvi}},\ and\ \bibinfo
  {author} {\bibfnamefont {S.}~\bibnamefont {Montangero}},\ }\bibfield  {title}
  {\bibinfo {title} {Lattice quantum electrodynamics in (3+1)-dimensions at
  finite density with tensor networks},\ }\href
  {https://doi.org/10.1038/s41467-021-23646-3} {\bibfield  {journal} {\bibinfo
  {journal} {Nature Communications}\ }\textbf {\bibinfo {volume} {12}},\
  \bibinfo {pages} {3600} (\bibinfo {year} {2021})}\BibitemShut {NoStop}%
\bibitem [{\citenamefont {Knaute}\ \emph {et~al.}(2024)\citenamefont {Knaute},
  \citenamefont {Feuerstein},\ and\ \citenamefont
  {Zohar}}]{Knaute2024TransferPEPS}%
  \BibitemOpen
  \bibfield  {author} {\bibinfo {author} {\bibfnamefont {J.}~\bibnamefont
  {Knaute}}, \bibinfo {author} {\bibfnamefont {M.}~\bibnamefont {Feuerstein}},\
  and\ \bibinfo {author} {\bibfnamefont {E.}~\bibnamefont {Zohar}},\ }\bibfield
   {title} {\bibinfo {title} {Entanglement and confinement in lattice gauge
  theory tensor networks},\ }\bibfield  {journal} {\bibinfo  {journal} {Journal
  of High Energy Physics}\ }\textbf {\bibinfo {volume} {2024}},\ \href
  {https://doi.org/10.1007/jhep02(2024)174} {10.1007/jhep02(2024)174} (\bibinfo
  {year} {2024})\BibitemShut {NoStop}%
\bibitem [{\citenamefont {Pradhan}\ \emph {et~al.}(2024)\citenamefont
  {Pradhan}, \citenamefont {Maroncelli},\ and\ \citenamefont
  {Ercolessi}}]{Pradhan2024QEDLadder}%
  \BibitemOpen
  \bibfield  {author} {\bibinfo {author} {\bibfnamefont {S.}~\bibnamefont
  {Pradhan}}, \bibinfo {author} {\bibfnamefont {A.}~\bibnamefont
  {Maroncelli}},\ and\ \bibinfo {author} {\bibfnamefont {E.}~\bibnamefont
  {Ercolessi}},\ }\bibfield  {title} {\bibinfo {title} {Discrete abelian
  lattice gauge theories on a ladder and their dualities with quantum clock
  models},\ }\bibfield  {journal} {\bibinfo  {journal} {Physical Review B}\
  }\textbf {\bibinfo {volume} {109}},\ \href
  {https://doi.org/10.1103/physrevb.109.064410} {10.1103/physrevb.109.064410}
  (\bibinfo {year} {2024})\BibitemShut {NoStop}%
\bibitem [{\citenamefont {Su}\ \emph {et~al.}(2024)\citenamefont {Su},
  \citenamefont {Osborne},\ and\ \citenamefont
  {Halimeh}}]{Su2024coldatomcollider}%
  \BibitemOpen
  \bibfield  {author} {\bibinfo {author} {\bibfnamefont {G.-X.}\ \bibnamefont
  {Su}}, \bibinfo {author} {\bibfnamefont {J.}~\bibnamefont {Osborne}},\ and\
  \bibinfo {author} {\bibfnamefont {J.~C.}\ \bibnamefont {Halimeh}},\
  }\href@noop {} {\bibinfo {title} {A cold-atom particle collider}} (\bibinfo
  {year} {2024}),\ \Eprint {https://arxiv.org/abs/2401.05489} {arXiv:2401.05489
  [cond-mat.quant-gas]} \BibitemShut {NoStop}%
\bibitem [{\citenamefont {Cataldi}\ \emph {et~al.}(2023)\citenamefont
  {Cataldi}, \citenamefont {Magnifico}, \citenamefont {Silvi},\ and\
  \citenamefont {Montangero}}]{Cataldi2023a}%
  \BibitemOpen
  \bibfield  {author} {\bibinfo {author} {\bibfnamefont {G.}~\bibnamefont
  {Cataldi}}, \bibinfo {author} {\bibfnamefont {G.}~\bibnamefont {Magnifico}},
  \bibinfo {author} {\bibfnamefont {P.}~\bibnamefont {Silvi}},\ and\ \bibinfo
  {author} {\bibfnamefont {S.}~\bibnamefont {Montangero}},\ }\href
  {https://doi.org/10.48550/arXiv.2307.09396} {\bibinfo {title} {(2+1){{D
  SU}}(2) {{Yang-Mills Lattice Gauge Theory}} at finite density via tensor
  networks}} (\bibinfo {year} {2023}),\ \Eprint
  {https://arxiv.org/abs/2307.09396} {arxiv:2307.09396 [cond-mat,
  physics:hep-lat, physics:hep-th, physics:quant-ph]} \BibitemShut {NoStop}%
\bibitem [{\citenamefont {Villalonga}\ \emph {et~al.}(2019)\citenamefont
  {Villalonga}, \citenamefont {Boixo}, \citenamefont {Nelson}, \citenamefont
  {Henze}, \citenamefont {Rieffel}, \citenamefont {Biswas},\ and\ \citenamefont
  {Mandr{\`{a}}}}]{Villalonga2019FlexibleHighPerformance}%
  \BibitemOpen
  \bibfield  {author} {\bibinfo {author} {\bibfnamefont {B.}~\bibnamefont
  {Villalonga}}, \bibinfo {author} {\bibfnamefont {S.}~\bibnamefont {Boixo}},
  \bibinfo {author} {\bibfnamefont {B.}~\bibnamefont {Nelson}}, \bibinfo
  {author} {\bibfnamefont {C.}~\bibnamefont {Henze}}, \bibinfo {author}
  {\bibfnamefont {E.}~\bibnamefont {Rieffel}}, \bibinfo {author} {\bibfnamefont
  {R.}~\bibnamefont {Biswas}},\ and\ \bibinfo {author} {\bibfnamefont
  {S.}~\bibnamefont {Mandr{\`{a}}}},\ }\bibfield  {title} {\bibinfo {title} {A
  flexible high-performance simulator for verifying and benchmarking quantum
  circuits implemented on real hardware},\ }\bibfield  {journal} {\bibinfo
  {journal} {npj Quantum Information}\ }\textbf {\bibinfo {volume} {5}},\ \href
  {https://doi.org/10.1038/s41534-019-0196-1} {10.1038/s41534-019-0196-1}
  (\bibinfo {year} {2019})\BibitemShut {NoStop}%
\bibitem [{\citenamefont {Mathis}\ \emph {et~al.}(2020)\citenamefont {Mathis},
  \citenamefont {Mazzola},\ and\ \citenamefont
  {Tavernelli}}]{Mathis2020ScalableSimulationsLattice}%
  \BibitemOpen
  \bibfield  {author} {\bibinfo {author} {\bibfnamefont {S.~V.}\ \bibnamefont
  {Mathis}}, \bibinfo {author} {\bibfnamefont {G.}~\bibnamefont {Mazzola}},\
  and\ \bibinfo {author} {\bibfnamefont {I.}~\bibnamefont {Tavernelli}},\
  }\bibfield  {title} {\bibinfo {title} {Toward scalable simulations of lattice
  gauge theories on quantum computers},\ }\href
  {https://doi.org/10.1103/PhysRevD.102.094501} {\bibfield  {journal} {\bibinfo
   {journal} {Physical Review D}\ }\textbf {\bibinfo {volume} {102}},\ \bibinfo
  {pages} {094501} (\bibinfo {year} {2020})}\BibitemShut {NoStop}%
\bibitem [{\citenamefont {Zhou}\ \emph {et~al.}(2020)\citenamefont {Zhou},
  \citenamefont {Stoudenmire},\ and\ \citenamefont
  {Waintal}}]{Zhou2020WhatLimitsSimulation}%
  \BibitemOpen
  \bibfield  {author} {\bibinfo {author} {\bibfnamefont {Y.}~\bibnamefont
  {Zhou}}, \bibinfo {author} {\bibfnamefont {E.~M.}\ \bibnamefont
  {Stoudenmire}},\ and\ \bibinfo {author} {\bibfnamefont {X.}~\bibnamefont
  {Waintal}},\ }\bibfield  {title} {\bibinfo {title} {What limits the
  simulation of quantum computers?},\ }\href
  {https://doi.org/10.1103/PhysRevX.10.041038} {\bibfield  {journal} {\bibinfo
  {journal} {Phys. Rev. X}\ }\textbf {\bibinfo {volume} {10}},\ \bibinfo
  {pages} {041038} (\bibinfo {year} {2020})}\BibitemShut {NoStop}%
\bibitem [{\citenamefont {Huang}\ \emph {et~al.}(2021)\citenamefont {Huang},
  \citenamefont {Zhang}, \citenamefont {Newman}, \citenamefont {Ni},
  \citenamefont {Ding}, \citenamefont {Cai}, \citenamefont {Gao}, \citenamefont
  {Wang}, \citenamefont {Wu}, \citenamefont {Zhang}, \citenamefont {Ku},
  \citenamefont {Tian}, \citenamefont {Wu}, \citenamefont {Xu}, \citenamefont
  {Yu}, \citenamefont {Yuan}, \citenamefont {Szegedy}, \citenamefont {Shi},
  \citenamefont {Zhao}, \citenamefont {Deng},\ and\ \citenamefont
  {Chen}}]{Huang2021EfficientParallelizationTensor}%
  \BibitemOpen
  \bibfield  {author} {\bibinfo {author} {\bibfnamefont {C.}~\bibnamefont
  {Huang}}, \bibinfo {author} {\bibfnamefont {F.}~\bibnamefont {Zhang}},
  \bibinfo {author} {\bibfnamefont {M.}~\bibnamefont {Newman}}, \bibinfo
  {author} {\bibfnamefont {X.}~\bibnamefont {Ni}}, \bibinfo {author}
  {\bibfnamefont {D.}~\bibnamefont {Ding}}, \bibinfo {author} {\bibfnamefont
  {J.}~\bibnamefont {Cai}}, \bibinfo {author} {\bibfnamefont {X.}~\bibnamefont
  {Gao}}, \bibinfo {author} {\bibfnamefont {T.}~\bibnamefont {Wang}}, \bibinfo
  {author} {\bibfnamefont {F.}~\bibnamefont {Wu}}, \bibinfo {author}
  {\bibfnamefont {G.}~\bibnamefont {Zhang}}, \bibinfo {author} {\bibfnamefont
  {H.-S.}\ \bibnamefont {Ku}}, \bibinfo {author} {\bibfnamefont
  {Z.}~\bibnamefont {Tian}}, \bibinfo {author} {\bibfnamefont {J.}~\bibnamefont
  {Wu}}, \bibinfo {author} {\bibfnamefont {H.}~\bibnamefont {Xu}}, \bibinfo
  {author} {\bibfnamefont {H.}~\bibnamefont {Yu}}, \bibinfo {author}
  {\bibfnamefont {B.}~\bibnamefont {Yuan}}, \bibinfo {author} {\bibfnamefont
  {M.}~\bibnamefont {Szegedy}}, \bibinfo {author} {\bibfnamefont
  {Y.}~\bibnamefont {Shi}}, \bibinfo {author} {\bibfnamefont {H.-H.}\
  \bibnamefont {Zhao}}, \bibinfo {author} {\bibfnamefont {C.}~\bibnamefont
  {Deng}},\ and\ \bibinfo {author} {\bibfnamefont {J.}~\bibnamefont {Chen}},\
  }\bibfield  {title} {\bibinfo {title} {Efficient parallelization of tensor
  network contraction for simulating quantum computation},\ }\href
  {https://doi.org/10.1038/s43588-021-00119-7} {\bibfield  {journal} {\bibinfo
  {journal} {Nature Computational Science}\ }\textbf {\bibinfo {volume} {1}},\
  \bibinfo {pages} {578} (\bibinfo {year} {2021})}\BibitemShut {NoStop}%
\bibitem [{\citenamefont {Haghshenas}\ \emph {et~al.}(2022)\citenamefont
  {Haghshenas}, \citenamefont {Gray}, \citenamefont {Potter},\ and\
  \citenamefont {Chan}}]{Haghshenas2022VariationalPowerQuantum}%
  \BibitemOpen
  \bibfield  {author} {\bibinfo {author} {\bibfnamefont {R.}~\bibnamefont
  {Haghshenas}}, \bibinfo {author} {\bibfnamefont {J.}~\bibnamefont {Gray}},
  \bibinfo {author} {\bibfnamefont {A.~C.}\ \bibnamefont {Potter}},\ and\
  \bibinfo {author} {\bibfnamefont {G.~K.-L.}\ \bibnamefont {Chan}},\
  }\bibfield  {title} {\bibinfo {title} {Variational power of quantum circuit
  tensor networks},\ }\href {https://doi.org/10.1103/PhysRevX.12.011047}
  {\bibfield  {journal} {\bibinfo  {journal} {Phys. Rev. X}\ }\textbf {\bibinfo
  {volume} {12}},\ \bibinfo {pages} {011047} (\bibinfo {year}
  {2022})}\BibitemShut {NoStop}%
\bibitem [{\citenamefont {Zohar}(2021)}]{Zohar2021QuantumSimulationLattice}%
  \BibitemOpen
  \bibfield  {author} {\bibinfo {author} {\bibfnamefont {E.}~\bibnamefont
  {Zohar}},\ }\bibfield  {title} {\bibinfo {title} {Quantum simulation of
  lattice gauge theories in more than one space
  dimension{\textemdash}requirements, challenges and methods},\ }\href
  {https://doi.org/10.1098/rsta.2021.0069} {\bibfield  {journal} {\bibinfo
  {journal} {Philosophical Transactions of the Royal Society A: Mathematical,
  Physical and Engineering Sciences}\ }\textbf {\bibinfo {volume} {380}},\
  \bibinfo {pages} {20210069} (\bibinfo {year} {2021})}\BibitemShut {NoStop}%
\bibitem [{\citenamefont {Catterall}\ \emph {et~al.}(2022)\citenamefont
  {Catterall}, \citenamefont {Harnik}, \citenamefont {Hubeny}, \citenamefont
  {Bauer}, \citenamefont {Berlin}, \citenamefont {Davoudi}, \citenamefont
  {Faulkner}, \citenamefont {Hartman}, \citenamefont {Headrick}, \citenamefont
  {Kahn}, \citenamefont {Lamm}, \citenamefont {Meurice}, \citenamefont
  {Rajendran}, \citenamefont {Rangamani},\ and\ \citenamefont
  {Swingle}}]{Catterall2022ReportSnowmass2021}%
  \BibitemOpen
  \bibfield  {author} {\bibinfo {author} {\bibfnamefont {S.}~\bibnamefont
  {Catterall}}, \bibinfo {author} {\bibfnamefont {R.}~\bibnamefont {Harnik}},
  \bibinfo {author} {\bibfnamefont {V.~E.}\ \bibnamefont {Hubeny}}, \bibinfo
  {author} {\bibfnamefont {C.~W.}\ \bibnamefont {Bauer}}, \bibinfo {author}
  {\bibfnamefont {A.}~\bibnamefont {Berlin}}, \bibinfo {author} {\bibfnamefont
  {Z.}~\bibnamefont {Davoudi}}, \bibinfo {author} {\bibfnamefont
  {T.}~\bibnamefont {Faulkner}}, \bibinfo {author} {\bibfnamefont
  {T.}~\bibnamefont {Hartman}}, \bibinfo {author} {\bibfnamefont
  {M.}~\bibnamefont {Headrick}}, \bibinfo {author} {\bibfnamefont {Y.~F.}\
  \bibnamefont {Kahn}}, \bibinfo {author} {\bibfnamefont {H.}~\bibnamefont
  {Lamm}}, \bibinfo {author} {\bibfnamefont {Y.}~\bibnamefont {Meurice}},
  \bibinfo {author} {\bibfnamefont {S.}~\bibnamefont {Rajendran}}, \bibinfo
  {author} {\bibfnamefont {M.}~\bibnamefont {Rangamani}},\ and\ \bibinfo
  {author} {\bibfnamefont {B.}~\bibnamefont {Swingle}},\ }\href
  {https://doi.org/10.2172/1892238} {\emph {\bibinfo {title} {Report of the
  {{Snowmass}} 2021 {{Theory Frontier Topical Group}} on {{Quantum Information
  Science}}}}},\ \bibinfo {type} {Tech. Rep.}\ \bibinfo {number}
  {FERMILAB-FN-1199-T; arXiv:2209.14839}\ (\bibinfo {year} {2022})\BibitemShut
  {NoStop}%
\bibitem [{\citenamefont {Pomarico}\ \emph {et~al.}(2023)\citenamefont
  {Pomarico}, \citenamefont {Cosmai}, \citenamefont {Facchi}, \citenamefont
  {Lupo}, \citenamefont {Pascazio},\ and\ \citenamefont
  {Pepe}}]{Pomarico2023DynamicalQuantumPhase}%
  \BibitemOpen
  \bibfield  {author} {\bibinfo {author} {\bibfnamefont {D.}~\bibnamefont
  {Pomarico}}, \bibinfo {author} {\bibfnamefont {L.}~\bibnamefont {Cosmai}},
  \bibinfo {author} {\bibfnamefont {P.}~\bibnamefont {Facchi}}, \bibinfo
  {author} {\bibfnamefont {C.}~\bibnamefont {Lupo}}, \bibinfo {author}
  {\bibfnamefont {S.}~\bibnamefont {Pascazio}},\ and\ \bibinfo {author}
  {\bibfnamefont {F.~V.}\ \bibnamefont {Pepe}},\ }\bibfield  {title} {\bibinfo
  {title} {Dynamical {{Quantum Phase Transitions}} of the {{Schwinger Model}}:
  {{Real-Time Dynamics}} on {{IBM Quantum}}},\ }\href
  {https://doi.org/10.3390/e25040608} {\bibfield  {journal} {\bibinfo
  {journal} {Entropy}\ }\textbf {\bibinfo {volume} {25}},\ \bibinfo {pages}
  {608} (\bibinfo {year} {2023})},\ \Eprint {https://arxiv.org/abs/2302.01151}
  {arxiv:2302.01151 [quant-ph]} \BibitemShut {NoStop}%
\bibitem [{\citenamefont {Funcke}\ \emph
  {et~al.}(2023{\natexlab{b}})\citenamefont {Funcke}, \citenamefont {Hartung},
  \citenamefont {Jansen},\ and\ \citenamefont
  {K{\"u}hn}}]{Funcke2023ReviewQuantumComputing}%
  \BibitemOpen
  \bibfield  {author} {\bibinfo {author} {\bibfnamefont {L.}~\bibnamefont
  {Funcke}}, \bibinfo {author} {\bibfnamefont {T.}~\bibnamefont {Hartung}},
  \bibinfo {author} {\bibfnamefont {K.}~\bibnamefont {Jansen}},\ and\ \bibinfo
  {author} {\bibfnamefont {S.}~\bibnamefont {K{\"u}hn}},\ }\bibfield  {title}
  {\bibinfo {title} {Review on {{Quantum Computing}} for {{Lattice Field
  Theory}}},\ }in\ \href {https://doi.org/10.22323/1.430.0228} {\emph {\bibinfo
  {booktitle} {Proceedings of {{The}} 39th {{International Symposium}} on
  {{Lattice Field Theory}} {\textemdash} {{PoS}}({{LATTICE2022}})}}},\ Vol.\
  \bibinfo {volume} {430}\ (\bibinfo  {publisher} {{SISSA Medialab}},\ \bibinfo
  {year} {2023})\ p.\ \bibinfo {pages} {228},\ \Eprint
  {https://arxiv.org/abs/2302.00467} {arxiv:2302.00467} \BibitemShut {NoStop}%
\bibitem [{\citenamefont {Mariani}\ \emph {et~al.}(2023)\citenamefont
  {Mariani}, \citenamefont {Pradhan},\ and\ \citenamefont
  {Ercolessi}}]{Mariani2023HamiltoniansGaugeInvariant}%
  \BibitemOpen
  \bibfield  {author} {\bibinfo {author} {\bibfnamefont {A.}~\bibnamefont
  {Mariani}}, \bibinfo {author} {\bibfnamefont {S.}~\bibnamefont {Pradhan}},\
  and\ \bibinfo {author} {\bibfnamefont {E.}~\bibnamefont {Ercolessi}},\
  }\bibfield  {title} {\bibinfo {title} {Hamiltonians and gauge-invariant
  {{Hilbert}} space for lattice {{Yang-Mills-like}} theories with finite gauge
  group},\ }\href {https://doi.org/10.1103/PhysRevD.107.114513} {\bibfield
  {journal} {\bibinfo  {journal} {Physical Review D}\ }\textbf {\bibinfo
  {volume} {107}},\ \bibinfo {pages} {114513} (\bibinfo {year} {2023})},\
  \Eprint {https://arxiv.org/abs/2301.12224} {arXiv:2301.12224 [quant-ph]}
  \BibitemShut {NoStop}%
\bibitem [{\citenamefont {Zache}\ \emph
  {et~al.}(2023{\natexlab{a}})\citenamefont {Zache}, \citenamefont
  {Gonz{\'a}lez-Cuadra},\ and\ \citenamefont
  {Zoller}}]{Zache2023FermionQuditQuantuma}%
  \BibitemOpen
  \bibfield  {author} {\bibinfo {author} {\bibfnamefont {T.~V.}\ \bibnamefont
  {Zache}}, \bibinfo {author} {\bibfnamefont {D.}~\bibnamefont
  {Gonz{\'a}lez-Cuadra}},\ and\ \bibinfo {author} {\bibfnamefont
  {P.}~\bibnamefont {Zoller}},\ }\bibfield  {title} {\bibinfo {title}
  {Fermion-qudit quantum processors for simulating lattice gauge theories with
  matter},\ }\href {https://doi.org/10.22331/q-2023-10-16-1140} {\bibfield
  {journal} {\bibinfo  {journal} {Quantum}\ }\textbf {\bibinfo {volume} {7}},\
  \bibinfo {pages} {1140} (\bibinfo {year} {2023}{\natexlab{a}})}\BibitemShut
  {NoStop}%
\bibitem [{\citenamefont {Eisert}\ \emph {et~al.}(2010)\citenamefont {Eisert},
  \citenamefont {Cramer},\ and\ \citenamefont
  {Plenio}}]{Eisert2010ColloquiumAreaLaws}%
  \BibitemOpen
  \bibfield  {author} {\bibinfo {author} {\bibfnamefont {J.}~\bibnamefont
  {Eisert}}, \bibinfo {author} {\bibfnamefont {M.}~\bibnamefont {Cramer}},\
  and\ \bibinfo {author} {\bibfnamefont {M.~B.}\ \bibnamefont {Plenio}},\
  }\bibfield  {title} {\bibinfo {title} {Colloquium: Area laws for the
  entanglement entropy},\ }\href {https://doi.org/10.1103/RevModPhys.82.277}
  {\bibfield  {journal} {\bibinfo  {journal} {Rev. Mod. Phys.}\ }\textbf
  {\bibinfo {volume} {82}},\ \bibinfo {pages} {277} (\bibinfo {year}
  {2010})}\BibitemShut {NoStop}%
\bibitem [{\citenamefont {Calabrese}\ and\ \citenamefont
  {Cardy}(2004)}]{Calabrese2004EntanglementEntropyQuantum}%
  \BibitemOpen
  \bibfield  {author} {\bibinfo {author} {\bibfnamefont {P.}~\bibnamefont
  {Calabrese}}\ and\ \bibinfo {author} {\bibfnamefont {J.}~\bibnamefont
  {Cardy}},\ }\bibfield  {title} {\bibinfo {title} {Entanglement entropy and
  quantum field theory},\ }\href
  {https://doi.org/10.1088/1742-5468/2004/06/P06002} {\bibfield  {journal}
  {\bibinfo  {journal} {Journal of Statistical Mechanics: Theory and
  Experiment}\ }\textbf {\bibinfo {volume} {2004}},\ \bibinfo {pages} {P06002}
  (\bibinfo {year} {2004})}\BibitemShut {NoStop}%
\bibitem [{\citenamefont {Hastings}(2007)}]{Hastings2007AreaLawOne}%
  \BibitemOpen
  \bibfield  {author} {\bibinfo {author} {\bibfnamefont {M.~B.}\ \bibnamefont
  {Hastings}},\ }\bibfield  {title} {\bibinfo {title} {An area law for
  one-dimensional quantum systems},\ }\href
  {https://doi.org/10.1088/1742-5468/2007/08/P08024} {\bibfield  {journal}
  {\bibinfo  {journal} {Journal of Statistical Mechanics: Theory and
  Experiment}\ }\textbf {\bibinfo {volume} {2007}},\ \bibinfo {pages} {P08024}
  (\bibinfo {year} {2007})}\BibitemShut {NoStop}%
\bibitem [{\citenamefont {Kuwahara}\ and\ \citenamefont
  {Saito}(2020)}]{Kuwahara2020AreaLawNoncritical}%
  \BibitemOpen
  \bibfield  {author} {\bibinfo {author} {\bibfnamefont {T.}~\bibnamefont
  {Kuwahara}}\ and\ \bibinfo {author} {\bibfnamefont {K.}~\bibnamefont
  {Saito}},\ }\bibfield  {title} {\bibinfo {title} {Area law of noncritical
  ground states in 1d long-range interacting systems},\ }\href
  {https://doi.org/10.1038/s41467-020-18055-x} {\bibfield  {journal} {\bibinfo
  {journal} {Nature Communications}\ }\textbf {\bibinfo {volume} {11}},\
  \bibinfo {pages} {4478} (\bibinfo {year} {2020})}\BibitemShut {NoStop}%
\bibitem [{\citenamefont {Cho}(2018)}]{Cho2018RealisticAreaLaw}%
  \BibitemOpen
  \bibfield  {author} {\bibinfo {author} {\bibfnamefont {J.}~\bibnamefont
  {Cho}},\ }\bibfield  {title} {\bibinfo {title} {Realistic area-law bound on
  entanglement from exponentially decaying correlations},\ }\href
  {https://doi.org/10.1103/PhysRevX.8.031009} {\bibfield  {journal} {\bibinfo
  {journal} {Phys. Rev. X}\ }\textbf {\bibinfo {volume} {8}},\ \bibinfo {pages}
  {031009} (\bibinfo {year} {2018})}\BibitemShut {NoStop}%
\bibitem [{\citenamefont {Wolf}\ \emph {et~al.}(2008)\citenamefont {Wolf},
  \citenamefont {Verstraete}, \citenamefont {Hastings},\ and\ \citenamefont
  {Cirac}}]{Wolf2008AreaLawsQuantum}%
  \BibitemOpen
  \bibfield  {author} {\bibinfo {author} {\bibfnamefont {M.~M.}\ \bibnamefont
  {Wolf}}, \bibinfo {author} {\bibfnamefont {F.}~\bibnamefont {Verstraete}},
  \bibinfo {author} {\bibfnamefont {M.~B.}\ \bibnamefont {Hastings}},\ and\
  \bibinfo {author} {\bibfnamefont {J.~I.}\ \bibnamefont {Cirac}},\ }\bibfield
  {title} {\bibinfo {title} {Area laws in quantum systems: Mutual information
  and correlations},\ }\href {https://doi.org/10.1103/PhysRevLett.100.070502}
  {\bibfield  {journal} {\bibinfo  {journal} {Phys. Rev. Lett.}\ }\textbf
  {\bibinfo {volume} {100}},\ \bibinfo {pages} {070502} (\bibinfo {year}
  {2008})}\BibitemShut {NoStop}%
\bibitem [{\citenamefont {Masanes}(2009)}]{Masanes2009AreaLaw}%
  \BibitemOpen
  \bibfield  {author} {\bibinfo {author} {\bibfnamefont {L.}~\bibnamefont
  {Masanes}},\ }\bibfield  {title} {\bibinfo {title} {Area law for the entropy
  of low-energy states},\ }\href {https://doi.org/10.1103/PhysRevA.80.052104}
  {\bibfield  {journal} {\bibinfo  {journal} {Physical Review A}\ }\textbf
  {\bibinfo {volume} {80}},\ \bibinfo {pages} {052104} (\bibinfo {year}
  {2009})}\BibitemShut {NoStop}%
\bibitem [{\citenamefont {Hamza}\ \emph {et~al.}(2009)\citenamefont {Hamza},
  \citenamefont {Michalakis}, \citenamefont {Nachtergaele},\ and\ \citenamefont
  {Sims}}]{Hamza2009ApproximatingtheGroundState}%
  \BibitemOpen
  \bibfield  {author} {\bibinfo {author} {\bibfnamefont {E.}~\bibnamefont
  {Hamza}}, \bibinfo {author} {\bibfnamefont {S.}~\bibnamefont {Michalakis}},
  \bibinfo {author} {\bibfnamefont {B.}~\bibnamefont {Nachtergaele}},\ and\
  \bibinfo {author} {\bibfnamefont {R.}~\bibnamefont {Sims}},\ }\bibfield
  {title} {\bibinfo {title} {Approximating the ground state of gapped quantum
  spin systems},\ }\href {https://doi.org/10.1063/1.3206662} {\bibfield
  {journal} {\bibinfo  {journal} {Journal of Mathematical Physics}\ }\textbf
  {\bibinfo {volume} {50}},\ \bibinfo {pages} {095213} (\bibinfo {year}
  {2009})}\BibitemShut {NoStop}%
\bibitem [{\citenamefont {Kastoryano}\ \emph {et~al.}(2019)\citenamefont
  {Kastoryano}, \citenamefont {Lucia},\ and\ \citenamefont
  {{Perez-Garcia}}}]{Kastoryano2019LocalityattheBoundary}%
  \BibitemOpen
  \bibfield  {author} {\bibinfo {author} {\bibfnamefont {M.~J.}\ \bibnamefont
  {Kastoryano}}, \bibinfo {author} {\bibfnamefont {A.}~\bibnamefont {Lucia}},\
  and\ \bibinfo {author} {\bibfnamefont {D.}~\bibnamefont {{Perez-Garcia}}},\
  }\bibfield  {title} {\bibinfo {title} {Locality at the {{Boundary Implies
  Gap}} in the {{Bulk}} for {{2D PEPS}}},\ }\href
  {https://doi.org/10.1007/s00220-019-03404-9} {\bibfield  {journal} {\bibinfo
  {journal} {Communications in Mathematical Physics}\ }\textbf {\bibinfo
  {volume} {366}},\ \bibinfo {pages} {895} (\bibinfo {year}
  {2019})}\BibitemShut {NoStop}%
\bibitem [{\citenamefont {Cirac}\ \emph {et~al.}(2019)\citenamefont {Cirac},
  \citenamefont {{Garre-Rubio}},\ and\ \citenamefont
  {{P{\'e}rez-Garc{\'i}a}}}]{Cirac2019MathematicalOpenProblems}%
  \BibitemOpen
  \bibfield  {author} {\bibinfo {author} {\bibfnamefont {J.~I.}\ \bibnamefont
  {Cirac}}, \bibinfo {author} {\bibfnamefont {J.}~\bibnamefont
  {{Garre-Rubio}}},\ and\ \bibinfo {author} {\bibfnamefont {D.}~\bibnamefont
  {{P{\'e}rez-Garc{\'i}a}}},\ }\bibfield  {title} {\bibinfo {title}
  {Mathematical open problems in projected entangled pair states},\ }\href
  {https://doi.org/10.1007/s13163-019-00318-x} {\bibfield  {journal} {\bibinfo
  {journal} {Revista Matem{\'a}tica Complutense}\ }\textbf {\bibinfo {volume}
  {32}},\ \bibinfo {pages} {579} (\bibinfo {year} {2019})}\BibitemShut
  {NoStop}%
\bibitem [{\citenamefont {Cirac}\ \emph {et~al.}(2021)\citenamefont {Cirac},
  \citenamefont {P\'erez-Garc\'{\i}a}, \citenamefont {Schuch},\ and\
  \citenamefont {Verstraete}}]{Cirac2021MatrixProductStates}%
  \BibitemOpen
  \bibfield  {author} {\bibinfo {author} {\bibfnamefont {J.~I.}\ \bibnamefont
  {Cirac}}, \bibinfo {author} {\bibfnamefont {D.}~\bibnamefont
  {P\'erez-Garc\'{\i}a}}, \bibinfo {author} {\bibfnamefont {N.}~\bibnamefont
  {Schuch}},\ and\ \bibinfo {author} {\bibfnamefont {F.}~\bibnamefont
  {Verstraete}},\ }\bibfield  {title} {\bibinfo {title} {Matrix product states
  and projected entangled pair states: Concepts, symmetries, theorems},\ }\href
  {https://doi.org/10.1103/RevModPhys.93.045003} {\bibfield  {journal}
  {\bibinfo  {journal} {Rev. Mod. Phys.}\ }\textbf {\bibinfo {volume} {93}},\
  \bibinfo {pages} {045003} (\bibinfo {year} {2021})}\BibitemShut {NoStop}%
\bibitem [{\citenamefont {Perez-Garcia}\ \emph {et~al.}(2007)\citenamefont
  {Perez-Garcia}, \citenamefont {Verstraete}, \citenamefont {Wolf},\ and\
  \citenamefont {Cirac}}]{PerezGarcia2007MatrixProductState}%
  \BibitemOpen
  \bibfield  {author} {\bibinfo {author} {\bibfnamefont {D.}~\bibnamefont
  {Perez-Garcia}}, \bibinfo {author} {\bibfnamefont {F.}~\bibnamefont
  {Verstraete}}, \bibinfo {author} {\bibfnamefont {M.~M.}\ \bibnamefont
  {Wolf}},\ and\ \bibinfo {author} {\bibfnamefont {J.~I.}\ \bibnamefont
  {Cirac}},\ }\bibfield  {title} {\bibinfo {title} {Matrix product state
  representations},\ }\href@noop {} {\bibfield  {journal} {\bibinfo  {journal}
  {Quantum Info. Comput.}\ }\textbf {\bibinfo {volume} {7}},\ \bibinfo {pages}
  {401} (\bibinfo {year} {2007})}\BibitemShut {NoStop}%
\bibitem [{\citenamefont
  {Schollw{\"{o}}ck}(2011)}]{Schollwoeck2011DensityMatrixRenormalization}%
  \BibitemOpen
  \bibfield  {author} {\bibinfo {author} {\bibfnamefont {U.}~\bibnamefont
  {Schollw{\"{o}}ck}},\ }\bibfield  {title} {\bibinfo {title} {The
  density-matrix renormalization group in the age of matrix product states},\
  }\href {https://doi.org/10.1016/j.aop.2010.09.012} {\bibfield  {journal}
  {\bibinfo  {journal} {Annals of Physics}\ }\textbf {\bibinfo {volume}
  {326}},\ \bibinfo {pages} {96} (\bibinfo {year} {2011})}\BibitemShut
  {NoStop}%
\bibitem [{\citenamefont {Chan}\ \emph {et~al.}(2016)\citenamefont {Chan},
  \citenamefont {Keselman}, \citenamefont {Nakatani}, \citenamefont {Li},\ and\
  \citenamefont {White}}]{Chan2016MatrixProductOperators}%
  \BibitemOpen
  \bibfield  {author} {\bibinfo {author} {\bibfnamefont {G.~K.-L.}\
  \bibnamefont {Chan}}, \bibinfo {author} {\bibfnamefont {A.}~\bibnamefont
  {Keselman}}, \bibinfo {author} {\bibfnamefont {N.}~\bibnamefont {Nakatani}},
  \bibinfo {author} {\bibfnamefont {Z.}~\bibnamefont {Li}},\ and\ \bibinfo
  {author} {\bibfnamefont {S.~R.}\ \bibnamefont {White}},\ }\href
  {https://doi.org/10.48550/arXiv.1605.02611} {\emph {\bibinfo {title} {Matrix
  {{Product Operators}}, {{Matrix Product States}}, and Ab Initio {{Density
  Matrix Renormalization Group}} Algorithms}}},\ \bibinfo {type} {Tech. Rep.}\
  \bibinfo {number} {arXiv:1605.02611}\ (\bibinfo {year} {2016})\ \Eprint
  {https://arxiv.org/abs/1605.02611} {arxiv:1605.02611 [cond-mat,
  physics:physics, physics:quant-ph]} \BibitemShut {NoStop}%
\bibitem [{Com()}]{CommentScalingA}%
  \BibitemOpen
  \href@noop {} {\bibinfo {title} {The scaling presented here does not consider
  the bond dimension of the mpo representing the hamiltonian. the bond
  dimension of the mpo depends on the dimensionality of the system. in a
  simplified argument holding for the ttn, the maximum number of interactions
  cut for any bipartition gives a first intuition.}}\BibitemShut {Stop}%
\bibitem [{\citenamefont {Gleis}\ \emph {et~al.}(2023)\citenamefont {Gleis},
  \citenamefont {Li},\ and\ \citenamefont {{\noopsort{delft}}{von
  Delft}}}]{Gleis2023ControlledBondExpansion}%
  \BibitemOpen
  \bibfield  {author} {\bibinfo {author} {\bibfnamefont {A.}~\bibnamefont
  {Gleis}}, \bibinfo {author} {\bibfnamefont {J.-W.}\ \bibnamefont {Li}},\ and\
  \bibinfo {author} {\bibfnamefont {J.}~\bibnamefont {{\noopsort{delft}}{von
  Delft}}},\ }\bibfield  {title} {\bibinfo {title} {Controlled {{Bond
  Expansion}} for {{Density Matrix Renormalization Group Ground State Search}}
  at {{Single-Site Costs}}},\ }\href
  {https://doi.org/10.1103/PhysRevLett.130.246402} {\bibfield  {journal}
  {\bibinfo  {journal} {Physical Review Letters}\ }\textbf {\bibinfo {volume}
  {130}},\ \bibinfo {pages} {246402} (\bibinfo {year} {2023})}\BibitemShut
  {NoStop}%
\bibitem [{\citenamefont {Schuch}\ \emph {et~al.}(2007)\citenamefont {Schuch},
  \citenamefont {Wolf}, \citenamefont {Verstraete},\ and\ \citenamefont
  {Cirac}}]{Schuch2007ComputationalComplexityProjected}%
  \BibitemOpen
  \bibfield  {author} {\bibinfo {author} {\bibfnamefont {N.}~\bibnamefont
  {Schuch}}, \bibinfo {author} {\bibfnamefont {M.~M.}\ \bibnamefont {Wolf}},
  \bibinfo {author} {\bibfnamefont {F.}~\bibnamefont {Verstraete}},\ and\
  \bibinfo {author} {\bibfnamefont {J.~I.}\ \bibnamefont {Cirac}},\ }\bibfield
  {title} {\bibinfo {title} {Computational complexity of projected entangled
  pair states},\ }\href {https://doi.org/10.1103/PhysRevLett.98.140506}
  {\bibfield  {journal} {\bibinfo  {journal} {Phys. Rev. Lett.}\ }\textbf
  {\bibinfo {volume} {98}},\ \bibinfo {pages} {140506} (\bibinfo {year}
  {2007})}\BibitemShut {NoStop}%
\bibitem [{\citenamefont {Eisert}(2013)}]{Eisert2013EntanglementTensorNetwork}%
  \BibitemOpen
  \bibfield  {author} {\bibinfo {author} {\bibfnamefont {J.}~\bibnamefont
  {Eisert}},\ }\href {https://doi.org/10.48550/arXiv.1308.3318} {\emph
  {\bibinfo {title} {Entanglement and tensor network states}}},\ \bibinfo
  {type} {Tech. Rep.}\ \bibinfo {number} {arXiv:1308.3318}\ (\bibinfo {year}
  {2013})\ \Eprint {https://arxiv.org/abs/1308.3318} {arxiv:1308.3318
  [cond-mat, physics:quant-ph]} \BibitemShut {NoStop}%
\bibitem [{\citenamefont {Vanderstraeten}\ \emph {et~al.}(2022)\citenamefont
  {Vanderstraeten}, \citenamefont {Burgelman}, \citenamefont {Ponsioen},
  \citenamefont {Van~Damme}, \citenamefont {Vanhecke}, \citenamefont {Corboz},
  \citenamefont {Haegeman},\ and\ \citenamefont
  {Verstraete}}]{Vanderstraeten2022VariationalMethodsContracting}%
  \BibitemOpen
  \bibfield  {author} {\bibinfo {author} {\bibfnamefont {L.}~\bibnamefont
  {Vanderstraeten}}, \bibinfo {author} {\bibfnamefont {L.}~\bibnamefont
  {Burgelman}}, \bibinfo {author} {\bibfnamefont {B.}~\bibnamefont {Ponsioen}},
  \bibinfo {author} {\bibfnamefont {M.}~\bibnamefont {Van~Damme}}, \bibinfo
  {author} {\bibfnamefont {B.}~\bibnamefont {Vanhecke}}, \bibinfo {author}
  {\bibfnamefont {P.}~\bibnamefont {Corboz}}, \bibinfo {author} {\bibfnamefont
  {J.}~\bibnamefont {Haegeman}},\ and\ \bibinfo {author} {\bibfnamefont
  {F.}~\bibnamefont {Verstraete}},\ }\bibfield  {title} {\bibinfo {title}
  {Variational methods for contracting projected entangled-pair states},\
  }\href {https://doi.org/10.1103/PhysRevB.105.195140} {\bibfield  {journal}
  {\bibinfo  {journal} {Phys. Rev. B}\ }\textbf {\bibinfo {volume} {105}},\
  \bibinfo {pages} {195140} (\bibinfo {year} {2022})}\BibitemShut {NoStop}%
\bibitem [{\citenamefont {Shi}\ \emph {et~al.}(2006)\citenamefont {Shi},
  \citenamefont {Duan},\ and\ \citenamefont
  {Vidal}}]{Shi2006ClassicalSimulationQuantum}%
  \BibitemOpen
  \bibfield  {author} {\bibinfo {author} {\bibfnamefont {Y.-Y.}\ \bibnamefont
  {Shi}}, \bibinfo {author} {\bibfnamefont {L.-M.}\ \bibnamefont {Duan}},\ and\
  \bibinfo {author} {\bibfnamefont {G.}~\bibnamefont {Vidal}},\ }\bibfield
  {title} {\bibinfo {title} {Classical simulation of quantum many-body systems
  with a tree tensor network},\ }\href
  {https://doi.org/10.1103/PhysRevA.74.022320} {\bibfield  {journal} {\bibinfo
  {journal} {Phys. Rev. A}\ }\textbf {\bibinfo {volume} {74}},\ \bibinfo
  {pages} {022320} (\bibinfo {year} {2006})}\BibitemShut {NoStop}%
\bibitem [{\citenamefont {Qian}\ and\ \citenamefont
  {Qin}(2022)}]{Qian2022TreeTensorNetwork}%
  \BibitemOpen
  \bibfield  {author} {\bibinfo {author} {\bibfnamefont {X.}~\bibnamefont
  {Qian}}\ and\ \bibinfo {author} {\bibfnamefont {M.}~\bibnamefont {Qin}},\
  }\bibfield  {title} {\bibinfo {title} {From tree tensor network to multiscale
  entanglement renormalization ansatz},\ }\href
  {https://doi.org/10.1103/PhysRevB.105.205102} {\bibfield  {journal} {\bibinfo
   {journal} {Phys. Rev. B}\ }\textbf {\bibinfo {volume} {105}},\ \bibinfo
  {pages} {205102} (\bibinfo {year} {2022})}\BibitemShut {NoStop}%
\bibitem [{\citenamefont {Ferris}(2013)}]{Ferris2013AreaLawReal}%
  \BibitemOpen
  \bibfield  {author} {\bibinfo {author} {\bibfnamefont {A.~J.}\ \bibnamefont
  {Ferris}},\ }\bibfield  {title} {\bibinfo {title} {Area law and real-space
  renormalization},\ }\href {https://doi.org/10.1103/PhysRevB.87.125139}
  {\bibfield  {journal} {\bibinfo  {journal} {Phys. Rev. B}\ }\textbf {\bibinfo
  {volume} {87}},\ \bibinfo {pages} {125139} (\bibinfo {year}
  {2013})}\BibitemShut {NoStop}%
\bibitem [{\citenamefont {Felser}\ \emph {et~al.}(2021)\citenamefont {Felser},
  \citenamefont {Notarnicola},\ and\ \citenamefont
  {Montangero}}]{Felser2021EfficientTensorNetwork}%
  \BibitemOpen
  \bibfield  {author} {\bibinfo {author} {\bibfnamefont {T.}~\bibnamefont
  {Felser}}, \bibinfo {author} {\bibfnamefont {S.}~\bibnamefont
  {Notarnicola}},\ and\ \bibinfo {author} {\bibfnamefont {S.}~\bibnamefont
  {Montangero}},\ }\bibfield  {title} {\bibinfo {title} {Efficient tensor
  network ansatz for high-dimensional quantum many-body problems},\ }\href
  {https://doi.org/10.1103/PhysRevLett.126.170603} {\bibfield  {journal}
  {\bibinfo  {journal} {Phys. Rev. Lett.}\ }\textbf {\bibinfo {volume} {126}},\
  \bibinfo {pages} {170603} (\bibinfo {year} {2021})}\BibitemShut {NoStop}%
\bibitem [{\citenamefont {Jaschke}\ \emph {et~al.}(2018)\citenamefont
  {Jaschke}, \citenamefont {Wall},\ and\ \citenamefont
  {Carr}}]{Jaschke2018OpenSourceMatrix}%
  \BibitemOpen
  \bibfield  {author} {\bibinfo {author} {\bibfnamefont {D.}~\bibnamefont
  {Jaschke}}, \bibinfo {author} {\bibfnamefont {M.~L.}\ \bibnamefont {Wall}},\
  and\ \bibinfo {author} {\bibfnamefont {L.~D.}\ \bibnamefont {Carr}},\
  }\bibfield  {title} {\bibinfo {title} {Open source matrix product states:
  Opening ways to simulate entangled many-body quantum systems in one
  dimension},\ }\href {https://doi.org/10.1016/j.cpc.2017.12.015} {\bibfield
  {journal} {\bibinfo  {journal} {Computer Physics Communications}\ }\textbf
  {\bibinfo {volume} {225}},\ \bibinfo {pages} {59} (\bibinfo {year}
  {2018})}\BibitemShut {NoStop}%
\bibitem [{\citenamefont {Vidal}(2004)}]{Vidal2004EfficientSimulationOne}%
  \BibitemOpen
  \bibfield  {author} {\bibinfo {author} {\bibfnamefont {G.}~\bibnamefont
  {Vidal}},\ }\bibfield  {title} {\bibinfo {title} {Efficient simulation of
  one-dimensional quantum many-body systems},\ }\href
  {https://doi.org/10.1103/PhysRevLett.93.040502} {\bibfield  {journal}
  {\bibinfo  {journal} {Phys. Rev. Lett.}\ }\textbf {\bibinfo {volume} {93}},\
  \bibinfo {pages} {040502} (\bibinfo {year} {2004})}\BibitemShut {NoStop}%
\bibitem [{\citenamefont {Haegeman}\ \emph {et~al.}(2011)\citenamefont
  {Haegeman}, \citenamefont {Cirac}, \citenamefont {Osborne}, \citenamefont
  {Pi\ifmmode~\check{z}\else \v{z}\fi{}orn}, \citenamefont {Verschelde},\ and\
  \citenamefont {Verstraete}}]{Haegeman2011TimeDependentVariational}%
  \BibitemOpen
  \bibfield  {author} {\bibinfo {author} {\bibfnamefont {J.}~\bibnamefont
  {Haegeman}}, \bibinfo {author} {\bibfnamefont {J.~I.}\ \bibnamefont {Cirac}},
  \bibinfo {author} {\bibfnamefont {T.~J.}\ \bibnamefont {Osborne}}, \bibinfo
  {author} {\bibfnamefont {I.}~\bibnamefont {Pi\ifmmode~\check{z}\else
  \v{z}\fi{}orn}}, \bibinfo {author} {\bibfnamefont {H.}~\bibnamefont
  {Verschelde}},\ and\ \bibinfo {author} {\bibfnamefont {F.}~\bibnamefont
  {Verstraete}},\ }\bibfield  {title} {\bibinfo {title} {Time-dependent
  variational principle for quantum lattices},\ }\href
  {https://doi.org/10.1103/PhysRevLett.107.070601} {\bibfield  {journal}
  {\bibinfo  {journal} {Phys. Rev. Lett.}\ }\textbf {\bibinfo {volume} {107}},\
  \bibinfo {pages} {070601} (\bibinfo {year} {2011})}\BibitemShut {NoStop}%
\bibitem [{\citenamefont {Bauernfeind}\ and\ \citenamefont
  {Aichhorn}(2020)}]{Bauernfeind2020TimeDependentVariational}%
  \BibitemOpen
  \bibfield  {author} {\bibinfo {author} {\bibfnamefont {D.}~\bibnamefont
  {Bauernfeind}}\ and\ \bibinfo {author} {\bibfnamefont {M.}~\bibnamefont
  {Aichhorn}},\ }\bibfield  {title} {\bibinfo {title} {{Time dependent
  variational principle for tree Tensor Networks}},\ }\href
  {https://doi.org/10.21468/SciPostPhys.8.2.024} {\bibfield  {journal}
  {\bibinfo  {journal} {SciPost Phys.}\ }\textbf {\bibinfo {volume} {8}},\
  \bibinfo {pages} {024} (\bibinfo {year} {2020})}\BibitemShut {NoStop}%
\bibitem [{\citenamefont {Kohn}\ \emph {et~al.}(2020)\citenamefont {Kohn},
  \citenamefont {Silvi}, \citenamefont {Gerster}, \citenamefont {Keck},
  \citenamefont {Fazio}, \citenamefont {Santoro},\ and\ \citenamefont
  {Montangero}}]{PhysRevA.101.023617}%
  \BibitemOpen
  \bibfield  {author} {\bibinfo {author} {\bibfnamefont {L.}~\bibnamefont
  {Kohn}}, \bibinfo {author} {\bibfnamefont {P.}~\bibnamefont {Silvi}},
  \bibinfo {author} {\bibfnamefont {M.}~\bibnamefont {Gerster}}, \bibinfo
  {author} {\bibfnamefont {M.}~\bibnamefont {Keck}}, \bibinfo {author}
  {\bibfnamefont {R.}~\bibnamefont {Fazio}}, \bibinfo {author} {\bibfnamefont
  {G.~E.}\ \bibnamefont {Santoro}},\ and\ \bibinfo {author} {\bibfnamefont
  {S.}~\bibnamefont {Montangero}},\ }\bibfield  {title} {\bibinfo {title}
  {Superfluid-to-mott transition in a bose-hubbard ring: Persistent currents
  and defect formation},\ }\href {https://doi.org/10.1103/PhysRevA.101.023617}
  {\bibfield  {journal} {\bibinfo  {journal} {Phys. Rev. A}\ }\textbf {\bibinfo
  {volume} {101}},\ \bibinfo {pages} {023617} (\bibinfo {year}
  {2020})}\BibitemShut {NoStop}%
\bibitem [{\citenamefont {Schuch}\ \emph {et~al.}(2008)\citenamefont {Schuch},
  \citenamefont {Wolf}, \citenamefont {Verstraete},\ and\ \citenamefont
  {Cirac}}]{Schuch2008EntropyScalingSimulability}%
  \BibitemOpen
  \bibfield  {author} {\bibinfo {author} {\bibfnamefont {N.}~\bibnamefont
  {Schuch}}, \bibinfo {author} {\bibfnamefont {M.~M.}\ \bibnamefont {Wolf}},
  \bibinfo {author} {\bibfnamefont {F.}~\bibnamefont {Verstraete}},\ and\
  \bibinfo {author} {\bibfnamefont {J.~I.}\ \bibnamefont {Cirac}},\ }\bibfield
  {title} {\bibinfo {title} {Entropy scaling and simulability by matrix product
  states},\ }\href {https://doi.org/10.1103/PhysRevLett.100.030504} {\bibfield
  {journal} {\bibinfo  {journal} {Phys. Rev. Lett.}\ }\textbf {\bibinfo
  {volume} {100}},\ \bibinfo {pages} {030504} (\bibinfo {year}
  {2008})}\BibitemShut {NoStop}%
\bibitem [{\citenamefont {Paeckel}\ \emph {et~al.}(2019)\citenamefont
  {Paeckel}, \citenamefont {K{\"{o}}hler}, \citenamefont {Swoboda},
  \citenamefont {Manmana}, \citenamefont {Schollw{\"{o}}ck},\ and\
  \citenamefont {Hubig}}]{Paeckel2019TimeEvolutionMethods}%
  \BibitemOpen
  \bibfield  {author} {\bibinfo {author} {\bibfnamefont {S.}~\bibnamefont
  {Paeckel}}, \bibinfo {author} {\bibfnamefont {T.}~\bibnamefont
  {K{\"{o}}hler}}, \bibinfo {author} {\bibfnamefont {A.}~\bibnamefont
  {Swoboda}}, \bibinfo {author} {\bibfnamefont {S.~R.}\ \bibnamefont
  {Manmana}}, \bibinfo {author} {\bibfnamefont {U.}~\bibnamefont
  {Schollw{\"{o}}ck}},\ and\ \bibinfo {author} {\bibfnamefont {C.}~\bibnamefont
  {Hubig}},\ }\bibfield  {title} {\bibinfo {title} {Time-evolution methods for
  matrix-product states},\ }\href {https://doi.org/10.1016/j.aop.2019.167998}
  {\bibfield  {journal} {\bibinfo  {journal} {Annals of Physics}\ }\textbf
  {\bibinfo {volume} {411}},\ \bibinfo {pages} {167998} (\bibinfo {year}
  {2019})}\BibitemShut {NoStop}%
\bibitem [{\citenamefont {White}\ \emph {et~al.}(2018)\citenamefont {White},
  \citenamefont {Zaletel}, \citenamefont {Mong},\ and\ \citenamefont
  {Refael}}]{White2018QuantumDynamicsThermalizing}%
  \BibitemOpen
  \bibfield  {author} {\bibinfo {author} {\bibfnamefont {C.~D.}\ \bibnamefont
  {White}}, \bibinfo {author} {\bibfnamefont {M.}~\bibnamefont {Zaletel}},
  \bibinfo {author} {\bibfnamefont {R.~S.~K.}\ \bibnamefont {Mong}},\ and\
  \bibinfo {author} {\bibfnamefont {G.}~\bibnamefont {Refael}},\ }\bibfield
  {title} {\bibinfo {title} {Quantum dynamics of thermalizing systems},\ }\href
  {https://doi.org/10.1103/PhysRevB.97.035127} {\bibfield  {journal} {\bibinfo
  {journal} {Phys. Rev. B}\ }\textbf {\bibinfo {volume} {97}},\ \bibinfo
  {pages} {035127} (\bibinfo {year} {2018})}\BibitemShut {NoStop}%
\bibitem [{\citenamefont {Surace}\ \emph {et~al.}(2019)\citenamefont {Surace},
  \citenamefont {Piani},\ and\ \citenamefont
  {Tagliacozzo}}]{Surace2019SimulatingOutEquilibrium}%
  \BibitemOpen
  \bibfield  {author} {\bibinfo {author} {\bibfnamefont {J.}~\bibnamefont
  {Surace}}, \bibinfo {author} {\bibfnamefont {M.}~\bibnamefont {Piani}},\ and\
  \bibinfo {author} {\bibfnamefont {L.}~\bibnamefont {Tagliacozzo}},\
  }\bibfield  {title} {\bibinfo {title} {Simulating the out-of-equilibrium
  dynamics of local observables by trading entanglement for mixture},\ }\href
  {https://doi.org/10.1103/PhysRevB.99.235115} {\bibfield  {journal} {\bibinfo
  {journal} {Phys. Rev. B}\ }\textbf {\bibinfo {volume} {99}},\ \bibinfo
  {pages} {235115} (\bibinfo {year} {2019})}\BibitemShut {NoStop}%
\bibitem [{\citenamefont {Susskind}(1977)}]{Susskind1977LatticeFermions}%
  \BibitemOpen
  \bibfield  {author} {\bibinfo {author} {\bibfnamefont {L.}~\bibnamefont
  {Susskind}},\ }\bibfield  {title} {\bibinfo {title} {Lattice fermions},\
  }\href {https://doi.org/10.1103/PhysRevD.16.3031} {\bibfield  {journal}
  {\bibinfo  {journal} {Physical Review D}\ }\textbf {\bibinfo {volume} {16}},\
  \bibinfo {pages} {3031} (\bibinfo {year} {1977})}\BibitemShut {NoStop}%
\bibitem [{\citenamefont {Zohar}\ and\ \citenamefont
  {Burrello}(2015)}]{Zohar2015FormulationLatticeGauge}%
  \BibitemOpen
  \bibfield  {author} {\bibinfo {author} {\bibfnamefont {E.}~\bibnamefont
  {Zohar}}\ and\ \bibinfo {author} {\bibfnamefont {M.}~\bibnamefont
  {Burrello}},\ }\bibfield  {title} {\bibinfo {title} {Formulation of lattice
  gauge theories for quantum simulations},\ }\href
  {https://doi.org/10.1103/PhysRevD.91.054506} {\bibfield  {journal} {\bibinfo
  {journal} {Physical Review D}\ }\textbf {\bibinfo {volume} {91}},\ \bibinfo
  {pages} {054506} (\bibinfo {year} {2015})}\BibitemShut {NoStop}%
\bibitem [{\citenamefont {Sala}\ \emph
  {et~al.}(2018{\natexlab{b}})\citenamefont {Sala}, \citenamefont {Shi},
  \citenamefont {K\"uhn}, \citenamefont {Ba\~nuls}, \citenamefont {Demler},\
  and\ \citenamefont {Cirac}}]{Sala2018VariationalStudyU1}%
  \BibitemOpen
  \bibfield  {author} {\bibinfo {author} {\bibfnamefont {P.}~\bibnamefont
  {Sala}}, \bibinfo {author} {\bibfnamefont {T.}~\bibnamefont {Shi}}, \bibinfo
  {author} {\bibfnamefont {S.}~\bibnamefont {K\"uhn}}, \bibinfo {author}
  {\bibfnamefont {M.~C.}\ \bibnamefont {Ba\~nuls}}, \bibinfo {author}
  {\bibfnamefont {E.}~\bibnamefont {Demler}},\ and\ \bibinfo {author}
  {\bibfnamefont {J.~I.}\ \bibnamefont {Cirac}},\ }\bibfield  {title} {\bibinfo
  {title} {Variational study of u(1) and su(2) lattice gauge theories with
  gaussian states in $1+1$ dimensions},\ }\href
  {https://doi.org/10.1103/PhysRevD.98.034505} {\bibfield  {journal} {\bibinfo
  {journal} {Phys. Rev. D}\ }\textbf {\bibinfo {volume} {98}},\ \bibinfo
  {pages} {034505} (\bibinfo {year} {2018}{\natexlab{b}})}\BibitemShut
  {NoStop}%
\bibitem [{\citenamefont {Bender}\ and\ \citenamefont
  {Zohar}(2020)}]{Bender2020GaugeRedundancyFree}%
  \BibitemOpen
  \bibfield  {author} {\bibinfo {author} {\bibfnamefont {J.}~\bibnamefont
  {Bender}}\ and\ \bibinfo {author} {\bibfnamefont {E.}~\bibnamefont {Zohar}},\
  }\bibfield  {title} {\bibinfo {title} {Gauge redundancy-free formulation of
  compact qed with dynamical matter for quantum and classical computations},\
  }\href {https://doi.org/10.1103/PhysRevD.102.114517} {\bibfield  {journal}
  {\bibinfo  {journal} {Phys. Rev. D}\ }\textbf {\bibinfo {volume} {102}},\
  \bibinfo {pages} {114517} (\bibinfo {year} {2020})}\BibitemShut {NoStop}%
\bibitem [{\citenamefont {Horn}(1981)}]{Horn1981FiniteMatrixModels}%
  \BibitemOpen
  \bibfield  {author} {\bibinfo {author} {\bibfnamefont {D.}~\bibnamefont
  {Horn}},\ }\bibfield  {title} {\bibinfo {title} {Finite matrix models with
  continuous local gauge invariance},\ }\href
  {https://doi.org/10.1016/0370-2693(81)90763-2} {\bibfield  {journal}
  {\bibinfo  {journal} {Physics Letters B}\ }\textbf {\bibinfo {volume}
  {100}},\ \bibinfo {pages} {149} (\bibinfo {year} {1981})}\BibitemShut
  {NoStop}%
\bibitem [{\citenamefont {Orland}\ and\ \citenamefont
  {Rohrlich}(1990)}]{Orland1990LatticeGaugeMagnets}%
  \BibitemOpen
  \bibfield  {author} {\bibinfo {author} {\bibfnamefont {P.}~\bibnamefont
  {Orland}}\ and\ \bibinfo {author} {\bibfnamefont {D.}~\bibnamefont
  {Rohrlich}},\ }\bibfield  {title} {\bibinfo {title} {Lattice gauge magnets:
  {{Local}} isospin from spin},\ }\href
  {https://doi.org/10.1016/0550-3213(90)90646-U} {\bibfield  {journal}
  {\bibinfo  {journal} {Nuclear Physics B}\ }\textbf {\bibinfo {volume}
  {338}},\ \bibinfo {pages} {647} (\bibinfo {year} {1990})}\BibitemShut
  {NoStop}%
\bibitem [{\citenamefont {Chandrasekharan}\ and\ \citenamefont
  {Wiese}(1997)}]{Chandrasekharan1997QuantumLinkModels}%
  \BibitemOpen
  \bibfield  {author} {\bibinfo {author} {\bibfnamefont {S.}~\bibnamefont
  {Chandrasekharan}}\ and\ \bibinfo {author} {\bibfnamefont {U.~J.}\
  \bibnamefont {Wiese}},\ }\bibfield  {title} {\bibinfo {title} {Quantum link
  models: {{A}} discrete approach to gauge theories},\ }\href
  {https://doi.org/10.1016/S0550-3213(97)80041-7} {\bibfield  {journal}
  {\bibinfo  {journal} {Nuclear Physics B}\ }\textbf {\bibinfo {volume}
  {492}},\ \bibinfo {pages} {455} (\bibinfo {year} {1997})}\BibitemShut
  {NoStop}%
\bibitem [{\citenamefont {Brower}\ \emph {et~al.}(1999)\citenamefont {Brower},
  \citenamefont {Chandrasekharan},\ and\ \citenamefont
  {Wiese}}]{Brower1999QcdAsQuantum}%
  \BibitemOpen
  \bibfield  {author} {\bibinfo {author} {\bibfnamefont {R.}~\bibnamefont
  {Brower}}, \bibinfo {author} {\bibfnamefont {S.}~\bibnamefont
  {Chandrasekharan}},\ and\ \bibinfo {author} {\bibfnamefont {U.-J.}\
  \bibnamefont {Wiese}},\ }\bibfield  {title} {\bibinfo {title} {{{QCD}} as a
  quantum link model},\ }\href {https://doi.org/10.1103/PhysRevD.60.094502}
  {\bibfield  {journal} {\bibinfo  {journal} {Physical Review D}\ }\textbf
  {\bibinfo {volume} {60}},\ \bibinfo {pages} {094502} (\bibinfo {year}
  {1999})}\BibitemShut {NoStop}%
\bibitem [{\citenamefont {Byrnes}\ and\ \citenamefont
  {Yamamoto}(2006)}]{Byrnes2006SimulatingLatticeGauge}%
  \BibitemOpen
  \bibfield  {author} {\bibinfo {author} {\bibfnamefont {T.}~\bibnamefont
  {Byrnes}}\ and\ \bibinfo {author} {\bibfnamefont {Y.}~\bibnamefont
  {Yamamoto}},\ }\bibfield  {title} {\bibinfo {title} {Simulating lattice gauge
  theories on a quantum computer},\ }\href
  {https://doi.org/10.1103/PhysRevA.73.022328} {\bibfield  {journal} {\bibinfo
  {journal} {Physical Review A}\ }\textbf {\bibinfo {volume} {73}},\ \bibinfo
  {pages} {022328} (\bibinfo {year} {2006})}\BibitemShut {NoStop}%
\bibitem [{\citenamefont {Davoudi}\ \emph {et~al.}(2020)\citenamefont
  {Davoudi}, \citenamefont {Hafezi}, \citenamefont {Monroe}, \citenamefont
  {Pagano}, \citenamefont {Seif},\ and\ \citenamefont
  {Shaw}}]{Davoudi2020TowardsAnalogQuantum}%
  \BibitemOpen
  \bibfield  {author} {\bibinfo {author} {\bibfnamefont {Z.}~\bibnamefont
  {Davoudi}}, \bibinfo {author} {\bibfnamefont {M.}~\bibnamefont {Hafezi}},
  \bibinfo {author} {\bibfnamefont {C.}~\bibnamefont {Monroe}}, \bibinfo
  {author} {\bibfnamefont {G.}~\bibnamefont {Pagano}}, \bibinfo {author}
  {\bibfnamefont {A.}~\bibnamefont {Seif}},\ and\ \bibinfo {author}
  {\bibfnamefont {A.}~\bibnamefont {Shaw}},\ }\bibfield  {title} {\bibinfo
  {title} {Towards analog quantum simulations of lattice gauge theories with
  trapped ions},\ }\href {https://doi.org/10.1103/PhysRevResearch.2.023015}
  {\bibfield  {journal} {\bibinfo  {journal} {Physical Review Research}\
  }\textbf {\bibinfo {volume} {2}},\ \bibinfo {pages} {023015} (\bibinfo {year}
  {2020})}\BibitemShut {NoStop}%
\bibitem [{\citenamefont {Mazzola}\ \emph {et~al.}(2021)\citenamefont
  {Mazzola}, \citenamefont {Mathis}, \citenamefont {Mazzola},\ and\
  \citenamefont {Tavernelli}}]{Mazzola2021GaugeInvariantQuantum}%
  \BibitemOpen
  \bibfield  {author} {\bibinfo {author} {\bibfnamefont {G.}~\bibnamefont
  {Mazzola}}, \bibinfo {author} {\bibfnamefont {S.~V.}\ \bibnamefont {Mathis}},
  \bibinfo {author} {\bibfnamefont {G.}~\bibnamefont {Mazzola}},\ and\ \bibinfo
  {author} {\bibfnamefont {I.}~\bibnamefont {Tavernelli}},\ }\bibfield  {title}
  {\bibinfo {title} {Gauge-invariant quantum circuits for \${{U}}\$(1) and
  {{Yang-Mills}} lattice gauge theories},\ }\href
  {https://doi.org/10.1103/PhysRevResearch.3.043209} {\bibfield  {journal}
  {\bibinfo  {journal} {Physical Review Research}\ }\textbf {\bibinfo {volume}
  {3}},\ \bibinfo {pages} {043209} (\bibinfo {year} {2021})}\BibitemShut
  {NoStop}%
\bibitem [{\citenamefont {Kan}\ \emph {et~al.}(2021)\citenamefont {Kan},
  \citenamefont {Funcke}, \citenamefont {K{\"u}hn}, \citenamefont
  {Dellantonio}, \citenamefont {Zhang}, \citenamefont {Haase}, \citenamefont
  {Muschik},\ and\ \citenamefont
  {Jansen}}]{Kan2021Investigating31mathrmdTopological}%
  \BibitemOpen
  \bibfield  {author} {\bibinfo {author} {\bibfnamefont {A.}~\bibnamefont
  {Kan}}, \bibinfo {author} {\bibfnamefont {L.}~\bibnamefont {Funcke}},
  \bibinfo {author} {\bibfnamefont {S.}~\bibnamefont {K{\"u}hn}}, \bibinfo
  {author} {\bibfnamefont {L.}~\bibnamefont {Dellantonio}}, \bibinfo {author}
  {\bibfnamefont {J.}~\bibnamefont {Zhang}}, \bibinfo {author} {\bibfnamefont
  {J.~F.}\ \bibnamefont {Haase}}, \bibinfo {author} {\bibfnamefont {C.~A.}\
  \bibnamefont {Muschik}},\ and\ \bibinfo {author} {\bibfnamefont
  {K.}~\bibnamefont {Jansen}},\ }\bibfield  {title} {\bibinfo {title}
  {Investigating a
  \$(3+1){\textbackslash}mathrm\{\vphantom\}{{D}}\vphantom\{\}\$ topological
  \${\textbackslash}ensuremath\{{\textbackslash}theta\}\$-term in the
  {{Hamiltonian}} formulation of lattice gauge theories for quantum and
  classical simulations},\ }\href {https://doi.org/10.1103/PhysRevD.104.034504}
  {\bibfield  {journal} {\bibinfo  {journal} {Physical Review D}\ }\textbf
  {\bibinfo {volume} {104}},\ \bibinfo {pages} {034504} (\bibinfo {year}
  {2021})}\BibitemShut {NoStop}%
\bibitem [{\citenamefont {Bauer}\ \emph
  {et~al.}(2023{\natexlab{a}})\citenamefont {Bauer}, \citenamefont {Davoudi},
  \citenamefont {Klco},\ and\ \citenamefont
  {Savage}}]{Bauer2023QuantumSimulationFundamental}%
  \BibitemOpen
  \bibfield  {author} {\bibinfo {author} {\bibfnamefont {C.~W.}\ \bibnamefont
  {Bauer}}, \bibinfo {author} {\bibfnamefont {Z.}~\bibnamefont {Davoudi}},
  \bibinfo {author} {\bibfnamefont {N.}~\bibnamefont {Klco}},\ and\ \bibinfo
  {author} {\bibfnamefont {M.~J.}\ \bibnamefont {Savage}},\ }\bibfield  {title}
  {\bibinfo {title} {Quantum simulation of fundamental particles and forces},\
  }\href {https://doi.org/10.1038/s42254-023-00599-8} {\bibfield  {journal}
  {\bibinfo  {journal} {Nature Reviews Physics}\ }\textbf {\bibinfo {volume}
  {5}},\ \bibinfo {pages} {420} (\bibinfo {year}
  {2023}{\natexlab{a}})}\BibitemShut {NoStop}%
\bibitem [{\citenamefont {Haase}\ \emph {et~al.}(2021)\citenamefont {Haase},
  \citenamefont {Dellantonio}, \citenamefont {Celi}, \citenamefont {Paulson},
  \citenamefont {Kan}, \citenamefont {Jansen},\ and\ \citenamefont
  {Muschik}}]{Haase2021ResourceEfficientApproach}%
  \BibitemOpen
  \bibfield  {author} {\bibinfo {author} {\bibfnamefont {J.~F.}\ \bibnamefont
  {Haase}}, \bibinfo {author} {\bibfnamefont {L.}~\bibnamefont {Dellantonio}},
  \bibinfo {author} {\bibfnamefont {A.}~\bibnamefont {Celi}}, \bibinfo {author}
  {\bibfnamefont {D.}~\bibnamefont {Paulson}}, \bibinfo {author} {\bibfnamefont
  {A.}~\bibnamefont {Kan}}, \bibinfo {author} {\bibfnamefont {K.}~\bibnamefont
  {Jansen}},\ and\ \bibinfo {author} {\bibfnamefont {C.~A.}\ \bibnamefont
  {Muschik}},\ }\bibfield  {title} {\bibinfo {title} {A resource efficient
  approach for quantum and classical simulations of gauge theories in particle
  physics},\ }\href {https://doi.org/10.22331/q-2021-02-04-393} {\bibfield
  {journal} {\bibinfo  {journal} {{Quantum}}\ }\textbf {\bibinfo {volume}
  {5}},\ \bibinfo {pages} {393} (\bibinfo {year} {2021})}\BibitemShut {NoStop}%
\bibitem [{\citenamefont {Hackett}\ \emph {et~al.}(2019)\citenamefont
  {Hackett}, \citenamefont {Howe}, \citenamefont {Hughes}, \citenamefont {Jay},
  \citenamefont {Neil},\ and\ \citenamefont
  {Simone}}]{Hackett2019DigitizingGaugeFields}%
  \BibitemOpen
  \bibfield  {author} {\bibinfo {author} {\bibfnamefont {D.~C.}\ \bibnamefont
  {Hackett}}, \bibinfo {author} {\bibfnamefont {K.}~\bibnamefont {Howe}},
  \bibinfo {author} {\bibfnamefont {C.}~\bibnamefont {Hughes}}, \bibinfo
  {author} {\bibfnamefont {W.}~\bibnamefont {Jay}}, \bibinfo {author}
  {\bibfnamefont {E.~T.}\ \bibnamefont {Neil}},\ and\ \bibinfo {author}
  {\bibfnamefont {J.~N.}\ \bibnamefont {Simone}},\ }\bibfield  {title}
  {\bibinfo {title} {Digitizing gauge fields: Lattice monte carlo results for
  future quantum computers},\ }\href
  {https://doi.org/10.1103/PhysRevA.99.062341} {\bibfield  {journal} {\bibinfo
  {journal} {Phys. Rev. A}\ }\textbf {\bibinfo {volume} {99}},\ \bibinfo
  {pages} {062341} (\bibinfo {year} {2019})}\BibitemShut {NoStop}%
\bibitem [{\citenamefont {Zache}\ \emph
  {et~al.}(2023{\natexlab{b}})\citenamefont {Zache}, \citenamefont
  {Gonz{\'a}lez-Cuadra},\ and\ \citenamefont {Zoller}}]{Zache2023}%
  \BibitemOpen
  \bibfield  {author} {\bibinfo {author} {\bibfnamefont {T.~V.}\ \bibnamefont
  {Zache}}, \bibinfo {author} {\bibfnamefont {D.}~\bibnamefont
  {Gonz{\'a}lez-Cuadra}},\ and\ \bibinfo {author} {\bibfnamefont
  {P.}~\bibnamefont {Zoller}},\ }\href
  {https://doi.org/10.48550/arXiv.2304.02527} {\bibinfo {title} {Quantum and
  classical spin network algorithms for \$q\$-deformed {{Kogut-Susskind}} gauge
  theories}} (\bibinfo {year} {2023}{\natexlab{b}}),\ \Eprint
  {https://arxiv.org/abs/2304.02527} {arxiv:2304.02527 [cond-mat,
  physics:hep-lat, physics:quant-ph]} \BibitemShut {NoStop}%
\bibitem [{\citenamefont {Rigobello}\ \emph {et~al.}(2023)\citenamefont
  {Rigobello}, \citenamefont {Magnifico}, \citenamefont {Silvi},\ and\
  \citenamefont {Montangero}}]{Rigobello2023a}%
  \BibitemOpen
  \bibfield  {author} {\bibinfo {author} {\bibfnamefont {M.}~\bibnamefont
  {Rigobello}}, \bibinfo {author} {\bibfnamefont {G.}~\bibnamefont
  {Magnifico}}, \bibinfo {author} {\bibfnamefont {P.}~\bibnamefont {Silvi}},\
  and\ \bibinfo {author} {\bibfnamefont {S.}~\bibnamefont {Montangero}},\
  }\href {https://doi.org/10.48550/arXiv.2308.04488} {\bibinfo {title} {Hadrons
  in (1+1){{D Hamiltonian}} hardcore lattice {{QCD}}}} (\bibinfo {year}
  {2023}),\ \Eprint {https://arxiv.org/abs/2308.04488} {arxiv:2308.04488
  [cond-mat, physics:hep-lat, physics:hep-ph, physics:physics,
  physics:quant-ph]} \BibitemShut {NoStop}%
\bibitem [{\citenamefont
  {Strocchi}(2013)}]{Strocchi2013IntroductionNonPerturbative}%
  \BibitemOpen
  \bibfield  {author} {\bibinfo {author} {\bibfnamefont {F.}~\bibnamefont
  {Strocchi}},\ }\href
  {https://doi.org/10.1093/acprof:oso/9780199671571.001.0001} {\emph {\bibinfo
  {title} {An {Introduction} to {Non-Perturbative Foundations} of {Quantum
  Field Theory}}}},\ International {Series} of {Monographs} on {Physics}\
  (\bibinfo  {publisher} {Oxford University Press},\ \bibinfo {address}
  {Oxford, New York},\ \bibinfo {year} {2013})\BibitemShut {NoStop}%
\bibitem [{\citenamefont {Zohar}\ and\ \citenamefont
  {Cirac}(2019)}]{Zohar2019RemovingStaggeredFermionic}%
  \BibitemOpen
  \bibfield  {author} {\bibinfo {author} {\bibfnamefont {E.}~\bibnamefont
  {Zohar}}\ and\ \bibinfo {author} {\bibfnamefont {J.~I.}\ \bibnamefont
  {Cirac}},\ }\bibfield  {title} {\bibinfo {title} {Removing staggered
  fermionic matter in ${U}({N})$ and ${SU}({N})$ lattice gauge theories},\
  }\href {https://doi.org/10.1103/PhysRevD.99.114511} {\bibfield  {journal}
  {\bibinfo  {journal} {Physical Review D}\ }\textbf {\bibinfo {volume} {99}},\
  \bibinfo {pages} {114511} (\bibinfo {year} {2019})}\BibitemShut {NoStop}%
\bibitem [{\citenamefont {Zohar}\ and\ \citenamefont
  {Cirac}(2018)}]{Zohar2018EliminatingFermionicMatter}%
  \BibitemOpen
  \bibfield  {author} {\bibinfo {author} {\bibfnamefont {E.}~\bibnamefont
  {Zohar}}\ and\ \bibinfo {author} {\bibfnamefont {J.~I.}\ \bibnamefont
  {Cirac}},\ }\bibfield  {title} {\bibinfo {title} {Eliminating fermionic
  matter fields in lattice gauge theories},\ }\href
  {https://doi.org/10.1103/PhysRevB.98.075119} {\bibfield  {journal} {\bibinfo
  {journal} {Physical Review B}\ }\textbf {\bibinfo {volume} {98}},\ \bibinfo
  {pages} {075119} (\bibinfo {year} {2018})}\BibitemShut {NoStop}%
\bibitem [{\citenamefont {Ballarin}\ \emph {et~al.}(2023)\citenamefont
  {Ballarin}, \citenamefont {Cataldi}, \citenamefont {Magnifico}, \citenamefont
  {Jaschke}, \citenamefont {Di~Liberto}, \citenamefont {Siloi}, \citenamefont
  {Montangero},\ and\ \citenamefont
  {Silvi}}]{Ballarin2023ScalableDigitalQuantum}%
  \BibitemOpen
  \bibfield  {author} {\bibinfo {author} {\bibfnamefont {M.}~\bibnamefont
  {Ballarin}}, \bibinfo {author} {\bibfnamefont {G.}~\bibnamefont {Cataldi}},
  \bibinfo {author} {\bibfnamefont {G.}~\bibnamefont {Magnifico}}, \bibinfo
  {author} {\bibfnamefont {D.}~\bibnamefont {Jaschke}}, \bibinfo {author}
  {\bibfnamefont {M.}~\bibnamefont {Di~Liberto}}, \bibinfo {author}
  {\bibfnamefont {I.}~\bibnamefont {Siloi}}, \bibinfo {author} {\bibfnamefont
  {S.}~\bibnamefont {Montangero}},\ and\ \bibinfo {author} {\bibfnamefont
  {P.}~\bibnamefont {Silvi}},\ }\href
  {https://doi.org/10.48550/arXiv.2310.15091} {\emph {\bibinfo {title}
  {Scalable digital quantum simulation of lattice fermion theories with local
  encoding}}},\ \bibinfo {type} {Tech. Rep.}\ (\bibinfo {year} {2023})\ \Eprint
  {https://arxiv.org/abs/2310.15091} {arxiv:2310.15091} \BibitemShut {NoStop}%
\bibitem [{\citenamefont {Verstraete}\ \emph {et~al.}(2008)\citenamefont
  {Verstraete}, \citenamefont {Murg},\ and\ \citenamefont
  {Cirac}}]{Verstraete2008MatrixProductStates}%
  \BibitemOpen
  \bibfield  {author} {\bibinfo {author} {\bibfnamefont {F.}~\bibnamefont
  {Verstraete}}, \bibinfo {author} {\bibfnamefont {V.}~\bibnamefont {Murg}},\
  and\ \bibinfo {author} {\bibfnamefont {J.~I.}\ \bibnamefont {Cirac}},\
  }\bibfield  {title} {\bibinfo {title} {Matrix product states, projected
  entangled pair states, and variational renormalization group methods for
  quantum spin systems},\ }\href {https://doi.org/10.1080/14789940801912366}
  {\bibfield  {journal} {\bibinfo  {journal} {Advances in Physics}\ }\textbf
  {\bibinfo {volume} {57}},\ \bibinfo {pages} {143} (\bibinfo {year}
  {2008})}\BibitemShut {NoStop}%
\bibitem [{\citenamefont {Ciavarella}\ \emph {et~al.}(2021)\citenamefont
  {Ciavarella}, \citenamefont {Klco},\ and\ \citenamefont
  {Savage}}]{Ciavarella2021TrailheadQuantumSimulation}%
  \BibitemOpen
  \bibfield  {author} {\bibinfo {author} {\bibfnamefont {A.}~\bibnamefont
  {Ciavarella}}, \bibinfo {author} {\bibfnamefont {N.}~\bibnamefont {Klco}},\
  and\ \bibinfo {author} {\bibfnamefont {M.~J.}\ \bibnamefont {Savage}},\
  }\bibfield  {title} {\bibinfo {title} {Trailhead for quantum simulation of
  su(3) yang-mills lattice gauge theory in the local multiplet basis},\ }\href
  {https://doi.org/10.1103/PhysRevD.103.094501} {\bibfield  {journal} {\bibinfo
   {journal} {Phys. Rev. D}\ }\textbf {\bibinfo {volume} {103}},\ \bibinfo
  {pages} {094501} (\bibinfo {year} {2021})}\BibitemShut {NoStop}%
\bibitem [{\citenamefont {Davoudi}\ \emph {et~al.}(2021)\citenamefont
  {Davoudi}, \citenamefont {Raychowdhury},\ and\ \citenamefont
  {Shaw}}]{Davoudi2021SearchEfficientFormulations}%
  \BibitemOpen
  \bibfield  {author} {\bibinfo {author} {\bibfnamefont {Z.}~\bibnamefont
  {Davoudi}}, \bibinfo {author} {\bibfnamefont {I.}~\bibnamefont
  {Raychowdhury}},\ and\ \bibinfo {author} {\bibfnamefont {A.}~\bibnamefont
  {Shaw}},\ }\bibfield  {title} {\bibinfo {title} {Search for efficient
  formulations for hamiltonian simulation of non-abelian lattice gauge
  theories},\ }\href {https://doi.org/10.1103/PhysRevD.104.074505} {\bibfield
  {journal} {\bibinfo  {journal} {Phys. Rev. D}\ }\textbf {\bibinfo {volume}
  {104}},\ \bibinfo {pages} {074505} (\bibinfo {year} {2021})}\BibitemShut
  {NoStop}%
\bibitem [{\citenamefont {Tong}\ \emph {et~al.}(2022)\citenamefont {Tong},
  \citenamefont {Albert}, \citenamefont {McClean}, \citenamefont {Preskill},\
  and\ \citenamefont {Su}}]{Tong2022ProvablyAccurateSimulation}%
  \BibitemOpen
  \bibfield  {author} {\bibinfo {author} {\bibfnamefont {Y.}~\bibnamefont
  {Tong}}, \bibinfo {author} {\bibfnamefont {V.~V.}\ \bibnamefont {Albert}},
  \bibinfo {author} {\bibfnamefont {J.~R.}\ \bibnamefont {McClean}}, \bibinfo
  {author} {\bibfnamefont {J.}~\bibnamefont {Preskill}},\ and\ \bibinfo
  {author} {\bibfnamefont {Y.}~\bibnamefont {Su}},\ }\bibfield  {title}
  {\bibinfo {title} {Provably accurate simulation of gauge theories and bosonic
  systems},\ }\href@noop {} {\bibfield  {journal} {\bibinfo  {journal}
  {Quantum}\ }\textbf {\bibinfo {volume} {6}},\ \bibinfo {pages} {816}
  (\bibinfo {year} {2022})}\BibitemShut {NoStop}%
\bibitem [{\citenamefont {Fiore}\ \emph {et~al.}(2005)\citenamefont {Fiore},
  \citenamefont {Giudice}, \citenamefont {Giuliano}, \citenamefont
  {Marmottini}, \citenamefont {Papa},\ and\ \citenamefont
  {Sodano}}]{Fiore2005Qed3SpaceTime}%
  \BibitemOpen
  \bibfield  {author} {\bibinfo {author} {\bibfnamefont {R.}~\bibnamefont
  {Fiore}}, \bibinfo {author} {\bibfnamefont {P.}~\bibnamefont {Giudice}},
  \bibinfo {author} {\bibfnamefont {D.}~\bibnamefont {Giuliano}}, \bibinfo
  {author} {\bibfnamefont {D.}~\bibnamefont {Marmottini}}, \bibinfo {author}
  {\bibfnamefont {A.}~\bibnamefont {Papa}},\ and\ \bibinfo {author}
  {\bibfnamefont {P.}~\bibnamefont {Sodano}},\ }\bibfield  {title} {\bibinfo
  {title} {{{QED}}\_3 on a space-time lattice: A comparison between compact and
  noncompact formulation},\ }in\ \href {https://doi.org/10.22323/1.020.0243}
  {\emph {\bibinfo {booktitle} {Proceedings of {{XXIIIrd International
  Symposium}} on {{Lattice Field Theory}} --- {{PoS}}({{LAT2005}})}}},\
  Vol.~\bibinfo {volume} {20}\ (\bibinfo  {publisher} {SISSA Medialab},\
  \bibinfo {year} {2005})\ p.\ \bibinfo {pages} {243}\BibitemShut {NoStop}%
\bibitem [{\citenamefont {Raviv}\ \emph {et~al.}(2014)\citenamefont {Raviv},
  \citenamefont {Shamir},\ and\ \citenamefont
  {Svetitsky}}]{Raviv2014NonperturbativeBetaFunction}%
  \BibitemOpen
  \bibfield  {author} {\bibinfo {author} {\bibfnamefont {O.}~\bibnamefont
  {Raviv}}, \bibinfo {author} {\bibfnamefont {Y.}~\bibnamefont {Shamir}},\ and\
  \bibinfo {author} {\bibfnamefont {B.}~\bibnamefont {Svetitsky}},\ }\bibfield
  {title} {\bibinfo {title} {Nonperturbative beta function in three-dimensional
  electrodynamics},\ }\href {https://doi.org/10.1103/PhysRevD.90.014512}
  {\bibfield  {journal} {\bibinfo  {journal} {Physical Review D}\ }\textbf
  {\bibinfo {volume} {90}},\ \bibinfo {pages} {014512} (\bibinfo {year}
  {2014})}\BibitemShut {NoStop}%
\bibitem [{\citenamefont {Svetitsky}\ \emph {et~al.}(2015)\citenamefont
  {Svetitsky}, \citenamefont {Raviv},\ and\ \citenamefont
  {Shamir}}]{Svetitsky2015BetaFunctionThree}%
  \BibitemOpen
  \bibfield  {author} {\bibinfo {author} {\bibfnamefont {B.}~\bibnamefont
  {Svetitsky}}, \bibinfo {author} {\bibfnamefont {O.}~\bibnamefont {Raviv}},\
  and\ \bibinfo {author} {\bibfnamefont {Y.}~\bibnamefont {Shamir}},\
  }\bibfield  {title} {\bibinfo {title} {Beta function of three-dimensional
  {{QED}}},\ }in\ \href {https://doi.org/10.22323/1.214.0051} {\emph {\bibinfo
  {booktitle} {Proceedings of {{The}} 32nd {{International Symposium}} on
  {{Lattice Field Theory}} --- {{PoS}}({{LATTICE2014}})}}},\ Vol.\ \bibinfo
  {volume} {214}\ (\bibinfo  {publisher} {SISSA Medialab},\ \bibinfo {year}
  {2015})\ p.\ \bibinfo {pages} {051}\BibitemShut {NoStop}%
\bibitem [{\citenamefont {Xu}\ \emph {et~al.}(2019)\citenamefont {Xu},
  \citenamefont {Qi}, \citenamefont {Zhang}, \citenamefont {Assaad},
  \citenamefont {Xu},\ and\ \citenamefont {Meng}}]{Xu2019MonteCarloStudy}%
  \BibitemOpen
  \bibfield  {author} {\bibinfo {author} {\bibfnamefont {X.~Y.}\ \bibnamefont
  {Xu}}, \bibinfo {author} {\bibfnamefont {Y.}~\bibnamefont {Qi}}, \bibinfo
  {author} {\bibfnamefont {L.}~\bibnamefont {Zhang}}, \bibinfo {author}
  {\bibfnamefont {F.~F.}\ \bibnamefont {Assaad}}, \bibinfo {author}
  {\bibfnamefont {C.}~\bibnamefont {Xu}},\ and\ \bibinfo {author}
  {\bibfnamefont {Z.~Y.}\ \bibnamefont {Meng}},\ }\bibfield  {title} {\bibinfo
  {title} {Monte {{Carlo Study}} of {{Lattice Compact Quantum Electrodynamics}}
  with {{Fermionic Matter}}: {{The Parent State}} of {{Quantum Phases}}},\
  }\href {https://doi.org/10.1103/PhysRevX.9.021022} {\bibfield  {journal}
  {\bibinfo  {journal} {Physical Review X}\ }\textbf {\bibinfo {volume} {9}},\
  \bibinfo {pages} {021022} (\bibinfo {year} {2019})}\BibitemShut {NoStop}%
\bibitem [{\citenamefont {Creutz}(1988)}]{Creutz1988LatticeGaugeTheory}%
  \BibitemOpen
  \bibfield  {author} {\bibinfo {author} {\bibfnamefont {M.}~\bibnamefont
  {Creutz}},\ }\href {https://doi.org/10.2172/6530895} {\emph {\bibinfo {title}
  {Lattice Gauge Theory and {{Monte Carlo}} Methods}}},\ \bibinfo {type} {Tech.
  Rep.}\ \bibinfo {number} {BNL-42086}\ (\bibinfo {year} {1988})\BibitemShut
  {NoStop}%
\bibitem [{\citenamefont {Creutz}(1989)}]{Creutz1989LatticeGaugeTheories}%
  \BibitemOpen
  \bibfield  {author} {\bibinfo {author} {\bibfnamefont {M.}~\bibnamefont
  {Creutz}},\ }\bibfield  {title} {\bibinfo {title} {Lattice gauge theories and
  {{Monte Carlo}} algorithms},\ }\href
  {https://doi.org/10.1016/0920-5632(89)90061-3} {\bibfield  {journal}
  {\bibinfo  {journal} {Nuclear Physics B - Proceedings Supplements}\ }\textbf
  {\bibinfo {volume} {10}},\ \bibinfo {pages} {1} (\bibinfo {year}
  {1989})}\BibitemShut {NoStop}%
\bibitem [{\citenamefont {Loan}\ \emph {et~al.}(2003)\citenamefont {Loan},
  \citenamefont {Brunner}, \citenamefont {Sloggett},\ and\ \citenamefont
  {Hamer}}]{Loan2003PathIntegralMonte}%
  \BibitemOpen
  \bibfield  {author} {\bibinfo {author} {\bibfnamefont {M.}~\bibnamefont
  {Loan}}, \bibinfo {author} {\bibfnamefont {M.}~\bibnamefont {Brunner}},
  \bibinfo {author} {\bibfnamefont {C.}~\bibnamefont {Sloggett}},\ and\
  \bibinfo {author} {\bibfnamefont {C.}~\bibnamefont {Hamer}},\ }\bibfield
  {title} {\bibinfo {title} {Path integral {{Monte Carlo}} approach to the
  {{U}}(1) lattice gauge theory in 2+1 dimensions},\ }\href
  {https://doi.org/10.1103/PhysRevD.68.034504} {\bibfield  {journal} {\bibinfo
  {journal} {Physical Review D}\ }\textbf {\bibinfo {volume} {68}},\ \bibinfo
  {pages} {034504} (\bibinfo {year} {2003})}\BibitemShut {NoStop}%
\bibitem [{\citenamefont {Rothe}(2012)}]{Rothe2012LatticeGaugeTheories}%
  \BibitemOpen
  \bibfield  {author} {\bibinfo {author} {\bibfnamefont {H.~J.}\ \bibnamefont
  {Rothe}},\ }\href {https://library.oapen.org/handle/20.500.12657/50492}
  {\emph {\bibinfo {title} {Lattice {{Gauge Theories}}: {{An Introduction}}
  ({{Fourth Edition}})}}}\ (\bibinfo  {publisher} {World Scientific Publishing
  Company},\ \bibinfo {year} {2012})\BibitemShut {NoStop}%
\bibitem [{\citenamefont {Funcke}\ \emph
  {et~al.}(2023{\natexlab{c}})\citenamefont {Funcke}, \citenamefont {Gro{\ss}},
  \citenamefont {Jansen}, \citenamefont {K{\"u}hn}, \citenamefont {Romiti},\
  and\ \citenamefont {Urbach}}]{Funcke2023HamiltonianLimitLattice}%
  \BibitemOpen
  \bibfield  {author} {\bibinfo {author} {\bibfnamefont {L.}~\bibnamefont
  {Funcke}}, \bibinfo {author} {\bibfnamefont {C.~F.}\ \bibnamefont
  {Gro{\ss}}}, \bibinfo {author} {\bibfnamefont {K.}~\bibnamefont {Jansen}},
  \bibinfo {author} {\bibfnamefont {S.}~\bibnamefont {K{\"u}hn}}, \bibinfo
  {author} {\bibfnamefont {S.}~\bibnamefont {Romiti}},\ and\ \bibinfo {author}
  {\bibfnamefont {C.}~\bibnamefont {Urbach}},\ }\bibfield  {title} {\bibinfo
  {title} {Hamiltonian limit of lattice {{QED}} in 2+1 dimensions},\ }in\ \href
  {https://doi.org/10.22323/1.430.0292} {\emph {\bibinfo {booktitle}
  {Proceedings of {{The}} 39th {{International Symposium}} on {{Lattice Field
  Theory}} --- {{PoS}}({{LATTICE2022}})}}},\ Vol.\ \bibinfo {volume} {430}\
  (\bibinfo  {publisher} {SISSA Medialab},\ \bibinfo {year} {2023})\ p.\
  \bibinfo {pages} {292}\BibitemShut {NoStop}%
\bibitem [{\citenamefont {Strouthos}\ and\ \citenamefont
  {Kogut}(2008)}]{Strouthos2008}%
  \BibitemOpen
  \bibfield  {author} {\bibinfo {author} {\bibfnamefont {C.}~\bibnamefont
  {Strouthos}}\ and\ \bibinfo {author} {\bibfnamefont {J.~B.}\ \bibnamefont
  {Kogut}},\ }\href {https://doi.org/10.48550/arXiv.0804.0300} {\bibinfo
  {title} {The {{Phases}} of {{Non-Compact QED}}(3)}} (\bibinfo {year}
  {2008}),\ \Eprint {https://arxiv.org/abs/0804.0300} {arxiv:0804.0300
  [cond-mat, physics:hep-lat, physics:hep-ph]} \BibitemShut {NoStop}%
\bibitem [{\citenamefont {Bender}(2023)}]{Bender2023QuantumClassicalMethods}%
  \BibitemOpen
  \bibfield  {author} {\bibinfo {author} {\bibfnamefont {J.}~\bibnamefont
  {Bender}},\ }\emph {\bibinfo {title} {Quantum and Classical Methods for
  Lattice Gauge Theories in Higher Dimensions}},\ \href
  {https://mediatum.ub.tum.de/?id=1709013} {Ph.D. thesis},\ \bibinfo  {school}
  {Technische Universit{\"a}t M{\"u}nchen} (\bibinfo {year} {2023})\BibitemShut
  {NoStop}%
\bibitem [{\citenamefont {Clemente}\ \emph {et~al.}(2022)\citenamefont
  {Clemente}, \citenamefont {Crippa},\ and\ \citenamefont
  {Jansen}}]{Clemente2022StrategiesDeterminationRunning}%
  \BibitemOpen
  \bibfield  {author} {\bibinfo {author} {\bibfnamefont {G.}~\bibnamefont
  {Clemente}}, \bibinfo {author} {\bibfnamefont {A.}~\bibnamefont {Crippa}},\
  and\ \bibinfo {author} {\bibfnamefont {K.}~\bibnamefont {Jansen}},\
  }\bibfield  {title} {\bibinfo {title} {Strategies for the determination of
  the running coupling of (\$2+1\$)-dimensional {{QED}} with quantum
  computing},\ }\href {https://doi.org/10.1103/PhysRevD.106.114511} {\bibfield
  {journal} {\bibinfo  {journal} {Physical Review D}\ }\textbf {\bibinfo
  {volume} {106}},\ \bibinfo {pages} {114511} (\bibinfo {year}
  {2022})}\BibitemShut {NoStop}%
\bibitem [{\citenamefont
  {Hern{\'{a}}ndez}(2011)}]{Hernandez2011LatticeFieldTheory}%
  \BibitemOpen
  \bibfield  {author} {\bibinfo {author} {\bibfnamefont {M.~P.}\ \bibnamefont
  {Hern{\'{a}}ndez}},\ }\bibfield  {title} {\bibinfo {title} {{Lattice} field
  theory fundamentals}\ }(\bibinfo  {publisher} {Oxford University Press},\
  \bibinfo {year} {2011})\ p.~\bibinfo {pages} {20}\BibitemShut {NoStop}%
\bibitem [{\citenamefont {Kang}\ \emph {et~al.}(2023)\citenamefont {Kang},
  \citenamefont {Soley}, \citenamefont {Crane}, \citenamefont {Girvin},\ and\
  \citenamefont {Wiebe}}]{kang2023leveraging}%
  \BibitemOpen
  \bibfield  {author} {\bibinfo {author} {\bibfnamefont {C.}~\bibnamefont
  {Kang}}, \bibinfo {author} {\bibfnamefont {M.~B.}\ \bibnamefont {Soley}},
  \bibinfo {author} {\bibfnamefont {E.}~\bibnamefont {Crane}}, \bibinfo
  {author} {\bibfnamefont {S.~M.}\ \bibnamefont {Girvin}},\ and\ \bibinfo
  {author} {\bibfnamefont {N.}~\bibnamefont {Wiebe}},\ }\href
  {https://doi.org/10.48550/arxiv.2303.15542} {\bibinfo {title} {Leveraging
  hamiltonian simulation techniques to compile operations on bosonic devices}}
  (\bibinfo {year} {2023}),\ \Eprint {https://arxiv.org/abs/2303.15542}
  {arXiv:2303.15542 [quant-ph]} \BibitemShut {NoStop}%
\bibitem [{\citenamefont {Workman}\ and\ \citenamefont
  {Others}(2022)}]{Workman2022}%
  \BibitemOpen
  \bibfield  {author} {\bibinfo {author} {\bibfnamefont {R.~L.}\ \bibnamefont
  {Workman}}\ and\ \bibinfo {author} {\bibnamefont {Others}} (\bibinfo
  {collaboration} {Particle Data Group}),\ }\bibfield  {title} {\bibinfo
  {title} {{Review of Particle Physics}},\ }\href
  {https://doi.org/10.1093/ptep/ptac097} {\bibfield  {journal} {\bibinfo
  {journal} {PTEP}\ }\textbf {\bibinfo {volume} {2022}},\ \bibinfo {pages}
  {083C01} (\bibinfo {year} {2022})}\BibitemShut {NoStop}%
\bibitem [{\citenamefont {Kelman}\ \emph {et~al.}(2024)\citenamefont {Kelman},
  \citenamefont {Borla}, \citenamefont {Gomelski}, \citenamefont {Elyovich},
  \citenamefont {Roose}, \citenamefont {Emonts},\ and\ \citenamefont
  {Zohar}}]{kelman2024gauged}%
  \BibitemOpen
  \bibfield  {author} {\bibinfo {author} {\bibfnamefont {A.}~\bibnamefont
  {Kelman}}, \bibinfo {author} {\bibfnamefont {U.}~\bibnamefont {Borla}},
  \bibinfo {author} {\bibfnamefont {I.}~\bibnamefont {Gomelski}}, \bibinfo
  {author} {\bibfnamefont {J.}~\bibnamefont {Elyovich}}, \bibinfo {author}
  {\bibfnamefont {G.}~\bibnamefont {Roose}}, \bibinfo {author} {\bibfnamefont
  {P.}~\bibnamefont {Emonts}},\ and\ \bibinfo {author} {\bibfnamefont
  {E.}~\bibnamefont {Zohar}},\ }\href@noop {} {\bibinfo {title} {Gauged
  gaussian peps -- a high dimensional tensor network formulation for lattice
  gauge theories}} (\bibinfo {year} {2024}),\ \Eprint
  {https://arxiv.org/abs/2404.13123} {arXiv:2404.13123 [hep-lat]} \BibitemShut
  {NoStop}%
\bibitem [{\citenamefont {Holstein}(1959)}]{Holstein1959StudiesPolaronMotion}%
  \BibitemOpen
  \bibfield  {author} {\bibinfo {author} {\bibfnamefont {T.}~\bibnamefont
  {Holstein}},\ }\bibfield  {title} {\bibinfo {title} {Studies of polaron
  motion: Part i. the molecular-crystal model},\ }\href
  {https://doi.org/10.1016/0003-4916(59)90002-8} {\bibfield  {journal}
  {\bibinfo  {journal} {Annals of Physics}\ }\textbf {\bibinfo {volume} {8}},\
  \bibinfo {pages} {325} (\bibinfo {year} {1959})}\BibitemShut {NoStop}%
\bibitem [{\citenamefont {Bloch}(2005)}]{Bloch2005UltracoldQuantumGases}%
  \BibitemOpen
  \bibfield  {author} {\bibinfo {author} {\bibfnamefont {I.}~\bibnamefont
  {Bloch}},\ }\bibfield  {title} {\bibinfo {title} {Ultracold quantum gases in
  optical lattices},\ }\href {https://doi.org/10.1038/nphys138} {\bibfield
  {journal} {\bibinfo  {journal} {Nature Physics}\ }\textbf {\bibinfo {volume}
  {1}},\ \bibinfo {pages} {23} (\bibinfo {year} {2005})}\BibitemShut {NoStop}%
\bibitem [{\citenamefont {Schlawin}\ \emph {et~al.}(2022)\citenamefont
  {Schlawin}, \citenamefont {Kennes},\ and\ \citenamefont
  {Sentef}}]{Schlawin2022CavityQuantumMaterials}%
  \BibitemOpen
  \bibfield  {author} {\bibinfo {author} {\bibfnamefont {F.}~\bibnamefont
  {Schlawin}}, \bibinfo {author} {\bibfnamefont {D.~M.}\ \bibnamefont
  {Kennes}},\ and\ \bibinfo {author} {\bibfnamefont {M.~A.}\ \bibnamefont
  {Sentef}},\ }\bibfield  {title} {\bibinfo {title} {Cavity quantum
  materials},\ }\href {https://doi.org/10.1063/5.0083825} {\bibfield  {journal}
  {\bibinfo  {journal} {Applied Physics Reviews}\ }\textbf {\bibinfo {volume}
  {9}},\ \bibinfo {pages} {011312} (\bibinfo {year} {2022})}\BibitemShut
  {NoStop}%
\bibitem [{\citenamefont {Stolpp}\ \emph {et~al.}(2021)\citenamefont {Stolpp},
  \citenamefont {K{\"{o}}hler}, \citenamefont {Manmana}, \citenamefont
  {Jeckelmann}, \citenamefont {Heidrich-Meisner},\ and\ \citenamefont
  {Paeckel}}]{Stolpp2021ComparativeStudyState}%
  \BibitemOpen
  \bibfield  {author} {\bibinfo {author} {\bibfnamefont {J.}~\bibnamefont
  {Stolpp}}, \bibinfo {author} {\bibfnamefont {T.}~\bibnamefont
  {K{\"{o}}hler}}, \bibinfo {author} {\bibfnamefont {S.~R.}\ \bibnamefont
  {Manmana}}, \bibinfo {author} {\bibfnamefont {E.}~\bibnamefont {Jeckelmann}},
  \bibinfo {author} {\bibfnamefont {F.}~\bibnamefont {Heidrich-Meisner}},\ and\
  \bibinfo {author} {\bibfnamefont {S.}~\bibnamefont {Paeckel}},\ }\bibfield
  {title} {\bibinfo {title} {Comparative study of state-of-the-art
  matrix-product-state methods for lattice models with large local hilbert
  spaces without u(1) symmetry},\ }\href
  {https://doi.org/10.1016/j.cpc.2021.108106} {\bibfield  {journal} {\bibinfo
  {journal} {Computer Physics Communications}\ }\textbf {\bibinfo {volume}
  {269}},\ \bibinfo {pages} {108106} (\bibinfo {year} {2021})}\BibitemShut
  {NoStop}%
\bibitem [{\citenamefont {Jeckelmann}\ and\ \citenamefont
  {White}(1998)}]{Jeckelmann1998DensityMatrixRenormalization}%
  \BibitemOpen
  \bibfield  {author} {\bibinfo {author} {\bibfnamefont {E.}~\bibnamefont
  {Jeckelmann}}\ and\ \bibinfo {author} {\bibfnamefont {S.~R.}\ \bibnamefont
  {White}},\ }\bibfield  {title} {\bibinfo {title} {Density-matrix
  renormalization-group study of the polaron problem in the holstein model},\
  }\href {https://doi.org/10.1103/PhysRevB.57.6376} {\bibfield  {journal}
  {\bibinfo  {journal} {Phys. Rev. B}\ }\textbf {\bibinfo {volume} {57}},\
  \bibinfo {pages} {6376} (\bibinfo {year} {1998})}\BibitemShut {NoStop}%
\bibitem [{\citenamefont {Guo}\ \emph {et~al.}(2012)\citenamefont {Guo},
  \citenamefont {Weichselbaum}, \citenamefont {von Delft},\ and\ \citenamefont
  {Vojta}}]{Guo2012CriticalStrongCoupling}%
  \BibitemOpen
  \bibfield  {author} {\bibinfo {author} {\bibfnamefont {C.}~\bibnamefont
  {Guo}}, \bibinfo {author} {\bibfnamefont {A.}~\bibnamefont {Weichselbaum}},
  \bibinfo {author} {\bibfnamefont {J.}~\bibnamefont {von Delft}},\ and\
  \bibinfo {author} {\bibfnamefont {M.}~\bibnamefont {Vojta}},\ }\bibfield
  {title} {\bibinfo {title} {Critical and strong-coupling phases in one- and
  two-bath spin-boson models},\ }\href
  {https://doi.org/10.1103/PhysRevLett.108.160401} {\bibfield  {journal}
  {\bibinfo  {journal} {Phys. Rev. Lett.}\ }\textbf {\bibinfo {volume} {108}},\
  \bibinfo {pages} {160401} (\bibinfo {year} {2012})}\BibitemShut {NoStop}%
\bibitem [{\citenamefont {Brockt}\ \emph {et~al.}(2015)\citenamefont {Brockt},
  \citenamefont {Dorfner}, \citenamefont {Vidmar}, \citenamefont
  {Heidrich-Meisner},\ and\ \citenamefont
  {Jeckelmann}}]{Brockt2015MatrixProductState}%
  \BibitemOpen
  \bibfield  {author} {\bibinfo {author} {\bibfnamefont {C.}~\bibnamefont
  {Brockt}}, \bibinfo {author} {\bibfnamefont {F.}~\bibnamefont {Dorfner}},
  \bibinfo {author} {\bibfnamefont {L.}~\bibnamefont {Vidmar}}, \bibinfo
  {author} {\bibfnamefont {F.}~\bibnamefont {Heidrich-Meisner}},\ and\ \bibinfo
  {author} {\bibfnamefont {E.}~\bibnamefont {Jeckelmann}},\ }\bibfield  {title}
  {\bibinfo {title} {Matrix-product-state method with a dynamical local basis
  optimization for bosonic systems out of equilibrium},\ }\href
  {https://doi.org/10.1103/PhysRevB.92.241106} {\bibfield  {journal} {\bibinfo
  {journal} {Phys. Rev. B}\ }\textbf {\bibinfo {volume} {92}},\ \bibinfo
  {pages} {241106} (\bibinfo {year} {2015})}\BibitemShut {NoStop}%
\bibitem [{\citenamefont {Stolpp}\ \emph {et~al.}(2020)\citenamefont {Stolpp},
  \citenamefont {Herbrych}, \citenamefont {Dorfner}, \citenamefont {Dagotto},\
  and\ \citenamefont {Heidrich-Meisner}}]{Stolpp2020ChargeDensityWave}%
  \BibitemOpen
  \bibfield  {author} {\bibinfo {author} {\bibfnamefont {J.}~\bibnamefont
  {Stolpp}}, \bibinfo {author} {\bibfnamefont {J.}~\bibnamefont {Herbrych}},
  \bibinfo {author} {\bibfnamefont {F.}~\bibnamefont {Dorfner}}, \bibinfo
  {author} {\bibfnamefont {E.}~\bibnamefont {Dagotto}},\ and\ \bibinfo {author}
  {\bibfnamefont {F.}~\bibnamefont {Heidrich-Meisner}},\ }\bibfield  {title}
  {\bibinfo {title} {Charge-density-wave melting in the one-dimensional
  holstein model},\ }\href {https://doi.org/10.1103/PhysRevB.101.035134}
  {\bibfield  {journal} {\bibinfo  {journal} {Phys. Rev. B}\ }\textbf {\bibinfo
  {volume} {101}},\ \bibinfo {pages} {035134} (\bibinfo {year}
  {2020})}\BibitemShut {NoStop}%
\bibitem [{\citenamefont {Zhang}\ \emph {et~al.}(1998)\citenamefont {Zhang},
  \citenamefont {Jeckelmann},\ and\ \citenamefont
  {White}}]{Zhang1998DensityMatrixApproach}%
  \BibitemOpen
  \bibfield  {author} {\bibinfo {author} {\bibfnamefont {C.}~\bibnamefont
  {Zhang}}, \bibinfo {author} {\bibfnamefont {E.}~\bibnamefont {Jeckelmann}},\
  and\ \bibinfo {author} {\bibfnamefont {S.~R.}\ \bibnamefont {White}},\
  }\bibfield  {title} {\bibinfo {title} {Density matrix approach to local
  hilbert space reduction},\ }\href
  {https://doi.org/10.1103/PhysRevLett.80.2661} {\bibfield  {journal} {\bibinfo
   {journal} {Phys. Rev. Lett.}\ }\textbf {\bibinfo {volume} {80}},\ \bibinfo
  {pages} {2661} (\bibinfo {year} {1998})}\BibitemShut {NoStop}%
\bibitem [{\citenamefont {Schr\"oder}\ and\ \citenamefont
  {Chin}(2016)}]{Schroeder2016SimulatingOpenQuantum}%
  \BibitemOpen
  \bibfield  {author} {\bibinfo {author} {\bibfnamefont {F.~A. Y.~N.}\
  \bibnamefont {Schr\"oder}}\ and\ \bibinfo {author} {\bibfnamefont {A.~W.}\
  \bibnamefont {Chin}},\ }\bibfield  {title} {\bibinfo {title} {Simulating open
  quantum dynamics with time-dependent variational matrix product states:
  Towards microscopic correlation of environment dynamics and reduced system
  evolution},\ }\href {https://doi.org/10.1103/PhysRevB.93.075105} {\bibfield
  {journal} {\bibinfo  {journal} {Phys. Rev. B}\ }\textbf {\bibinfo {volume}
  {93}},\ \bibinfo {pages} {075105} (\bibinfo {year} {2016})}\BibitemShut
  {NoStop}%
\bibitem [{\citenamefont {Mathur}\ and\ \citenamefont
  {Rathor}(2023)}]{mathur2023exact}%
  \BibitemOpen
  \bibfield  {author} {\bibinfo {author} {\bibfnamefont {M.}~\bibnamefont
  {Mathur}}\ and\ \bibinfo {author} {\bibfnamefont {A.}~\bibnamefont
  {Rathor}},\ }\href@noop {} {\bibinfo {title} {Exact duality and local
  dynamics in su(n) lattice gauge theory}} (\bibinfo {year} {2023}),\ \Eprint
  {https://arxiv.org/abs/2109.00992} {arXiv:2109.00992 [hep-lat]} \BibitemShut
  {NoStop}%
\bibitem [{\citenamefont {Bauer}\ \emph
  {et~al.}(2023{\natexlab{b}})\citenamefont {Bauer}, \citenamefont {D'Andrea},
  \citenamefont {Freytsis},\ and\ \citenamefont {Grabowska}}]{bauer2023new}%
  \BibitemOpen
  \bibfield  {author} {\bibinfo {author} {\bibfnamefont {C.~W.}\ \bibnamefont
  {Bauer}}, \bibinfo {author} {\bibfnamefont {I.}~\bibnamefont {D'Andrea}},
  \bibinfo {author} {\bibfnamefont {M.}~\bibnamefont {Freytsis}},\ and\
  \bibinfo {author} {\bibfnamefont {D.~M.}\ \bibnamefont {Grabowska}},\
  }\href@noop {} {\bibinfo {title} {A new basis for hamiltonian su(2)
  simulations}} (\bibinfo {year} {2023}{\natexlab{b}}),\ \Eprint
  {https://arxiv.org/abs/2307.11829} {arXiv:2307.11829 [hep-ph]} \BibitemShut
  {NoStop}%
\bibitem [{\citenamefont {Kessler}\ \emph {et~al.}(2021)\citenamefont
  {Kessler}, \citenamefont {Calcavecchia},\ and\ \citenamefont
  {K{\"u}hne}}]{Kessler2021ArtificialNeuralNetworks}%
  \BibitemOpen
  \bibfield  {author} {\bibinfo {author} {\bibfnamefont {J.}~\bibnamefont
  {Kessler}}, \bibinfo {author} {\bibfnamefont {F.}~\bibnamefont
  {Calcavecchia}},\ and\ \bibinfo {author} {\bibfnamefont {T.~D.}\ \bibnamefont
  {K{\"u}hne}},\ }\bibfield  {title} {\bibinfo {title} {Artificial {{Neural
  Networks}} as {{Trial Wave Functions}} for {{Quantum Monte Carlo}}},\ }\href
  {https://doi.org/10.1002/adts.202000269} {\bibfield  {journal} {\bibinfo
  {journal} {Advanced Theory and Simulations}\ }\textbf {\bibinfo {volume}
  {4}},\ \bibinfo {pages} {2000269} (\bibinfo {year} {2021})}\BibitemShut
  {NoStop}%
\bibitem [{\citenamefont {Schr{\"o}der}\ \emph {et~al.}(2019)\citenamefont
  {Schr{\"o}der}, \citenamefont {Turban}, \citenamefont {Musser}, \citenamefont
  {Hine},\ and\ \citenamefont {Chin}}]{Schroeder2019TensorNetworkSimulation}%
  \BibitemOpen
  \bibfield  {author} {\bibinfo {author} {\bibfnamefont {F.~A. Y.~N.}\
  \bibnamefont {Schr{\"o}der}}, \bibinfo {author} {\bibfnamefont {D.~H.~P.}\
  \bibnamefont {Turban}}, \bibinfo {author} {\bibfnamefont {A.~J.}\
  \bibnamefont {Musser}}, \bibinfo {author} {\bibfnamefont {N.~D.~M.}\
  \bibnamefont {Hine}},\ and\ \bibinfo {author} {\bibfnamefont {A.~W.}\
  \bibnamefont {Chin}},\ }\bibfield  {title} {\bibinfo {title} {Tensor network
  simulation of multi-environmental open quantum dynamics via machine learning
  and entanglement renormalisation},\ }\href
  {https://doi.org/10.1038/s41467-019-09039-7} {\bibfield  {journal} {\bibinfo
  {journal} {Nature Communications}\ }\textbf {\bibinfo {volume} {10}},\
  \bibinfo {pages} {1062} (\bibinfo {year} {2019})}\BibitemShut {NoStop}%
\bibitem [{\citenamefont
  {Arnoldi}(1951)}]{Arnoldi1951PrincipleMinimizedIterations}%
  \BibitemOpen
  \bibfield  {author} {\bibinfo {author} {\bibfnamefont {W.~E.}\ \bibnamefont
  {Arnoldi}},\ }\bibfield  {title} {\bibinfo {title} {{The principle of
  minimized iterations in the solution of the matrix eigenvalue problem}},\
  }\href {https://doi.org/10.1090/qam/42792} {\bibfield  {journal} {\bibinfo
  {journal} {{Quarterly of Applied Mathematics}}\ }\textbf {\bibinfo {volume}
  {9}},\ \bibinfo {pages} {17} (\bibinfo {year} {1951})}\BibitemShut {NoStop}%
\bibitem [{\citenamefont {Lehoucq}\ \emph {et~al.}(1997)\citenamefont
  {Lehoucq}, \citenamefont {Sorensen},\ and\ \citenamefont
  {Yang}}]{Lehoucq1997ArpackSolutionLarge}%
  \BibitemOpen
  \bibfield  {author} {\bibinfo {author} {\bibfnamefont {R.~B.}\ \bibnamefont
  {Lehoucq}}, \bibinfo {author} {\bibfnamefont {D.~C.}\ \bibnamefont
  {Sorensen}},\ and\ \bibinfo {author} {\bibfnamefont {C.}~\bibnamefont
  {Yang}},\ }\href@noop {} {\emph {\bibinfo {title} {ARPACK: Solution of Large
  Scale Eigenvalue Problems by Implicitly Restarted Arnoldi Methods}}},\
  \bibinfo {address} {Available from netlib@ornl.gov} (\bibinfo {year}
  {1997})\BibitemShut {NoStop}%
\bibitem [{\citenamefont {Dagum}\ and\ \citenamefont
  {Menon}(1998)}]{Dagum1998OpenmpIndustryStandard}%
  \BibitemOpen
  \bibfield  {author} {\bibinfo {author} {\bibfnamefont {L.}~\bibnamefont
  {Dagum}}\ and\ \bibinfo {author} {\bibfnamefont {R.}~\bibnamefont {Menon}},\
  }\bibfield  {title} {\bibinfo {title} {Openmp: an industry standard api for
  shared-memory programming},\ }\href@noop {} {\bibfield  {journal} {\bibinfo
  {journal} {Computational Science \& Engineering, IEEE}\ }\textbf {\bibinfo
  {volume} {5}},\ \bibinfo {pages} {46} (\bibinfo {year} {1998})}\BibitemShut
  {NoStop}%
\bibitem [{\citenamefont {Shi}\ \emph {et~al.}(2016)\citenamefont {Shi},
  \citenamefont {Niranjan}, \citenamefont {Anandkumar},\ and\ \citenamefont
  {Cecka}}]{Shi2016TensorContractionsExtended}%
  \BibitemOpen
  \bibfield  {author} {\bibinfo {author} {\bibfnamefont {Y.}~\bibnamefont
  {Shi}}, \bibinfo {author} {\bibfnamefont {U.~N.}\ \bibnamefont {Niranjan}},
  \bibinfo {author} {\bibfnamefont {A.}~\bibnamefont {Anandkumar}},\ and\
  \bibinfo {author} {\bibfnamefont {C.}~\bibnamefont {Cecka}},\ }\bibfield
  {title} {\bibinfo {title} {Tensor {{Contractions}} with {{Extended BLAS
  Kernels}} on {{CPU}} and {{GPU}}},\ }in\ \href
  {https://doi.org/10.1109/hipc.2016.031} {\emph {\bibinfo {booktitle} {2016
  {IEEE} 23rd International Conference on High Performance Computing
  ({HiPC})}}}\ (\bibinfo  {publisher} {{IEEE}},\ \bibinfo {year} {2016})\ pp.\
  \bibinfo {pages} {193--202}\BibitemShut {NoStop}%
\bibitem [{\citenamefont {Abdelfattah}\ \emph {et~al.}(2016)\citenamefont
  {Abdelfattah}, \citenamefont {Baboulin}, \citenamefont {Dobrev},
  \citenamefont {Dongarra}, \citenamefont {Earl}, \citenamefont {Falcou},
  \citenamefont {Haidar}, \citenamefont {Karlin}, \citenamefont {Kolev},
  \citenamefont {Masliah},\ and\ \citenamefont
  {Tomov}}]{Abdelfattah2016HighPerformanceTensor}%
  \BibitemOpen
  \bibfield  {author} {\bibinfo {author} {\bibfnamefont {A.}~\bibnamefont
  {Abdelfattah}}, \bibinfo {author} {\bibfnamefont {M.}~\bibnamefont
  {Baboulin}}, \bibinfo {author} {\bibfnamefont {V.}~\bibnamefont {Dobrev}},
  \bibinfo {author} {\bibfnamefont {J.}~\bibnamefont {Dongarra}}, \bibinfo
  {author} {\bibfnamefont {C.}~\bibnamefont {Earl}}, \bibinfo {author}
  {\bibfnamefont {J.}~\bibnamefont {Falcou}}, \bibinfo {author} {\bibfnamefont
  {A.}~\bibnamefont {Haidar}}, \bibinfo {author} {\bibfnamefont
  {I.}~\bibnamefont {Karlin}}, \bibinfo {author} {\bibfnamefont {{\relax
  Tz}.}~\bibnamefont {Kolev}}, \bibinfo {author} {\bibfnamefont
  {I.}~\bibnamefont {Masliah}},\ and\ \bibinfo {author} {\bibfnamefont
  {S.}~\bibnamefont {Tomov}},\ }\bibfield  {title} {\bibinfo {title}
  {High-performance {{Tensor Contractions}} for {{GPUs}}},\ }\href
  {https://doi.org/10.1016/j.procs.2016.05.302} {\bibfield  {journal} {\bibinfo
   {journal} {Procedia Computer Science}\ }\bibinfo {series} {International
  {{Conference}} on {{Computational Science}} 2016, {{ICCS}} 2016, 6-8 {{June}}
  2016, {{San Diego}}, {{California}}, {{USA}}},\ \textbf {\bibinfo {volume}
  {80}},\ \bibinfo {pages} {108} (\bibinfo {year} {2016})}\BibitemShut
  {NoStop}%
\bibitem [{\citenamefont {Vincent}\ \emph {et~al.}(2022)\citenamefont
  {Vincent}, \citenamefont {O'Riordan}, \citenamefont {Andrenkov},
  \citenamefont {Brown}, \citenamefont {Killoran}, \citenamefont {Qi},\ and\
  \citenamefont {Dhand}}]{Vincent2022JetFastQuantum}%
  \BibitemOpen
  \bibfield  {author} {\bibinfo {author} {\bibfnamefont {T.}~\bibnamefont
  {Vincent}}, \bibinfo {author} {\bibfnamefont {L.~J.}\ \bibnamefont
  {O'Riordan}}, \bibinfo {author} {\bibfnamefont {M.}~\bibnamefont
  {Andrenkov}}, \bibinfo {author} {\bibfnamefont {J.}~\bibnamefont {Brown}},
  \bibinfo {author} {\bibfnamefont {N.}~\bibnamefont {Killoran}}, \bibinfo
  {author} {\bibfnamefont {H.}~\bibnamefont {Qi}},\ and\ \bibinfo {author}
  {\bibfnamefont {I.}~\bibnamefont {Dhand}},\ }\bibfield  {title} {\bibinfo
  {title} {Jet: {F}ast quantum circuit simulations with parallel task-based
  tensor-network contraction},\ }\href
  {https://doi.org/10.22331/q-2022-05-09-709} {\bibfield  {journal} {\bibinfo
  {journal} {{Quantum}}\ }\textbf {\bibinfo {volume} {6}},\ \bibinfo {pages}
  {709} (\bibinfo {year} {2022})}\BibitemShut {NoStop}%
\bibitem [{\citenamefont {Pan}\ and\ \citenamefont
  {Zhang}(2022)}]{Pan2022SimulationQuantumCircuits}%
  \BibitemOpen
  \bibfield  {author} {\bibinfo {author} {\bibfnamefont {F.}~\bibnamefont
  {Pan}}\ and\ \bibinfo {author} {\bibfnamefont {P.}~\bibnamefont {Zhang}},\
  }\bibfield  {title} {\bibinfo {title} {Simulation of quantum circuits using
  the big-batch tensor network method},\ }\href
  {https://doi.org/10.1103/PhysRevLett.128.030501} {\bibfield  {journal}
  {\bibinfo  {journal} {Phys. Rev. Lett.}\ }\textbf {\bibinfo {volume} {128}},\
  \bibinfo {pages} {030501} (\bibinfo {year} {2022})}\BibitemShut {NoStop}%
\bibitem [{\citenamefont {Jouppi}\ \emph {et~al.}(2017)\citenamefont {Jouppi},
  \citenamefont {Young}, \citenamefont {Patil}, \citenamefont {Patterson},
  \citenamefont {Agrawal}, \citenamefont {Bajwa}, \citenamefont {Bates},
  \citenamefont {Bhatia}, \citenamefont {Boden}, \citenamefont {Borchers},
  \citenamefont {Boyle}, \citenamefont {Cantin}, \citenamefont {Chao},
  \citenamefont {Clark}, \citenamefont {Coriell}, \citenamefont {Daley},
  \citenamefont {Dau}, \citenamefont {Dean}, \citenamefont {Gelb},
  \citenamefont {Ghaemmaghami}, \citenamefont {Gottipati}, \citenamefont
  {Gulland}, \citenamefont {Hagmann}, \citenamefont {Ho}, \citenamefont
  {Hogberg}, \citenamefont {Hu}, \citenamefont {Hundt}, \citenamefont {Hurt},
  \citenamefont {Ibarz}, \citenamefont {Jaffey}, \citenamefont {Jaworski},
  \citenamefont {Kaplan}, \citenamefont {Khaitan}, \citenamefont {Koch},
  \citenamefont {Kumar}, \citenamefont {Lacy}, \citenamefont {Laudon},
  \citenamefont {Law}, \citenamefont {Le}, \citenamefont {Leary}, \citenamefont
  {Liu}, \citenamefont {Lucke}, \citenamefont {Lundin}, \citenamefont
  {MacKean}, \citenamefont {Maggiore}, \citenamefont {Mahony}, \citenamefont
  {Miller}, \citenamefont {Nagarajan}, \citenamefont {Narayanaswami},
  \citenamefont {Ni}, \citenamefont {Nix}, \citenamefont {Norrie},
  \citenamefont {Omernick}, \citenamefont {Penukonda}, \citenamefont {Phelps},
  \citenamefont {Ross}, \citenamefont {Ross}, \citenamefont {Salek},
  \citenamefont {Samadiani}, \citenamefont {Severn}, \citenamefont {Sizikov},
  \citenamefont {Snelham}, \citenamefont {Souter}, \citenamefont {Steinberg},
  \citenamefont {Swing}, \citenamefont {Tan}, \citenamefont {Thorson},
  \citenamefont {Tian}, \citenamefont {Toma}, \citenamefont {Tuttle},
  \citenamefont {Vasudevan}, \citenamefont {Walter}, \citenamefont {Wang},
  \citenamefont {Wilcox},\ and\ \citenamefont {Yoon}}]{jouppi2017indatacenter}%
  \BibitemOpen
  \bibfield  {author} {\bibinfo {author} {\bibfnamefont {N.~P.}\ \bibnamefont
  {Jouppi}}, \bibinfo {author} {\bibfnamefont {C.}~\bibnamefont {Young}},
  \bibinfo {author} {\bibfnamefont {N.}~\bibnamefont {Patil}}, \bibinfo
  {author} {\bibfnamefont {D.}~\bibnamefont {Patterson}}, \bibinfo {author}
  {\bibfnamefont {G.}~\bibnamefont {Agrawal}}, \bibinfo {author} {\bibfnamefont
  {R.}~\bibnamefont {Bajwa}}, \bibinfo {author} {\bibfnamefont
  {S.}~\bibnamefont {Bates}}, \bibinfo {author} {\bibfnamefont
  {S.}~\bibnamefont {Bhatia}}, \bibinfo {author} {\bibfnamefont
  {N.}~\bibnamefont {Boden}}, \bibinfo {author} {\bibfnamefont
  {A.}~\bibnamefont {Borchers}}, \bibinfo {author} {\bibfnamefont
  {R.}~\bibnamefont {Boyle}}, \bibinfo {author} {\bibfnamefont {P.-l.}\
  \bibnamefont {Cantin}}, \bibinfo {author} {\bibfnamefont {C.}~\bibnamefont
  {Chao}}, \bibinfo {author} {\bibfnamefont {C.}~\bibnamefont {Clark}},
  \bibinfo {author} {\bibfnamefont {J.}~\bibnamefont {Coriell}}, \bibinfo
  {author} {\bibfnamefont {M.}~\bibnamefont {Daley}}, \bibinfo {author}
  {\bibfnamefont {M.}~\bibnamefont {Dau}}, \bibinfo {author} {\bibfnamefont
  {J.}~\bibnamefont {Dean}}, \bibinfo {author} {\bibfnamefont {B.}~\bibnamefont
  {Gelb}}, \bibinfo {author} {\bibfnamefont {T.~V.}\ \bibnamefont
  {Ghaemmaghami}}, \bibinfo {author} {\bibfnamefont {R.}~\bibnamefont
  {Gottipati}}, \bibinfo {author} {\bibfnamefont {W.}~\bibnamefont {Gulland}},
  \bibinfo {author} {\bibfnamefont {R.}~\bibnamefont {Hagmann}}, \bibinfo
  {author} {\bibfnamefont {C.~R.}\ \bibnamefont {Ho}}, \bibinfo {author}
  {\bibfnamefont {D.}~\bibnamefont {Hogberg}}, \bibinfo {author} {\bibfnamefont
  {J.}~\bibnamefont {Hu}}, \bibinfo {author} {\bibfnamefont {R.}~\bibnamefont
  {Hundt}}, \bibinfo {author} {\bibfnamefont {D.}~\bibnamefont {Hurt}},
  \bibinfo {author} {\bibfnamefont {J.}~\bibnamefont {Ibarz}}, \bibinfo
  {author} {\bibfnamefont {A.}~\bibnamefont {Jaffey}}, \bibinfo {author}
  {\bibfnamefont {A.}~\bibnamefont {Jaworski}}, \bibinfo {author}
  {\bibfnamefont {A.}~\bibnamefont {Kaplan}}, \bibinfo {author} {\bibfnamefont
  {H.}~\bibnamefont {Khaitan}}, \bibinfo {author} {\bibfnamefont
  {A.}~\bibnamefont {Koch}}, \bibinfo {author} {\bibfnamefont {N.}~\bibnamefont
  {Kumar}}, \bibinfo {author} {\bibfnamefont {S.}~\bibnamefont {Lacy}},
  \bibinfo {author} {\bibfnamefont {J.}~\bibnamefont {Laudon}}, \bibinfo
  {author} {\bibfnamefont {J.}~\bibnamefont {Law}}, \bibinfo {author}
  {\bibfnamefont {D.}~\bibnamefont {Le}}, \bibinfo {author} {\bibfnamefont
  {C.}~\bibnamefont {Leary}}, \bibinfo {author} {\bibfnamefont
  {Z.}~\bibnamefont {Liu}}, \bibinfo {author} {\bibfnamefont {K.}~\bibnamefont
  {Lucke}}, \bibinfo {author} {\bibfnamefont {A.}~\bibnamefont {Lundin}},
  \bibinfo {author} {\bibfnamefont {G.}~\bibnamefont {MacKean}}, \bibinfo
  {author} {\bibfnamefont {A.}~\bibnamefont {Maggiore}}, \bibinfo {author}
  {\bibfnamefont {M.}~\bibnamefont {Mahony}}, \bibinfo {author} {\bibfnamefont
  {K.}~\bibnamefont {Miller}}, \bibinfo {author} {\bibfnamefont
  {R.}~\bibnamefont {Nagarajan}}, \bibinfo {author} {\bibfnamefont
  {R.}~\bibnamefont {Narayanaswami}}, \bibinfo {author} {\bibfnamefont
  {R.}~\bibnamefont {Ni}}, \bibinfo {author} {\bibfnamefont {K.}~\bibnamefont
  {Nix}}, \bibinfo {author} {\bibfnamefont {T.}~\bibnamefont {Norrie}},
  \bibinfo {author} {\bibfnamefont {M.}~\bibnamefont {Omernick}}, \bibinfo
  {author} {\bibfnamefont {N.}~\bibnamefont {Penukonda}}, \bibinfo {author}
  {\bibfnamefont {A.}~\bibnamefont {Phelps}}, \bibinfo {author} {\bibfnamefont
  {J.}~\bibnamefont {Ross}}, \bibinfo {author} {\bibfnamefont {M.}~\bibnamefont
  {Ross}}, \bibinfo {author} {\bibfnamefont {A.}~\bibnamefont {Salek}},
  \bibinfo {author} {\bibfnamefont {E.}~\bibnamefont {Samadiani}}, \bibinfo
  {author} {\bibfnamefont {C.}~\bibnamefont {Severn}}, \bibinfo {author}
  {\bibfnamefont {G.}~\bibnamefont {Sizikov}}, \bibinfo {author} {\bibfnamefont
  {M.}~\bibnamefont {Snelham}}, \bibinfo {author} {\bibfnamefont
  {J.}~\bibnamefont {Souter}}, \bibinfo {author} {\bibfnamefont
  {D.}~\bibnamefont {Steinberg}}, \bibinfo {author} {\bibfnamefont
  {A.}~\bibnamefont {Swing}}, \bibinfo {author} {\bibfnamefont
  {M.}~\bibnamefont {Tan}}, \bibinfo {author} {\bibfnamefont {G.}~\bibnamefont
  {Thorson}}, \bibinfo {author} {\bibfnamefont {B.}~\bibnamefont {Tian}},
  \bibinfo {author} {\bibfnamefont {H.}~\bibnamefont {Toma}}, \bibinfo {author}
  {\bibfnamefont {E.}~\bibnamefont {Tuttle}}, \bibinfo {author} {\bibfnamefont
  {V.}~\bibnamefont {Vasudevan}}, \bibinfo {author} {\bibfnamefont
  {R.}~\bibnamefont {Walter}}, \bibinfo {author} {\bibfnamefont
  {W.}~\bibnamefont {Wang}}, \bibinfo {author} {\bibfnamefont {E.}~\bibnamefont
  {Wilcox}},\ and\ \bibinfo {author} {\bibfnamefont {D.~H.}\ \bibnamefont
  {Yoon}},\ }\href {https://doi.org/10.48550/arXiv.1704.04760} {\bibinfo
  {title} {In-{{Datacenter Performance Analysis}} of a {{Tensor Processing
  Unit}}}} (\bibinfo {year} {2017}),\ \Eprint
  {https://arxiv.org/abs/1704.04760} {arXiv:1704.04760 [cs.AR]} \BibitemShut
  {NoStop}%
\bibitem [{\citenamefont {Hauru}\ \emph {et~al.}(2021)\citenamefont {Hauru},
  \citenamefont {Morningstar}, \citenamefont {Beall}, \citenamefont {Ganahl},
  \citenamefont {Lewis},\ and\ \citenamefont {Vidal}}]{hauru2021simulation}%
  \BibitemOpen
  \bibfield  {author} {\bibinfo {author} {\bibfnamefont {M.}~\bibnamefont
  {Hauru}}, \bibinfo {author} {\bibfnamefont {A.}~\bibnamefont {Morningstar}},
  \bibinfo {author} {\bibfnamefont {J.}~\bibnamefont {Beall}}, \bibinfo
  {author} {\bibfnamefont {M.}~\bibnamefont {Ganahl}}, \bibinfo {author}
  {\bibfnamefont {A.}~\bibnamefont {Lewis}},\ and\ \bibinfo {author}
  {\bibfnamefont {G.}~\bibnamefont {Vidal}},\ }\href
  {https://doi.org/10.48550/arXiv.2111.10466} {\bibinfo {title} {Simulation of
  quantum physics with tensor processing units: brute-force computation of
  ground states and time evolution}} (\bibinfo {year} {2021}),\ \Eprint
  {https://arxiv.org/abs/2111.10466} {arXiv:2111.10466 [quant-ph]} \BibitemShut
  {NoStop}%
\bibitem [{\citenamefont {Morningstar}\ \emph {et~al.}(2022)\citenamefont
  {Morningstar}, \citenamefont {Hauru}, \citenamefont {Beall}, \citenamefont
  {Ganahl}, \citenamefont {Lewis}, \citenamefont {Khemani},\ and\ \citenamefont
  {Vidal}}]{Morningstar2022SimulationQuantumMany}%
  \BibitemOpen
  \bibfield  {author} {\bibinfo {author} {\bibfnamefont {A.}~\bibnamefont
  {Morningstar}}, \bibinfo {author} {\bibfnamefont {M.}~\bibnamefont {Hauru}},
  \bibinfo {author} {\bibfnamefont {J.}~\bibnamefont {Beall}}, \bibinfo
  {author} {\bibfnamefont {M.}~\bibnamefont {Ganahl}}, \bibinfo {author}
  {\bibfnamefont {A.~G.~M.}\ \bibnamefont {Lewis}}, \bibinfo {author}
  {\bibfnamefont {V.}~\bibnamefont {Khemani}},\ and\ \bibinfo {author}
  {\bibfnamefont {G.}~\bibnamefont {Vidal}},\ }\bibfield  {title} {\bibinfo
  {title} {Simulation of {{Quantum Many-Body Dynamics}} with {{Tensor
  Processing Units}}: {{Floquet Prethermalization}}},\ }\href
  {https://doi.org/10.1103/PRXQuantum.3.020331} {\bibfield  {journal} {\bibinfo
   {journal} {PRX Quantum}\ }\textbf {\bibinfo {volume} {3}},\ \bibinfo {pages}
  {020331} (\bibinfo {year} {2022})}\BibitemShut {NoStop}%
\bibitem [{\citenamefont {Ganahl}\ \emph {et~al.}(2023)\citenamefont {Ganahl},
  \citenamefont {Beall}, \citenamefont {Hauru}, \citenamefont {Lewis},
  \citenamefont {Wojno}, \citenamefont {Yoo}, \citenamefont {Zou},\ and\
  \citenamefont {Vidal}}]{Ganahl2023DensityMatrixRenormalization}%
  \BibitemOpen
  \bibfield  {author} {\bibinfo {author} {\bibfnamefont {M.}~\bibnamefont
  {Ganahl}}, \bibinfo {author} {\bibfnamefont {J.}~\bibnamefont {Beall}},
  \bibinfo {author} {\bibfnamefont {M.}~\bibnamefont {Hauru}}, \bibinfo
  {author} {\bibfnamefont {A.~G.~M.}\ \bibnamefont {Lewis}}, \bibinfo {author}
  {\bibfnamefont {T.}~\bibnamefont {Wojno}}, \bibinfo {author} {\bibfnamefont
  {J.~H.}\ \bibnamefont {Yoo}}, \bibinfo {author} {\bibfnamefont
  {Y.}~\bibnamefont {Zou}},\ and\ \bibinfo {author} {\bibfnamefont
  {G.}~\bibnamefont {Vidal}},\ }\bibfield  {title} {\bibinfo {title} {Density
  {{Matrix Renormalization Group}} with {{Tensor Processing Units}}},\ }\href
  {https://doi.org/10.1103/PRXQuantum.4.010317} {\bibfield  {journal} {\bibinfo
   {journal} {PRX Quantum}\ }\textbf {\bibinfo {volume} {4}},\ \bibinfo {pages}
  {010317} (\bibinfo {year} {2023})}\BibitemShut {NoStop}%
\bibitem [{\citenamefont {Blackford}\ \emph {et~al.}(1997)\citenamefont
  {Blackford}, \citenamefont {Choi}, \citenamefont {Cleary}, \citenamefont
  {D'Azevedo}, \citenamefont {Demmel}, \citenamefont {Dhillon}, \citenamefont
  {Dongarra}, \citenamefont {Hammarling}, \citenamefont {Henry}, \citenamefont
  {Petitet}, \citenamefont {Stanley}, \citenamefont {Walker},\ and\
  \citenamefont {Whaley}}]{Blackford1997ScalapackUsersGuide}%
  \BibitemOpen
  \bibfield  {author} {\bibinfo {author} {\bibfnamefont {L.~S.}\ \bibnamefont
  {Blackford}}, \bibinfo {author} {\bibfnamefont {J.}~\bibnamefont {Choi}},
  \bibinfo {author} {\bibfnamefont {A.}~\bibnamefont {Cleary}}, \bibinfo
  {author} {\bibfnamefont {E.}~\bibnamefont {D'Azevedo}}, \bibinfo {author}
  {\bibfnamefont {J.}~\bibnamefont {Demmel}}, \bibinfo {author} {\bibfnamefont
  {I.}~\bibnamefont {Dhillon}}, \bibinfo {author} {\bibfnamefont
  {J.}~\bibnamefont {Dongarra}}, \bibinfo {author} {\bibfnamefont
  {S.}~\bibnamefont {Hammarling}}, \bibinfo {author} {\bibfnamefont
  {G.}~\bibnamefont {Henry}}, \bibinfo {author} {\bibfnamefont
  {A.}~\bibnamefont {Petitet}}, \bibinfo {author} {\bibfnamefont
  {K.}~\bibnamefont {Stanley}}, \bibinfo {author} {\bibfnamefont
  {D.}~\bibnamefont {Walker}},\ and\ \bibinfo {author} {\bibfnamefont {R.~C.}\
  \bibnamefont {Whaley}},\ }\href {https://doi.org/10.1137/1.9780898719642}
  {\emph {\bibinfo {title} {{{ScaLAPACK Users}}' {{Guide}}}}},\ Software,
  {{Environments}}, and {{Tools}}\ (\bibinfo  {publisher} {Society for
  Industrial and Applied Mathematics},\ \bibinfo {address} {Philadelphia, PA},\
  \bibinfo {year} {1997})\BibitemShut {NoStop}%
\bibitem [{\citenamefont {Lu}\ \emph {et~al.}(2017)\citenamefont {Lu},
  \citenamefont {Ino},\ and\ \citenamefont
  {Matsushita}}]{Lu2017HighPerformance}%
  \BibitemOpen
  \bibfield  {author} {\bibinfo {author} {\bibfnamefont {Y.}~\bibnamefont
  {Lu}}, \bibinfo {author} {\bibfnamefont {F.}~\bibnamefont {Ino}},\ and\
  \bibinfo {author} {\bibfnamefont {Y.}~\bibnamefont {Matsushita}},\ }\href
  {https://doi.org/10.48550/arXiv.1706.07191} {\bibinfo {title}
  {High-performance out-of-core block randomized singular value decomposition
  on gpu}} (\bibinfo {year} {2017}),\ \Eprint
  {https://arxiv.org/abs/1706.07191} {arXiv:1706.07191 [cs]} \BibitemShut
  {NoStop}%
\bibitem [{\citenamefont {Gabriel}\ \emph {et~al.}(2004)\citenamefont
  {Gabriel}, \citenamefont {Fagg}, \citenamefont {Bosilca}, \citenamefont
  {Angskun}, \citenamefont {Dongarra}, \citenamefont {Squyres}, \citenamefont
  {Sahay}, \citenamefont {Kambadur}, \citenamefont {Barrett}, \citenamefont
  {Lumsdaine}, \citenamefont {Castain}, \citenamefont {Daniel}, \citenamefont
  {Graham},\ and\ \citenamefont {Woodall}}]{Gabriel2004OpenMpiGoals}%
  \BibitemOpen
  \bibfield  {author} {\bibinfo {author} {\bibfnamefont {E.}~\bibnamefont
  {Gabriel}}, \bibinfo {author} {\bibfnamefont {G.~E.}\ \bibnamefont {Fagg}},
  \bibinfo {author} {\bibfnamefont {G.}~\bibnamefont {Bosilca}}, \bibinfo
  {author} {\bibfnamefont {T.}~\bibnamefont {Angskun}}, \bibinfo {author}
  {\bibfnamefont {J.~J.}\ \bibnamefont {Dongarra}}, \bibinfo {author}
  {\bibfnamefont {J.~M.}\ \bibnamefont {Squyres}}, \bibinfo {author}
  {\bibfnamefont {V.}~\bibnamefont {Sahay}}, \bibinfo {author} {\bibfnamefont
  {P.}~\bibnamefont {Kambadur}}, \bibinfo {author} {\bibfnamefont
  {B.}~\bibnamefont {Barrett}}, \bibinfo {author} {\bibfnamefont
  {A.}~\bibnamefont {Lumsdaine}}, \bibinfo {author} {\bibfnamefont {R.~H.}\
  \bibnamefont {Castain}}, \bibinfo {author} {\bibfnamefont {D.~J.}\
  \bibnamefont {Daniel}}, \bibinfo {author} {\bibfnamefont {R.~L.}\
  \bibnamefont {Graham}},\ and\ \bibinfo {author} {\bibfnamefont {T.~S.}\
  \bibnamefont {Woodall}},\ }\bibfield  {title} {\bibinfo {title} {Open {MPI}:
  Goals, concept, and design of a next generation {MPI} implementation},\ }in\
  \href@noop {} {\emph {\bibinfo {booktitle} {Proceedings, 11th European
  PVM/MPI Users' Group Meeting}}}\ (\bibinfo {address} {Budapest, Hungary},\
  \bibinfo {year} {2004})\ pp.\ \bibinfo {pages} {97--104}\BibitemShut
  {NoStop}%
\bibitem [{\citenamefont {Stoudenmire}\ and\ \citenamefont
  {White}(2013)}]{Stoudenmire2013RealSpaceParallel}%
  \BibitemOpen
  \bibfield  {author} {\bibinfo {author} {\bibfnamefont {E.~M.}\ \bibnamefont
  {Stoudenmire}}\ and\ \bibinfo {author} {\bibfnamefont {S.~R.}\ \bibnamefont
  {White}},\ }\bibfield  {title} {\bibinfo {title} {Real-space parallel density
  matrix renormalization group},\ }\href
  {https://doi.org/10.1103/PhysRevB.87.155137} {\bibfield  {journal} {\bibinfo
  {journal} {Phys. Rev. B}\ }\textbf {\bibinfo {volume} {87}},\ \bibinfo
  {pages} {155137} (\bibinfo {year} {2013})}\BibitemShut {NoStop}%
\bibitem [{\citenamefont {Secular}\ \emph {et~al.}(2020)\citenamefont
  {Secular}, \citenamefont {Gourianov}, \citenamefont {Lubasch}, \citenamefont
  {Dolgov}, \citenamefont {Clark},\ and\ \citenamefont
  {Jaksch}}]{Secular2020ParallelTimeDependent}%
  \BibitemOpen
  \bibfield  {author} {\bibinfo {author} {\bibfnamefont {P.}~\bibnamefont
  {Secular}}, \bibinfo {author} {\bibfnamefont {N.}~\bibnamefont {Gourianov}},
  \bibinfo {author} {\bibfnamefont {M.}~\bibnamefont {Lubasch}}, \bibinfo
  {author} {\bibfnamefont {S.}~\bibnamefont {Dolgov}}, \bibinfo {author}
  {\bibfnamefont {S.~R.}\ \bibnamefont {Clark}},\ and\ \bibinfo {author}
  {\bibfnamefont {D.}~\bibnamefont {Jaksch}},\ }\bibfield  {title} {\bibinfo
  {title} {Parallel time-dependent variational principle algorithm for matrix
  product states},\ }\href {https://doi.org/10.1103/PhysRevB.101.235123}
  {\bibfield  {journal} {\bibinfo  {journal} {Phys. Rev. B}\ }\textbf {\bibinfo
  {volume} {101}},\ \bibinfo {pages} {235123} (\bibinfo {year}
  {2020})}\BibitemShut {NoStop}%
\bibitem [{\citenamefont {Cataldi}\ \emph {et~al.}(2021)\citenamefont
  {Cataldi}, \citenamefont {Abedi}, \citenamefont {Magnifico}, \citenamefont
  {Notarnicola}, \citenamefont {Pozza}, \citenamefont {Giovannetti},\ and\
  \citenamefont {Montangero}}]{Cataldi2021hilbertcurvevs}%
  \BibitemOpen
  \bibfield  {author} {\bibinfo {author} {\bibfnamefont {G.}~\bibnamefont
  {Cataldi}}, \bibinfo {author} {\bibfnamefont {A.}~\bibnamefont {Abedi}},
  \bibinfo {author} {\bibfnamefont {G.}~\bibnamefont {Magnifico}}, \bibinfo
  {author} {\bibfnamefont {S.}~\bibnamefont {Notarnicola}}, \bibinfo {author}
  {\bibfnamefont {N.~D.}\ \bibnamefont {Pozza}}, \bibinfo {author}
  {\bibfnamefont {V.}~\bibnamefont {Giovannetti}},\ and\ \bibinfo {author}
  {\bibfnamefont {S.}~\bibnamefont {Montangero}},\ }\bibfield  {title}
  {\bibinfo {title} {Hilbert curve vs {H}ilbert space: exploiting fractal 2{D}
  covering to increase tensor network efficiency},\ }\href
  {https://doi.org/10.22331/q-2021-09-29-556} {\bibfield  {journal} {\bibinfo
  {journal} {{Quantum}}\ }\textbf {\bibinfo {volume} {5}},\ \bibinfo {pages}
  {556} (\bibinfo {year} {2021})}\BibitemShut {NoStop}%
\bibitem [{qte()}]{qtealeaves}%
  \BibitemOpen
  \href@noop {} {\bibinfo {title} {Quantum tea leaves: Customized version
  b5bc166c which is not public yet; optimized towards large bond dimensions
  moving data between {CPU} and {GPU} on more levels than available
  versions.}},\ \bibinfo {howpublished}
  {\url{https://baltig.infn.it/quantum\_tea\_leaves/py\_api\_quantum\_tea\_leaves}}\BibitemShut
  {NoStop}%
\bibitem [{dib()}]{dibona}%
  \BibitemOpen
  \href@noop {} {\bibinfo {title} {{Textarossa Dibona: using node of the
  EuroHPC project with Nvidia A100 GPU with 40GB memory.}}}\BibitemShut {Stop}%
\bibitem [{\citenamefont {Verstraete}\ \emph {et~al.}(2004)\citenamefont
  {Verstraete}, \citenamefont {Garc\'{\i}a-Ripoll},\ and\ \citenamefont
  {Cirac}}]{Verstraete2004MatrixProductDensity}%
  \BibitemOpen
  \bibfield  {author} {\bibinfo {author} {\bibfnamefont {F.}~\bibnamefont
  {Verstraete}}, \bibinfo {author} {\bibfnamefont {J.~J.}\ \bibnamefont
  {Garc\'{\i}a-Ripoll}},\ and\ \bibinfo {author} {\bibfnamefont {J.~I.}\
  \bibnamefont {Cirac}},\ }\bibfield  {title} {\bibinfo {title} {Matrix product
  density operators: Simulation of finite-temperature and dissipative
  systems},\ }\href {https://doi.org/10.1103/PhysRevLett.93.207204} {\bibfield
  {journal} {\bibinfo  {journal} {Phys. Rev. Lett.}\ }\textbf {\bibinfo
  {volume} {93}},\ \bibinfo {pages} {207204} (\bibinfo {year}
  {2004})}\BibitemShut {NoStop}%
\bibitem [{\citenamefont {Zwolak}\ and\ \citenamefont
  {Vidal}(2004)}]{Zwolak2004MixedStateDynamics}%
  \BibitemOpen
  \bibfield  {author} {\bibinfo {author} {\bibfnamefont {M.}~\bibnamefont
  {Zwolak}}\ and\ \bibinfo {author} {\bibfnamefont {G.}~\bibnamefont {Vidal}},\
  }\bibfield  {title} {\bibinfo {title} {Mixed-state dynamics in
  one-dimensional quantum lattice systems: A time-dependent superoperator
  renormalization algorithm},\ }\href
  {https://doi.org/10.1103/PhysRevLett.93.207205} {\bibfield  {journal}
  {\bibinfo  {journal} {Phys. Rev. Lett.}\ }\textbf {\bibinfo {volume} {93}},\
  \bibinfo {pages} {207205} (\bibinfo {year} {2004})}\BibitemShut {NoStop}%
\bibitem [{\citenamefont {Werner}\ \emph {et~al.}(2016)\citenamefont {Werner},
  \citenamefont {Jaschke}, \citenamefont {Silvi}, \citenamefont {Kliesch},
  \citenamefont {Calarco}, \citenamefont {Eisert},\ and\ \citenamefont
  {Montangero}}]{Werner2016PositiveTensorNetwork}%
  \BibitemOpen
  \bibfield  {author} {\bibinfo {author} {\bibfnamefont {A.~H.}\ \bibnamefont
  {Werner}}, \bibinfo {author} {\bibfnamefont {D.}~\bibnamefont {Jaschke}},
  \bibinfo {author} {\bibfnamefont {P.}~\bibnamefont {Silvi}}, \bibinfo
  {author} {\bibfnamefont {M.}~\bibnamefont {Kliesch}}, \bibinfo {author}
  {\bibfnamefont {T.}~\bibnamefont {Calarco}}, \bibinfo {author} {\bibfnamefont
  {J.}~\bibnamefont {Eisert}},\ and\ \bibinfo {author} {\bibfnamefont
  {S.}~\bibnamefont {Montangero}},\ }\bibfield  {title} {\bibinfo {title}
  {Positive tensor network approach for simulating open quantum many-body
  systems},\ }\href {https://doi.org/10.1103/PhysRevLett.116.237201} {\bibfield
   {journal} {\bibinfo  {journal} {Phys. Rev. Lett.}\ }\textbf {\bibinfo
  {volume} {116}},\ \bibinfo {pages} {237201} (\bibinfo {year}
  {2016})}\BibitemShut {NoStop}%
\bibitem [{\citenamefont {Kliesch}\ \emph {et~al.}(2014)\citenamefont
  {Kliesch}, \citenamefont {Gross},\ and\ \citenamefont
  {Eisert}}]{Kliesch2014MatrixProductOperators}%
  \BibitemOpen
  \bibfield  {author} {\bibinfo {author} {\bibfnamefont {M.}~\bibnamefont
  {Kliesch}}, \bibinfo {author} {\bibfnamefont {D.}~\bibnamefont {Gross}},\
  and\ \bibinfo {author} {\bibfnamefont {J.}~\bibnamefont {Eisert}},\
  }\bibfield  {title} {\bibinfo {title} {Matrix-product operators and states:
  Np-hardness and undecidability},\ }\href
  {https://doi.org/10.1103/PhysRevLett.113.160503} {\bibfield  {journal}
  {\bibinfo  {journal} {Phys. Rev. Lett.}\ }\textbf {\bibinfo {volume} {113}},\
  \bibinfo {pages} {160503} (\bibinfo {year} {2014})}\BibitemShut {NoStop}%
\bibitem [{\citenamefont {Arceci}\ \emph {et~al.}(2022)\citenamefont {Arceci},
  \citenamefont {Silvi},\ and\ \citenamefont
  {Montangero}}]{Arceci2022EntanglementFormationMixed}%
  \BibitemOpen
  \bibfield  {author} {\bibinfo {author} {\bibfnamefont {L.}~\bibnamefont
  {Arceci}}, \bibinfo {author} {\bibfnamefont {P.}~\bibnamefont {Silvi}},\ and\
  \bibinfo {author} {\bibfnamefont {S.}~\bibnamefont {Montangero}},\ }\bibfield
   {title} {\bibinfo {title} {Entanglement of {{Formation}} of {{Mixed
  Many-Body Quantum States}} via {{Tree Tensor Operators}}},\ }\href
  {https://doi.org/10.1103/PhysRevLett.128.040501} {\bibfield  {journal}
  {\bibinfo  {journal} {Physical Review Letters}\ }\textbf {\bibinfo {volume}
  {128}},\ \bibinfo {pages} {040501} (\bibinfo {year} {2022})}\BibitemShut
  {NoStop}%
\end{thebibliography}%
\end{document}